\documentclass[fleqn,usenatbib]{mnras}

\usepackage{newtxtext,newtxmath}
\usepackage[normalem]{ulem}

\usepackage[T1]{fontenc}
\usepackage{ae,aecompl}
\usepackage[dvipsnames]{xcolor}
\usepackage[toc,page]{appendix}

\usepackage{graphicx}	
\usepackage{amsmath}	
\usepackage{booktabs}
\usepackage{longtable}
\usepackage{pdflscape}
\usepackage{caption}
\usepackage{soul}

\title[2024/25 Outburst from EXO 0748--676]{A Multi-Wavelength Picture of the Surprisingly Short 2024/25 Outburst from EXO 0748--676}

\author[A. H. Knight \& L. Rhodes]{
Amy H. Knight,$^{1}$ \thanks{E-mail: amy.h.knight@durham.ac.uk} \footnote[3]{}
Lauren Rhodes,$^{2,3}$\thanks{E-mail: lauren.rhodes@mcgill.ca} \thanks{These authors contributed equally to this work.}
Noel Castro Segura,$^{4}$
Douglas J. K. Buisson,$^{5}$
Aru Beri,$^{6, 7}$
\and \ David M. Russell,$^{8}$
Kevin Alabarta,$^{9}$
Maria C. Baglio,$^{9}$
Jakob van den Eijnden,$^{10}$
 Matthew J. Middleton,$^{7}$
\and \ Payaswini Saikia,$^{11}$
Chris Done, $^{1}$
Rob Fender,$^{12, 13}$
and Fraser Lewis,$^{14, 15}$
\\
$^{1}$Centre for Extragalactic Astronomy, Department of Physics, Durham University, South Road, Durham DH1 3LE, UK\\
$^{2}$Trottier Space Institute at McGill, 3550 Rue University, Montreal, Quebec H3A 2A7, Canada\\
$^{3}$Department of Physics, McGill University, 3600 Rue University, Montreal, Quebec H3A 2T8, Canada\\
$^{4}$Department of Physics, Gibbet Hill Road, University of Warwick, Coventry, CV4 7AL, United Kingdom\\
$^{5}$Independent Researcher\\
$^{6}$Indian Institute of Astrophysics, Koramangala II Block, Bangalore 560034, India\\
$^{7}$Department of Physics and Astronomy, University of Southampton, Highfield, Southampton SO17 1BJ, UK\\
$^{8}$Center for Astrophysics and Space Science (CASS), New York University Abu Dhabi, PO Box 129188 Abu Dhabi, UAE\\
$^{9}$INAF, Osservatorio Astronomico di Brera, Via E. Bianchi 46, I-23807 Merate (LC), Italy\\
$^{10}$Anton Pannekoek Institute for Astronomy, Universiteit van Amsterdam, Science Park 904, NL-1098 XH Amsterdam, the Netherlands\\
$^{11}$Department of Astronomy, Yale University, PO Box 208101, New Haven, CT 06520-8101, USA\\
$^{12}$Astrophysics, Department of Physics, The University of Oxford, Keble Road, Oxford, OX1 3RH, UK.\\
$^{13}$Department of Astronomy, University of Cape Town, Private Bag X3, Rondebosch 7701, South Africa\\
$^{14}$ Faulkes Telescope Project, Cardiff, Wales, UK\\
$^{15}$ The Schools' Observatory, Astrophysics Research Institute, Liverpool John Moores University, 146 Brownlow Hill, Liverpool L3 5RF, UK\\
}
\date{Accepted XXX. Received YYY; in original form ZZZ}

\pubyear{\the\year{}}

\begin{document}
\label{firstpage}
\pagerange{\pageref{firstpage}--\pageref{lastpage}}
\maketitle

\begin{abstract}
After sixteen years in quiescence, EXO 0748--676 unexpectedly returned to outburst in March 2024. However, in stark contrast to its previous outburst (1985 - 2008), EXO 0748--676 returned to quiescence in January 2025, after $\sim10$ months in outburst. Here, we collate data from nine observing facilities, covering wavelengths from radio to X-ray, to provide the first multi-wavelength study of the 2024/25 outburst. Unlike the 1985 outburst, both the rise and decay are fully captured at X-ray and optical wavelengths. The light curves do not exhibit the hallmarks of a canonical transient low-mass X-ray binary outburst. Instead, the X-ray and optical fluxes remain reasonably stable during the 2024/25 outburst. The wealth of optical data of EXO 0748-676 reveals a skewed optical phase curve and an irradiation-dominated optical light curve which lasts until the onset of hydrogen recombination in the outer disc at $\sim 11000\mathrm{K}$, triggering a rapid fade at the end of the outburst. X-ray observations by \textit{XMM-Newton} and \textit{NICER} revealed that EXO 0748--676 was heavily obscured early on, with spectra exhibiting a dampened soft X-ray contribution. We find that a spectral model including a clumpy, ionised absorber describes the obscuration, although the exact cause of the early-time obscuration remains uncertain. X-ray colour analysis indicates that EXO 0748--676 spent most of the $\sim 10$ month outburst in a faint, hard state. EXO 0748-676 increased marginally in brightness towards the end of the outburst, but never entered a soft spectral state. This behaviour is characteristic of a failed transition outburst.

\end{abstract}

\begin{keywords}
X-rays: binaries -- accretion, accretion discs -- stars: neutron
\end{keywords}


\section{Introduction}
Since its discovery in 1985 by the European X-ray Observatory Satellite (\textit{EXOSAT}; \citealt{Parmar1986}), the neutron star (NS) low-mass X-ray binary (LMXB) EXO 0748--676 (hereafter EXO 0748) has been the focus of numerous studies both during its first recorded X-ray outburst, which lasted $\sim 24$ years, and during its first observed quiescence (late 2008 - June 2024; see e.g. \citealt{Hynes2009, Degenaar2011}). Early \textit{EXOSAT} observations revealed type I X-ray bursts \citep{Gottwald1986} and X-ray eclipses \citep{Parmar1991}, confirming that EXO 0748 hosts an accreting NS and that the system is highly inclined. EXO 0748's optical counterpart, UY Vol, was discovered shortly after the initial X-ray detection \citep{Wade1985}. To the authors' knowledge, there were no radio observations during the 1985 outburst.

Between 1985 and 2008, EXO 0748 was regularly observed at X-ray wavelengths by the Rossi X-ray Timing Explorer (\textit{RXTE}; see e.g. \citealt{Wolff2009, Knight2023}) and the X-ray Multi-Mirror Mission (\textit{XMM-Newton}; see e.g. \citealt{Bonnet-Bidaud2001, Homan2003}). As with the \textit{EXOSAT} observations, the \textit{RXTE} monitoring captured X-ray eclipses lasting $\approx 500$ s that recur on the orbital period of $P=3.824$ hrs (see e.g \citealt{Wolff2009, Knight2023} for discussion of the X-ray eclipses) and a variety of type I X-ray bursts \citep[e.g.][]{Galloway2008, Galloway2020}. Corresponding eclipses were also observed at optical frequencies, as reported by \citet{Crampton1986}. Among the detected bursts are photospheric radius expansion bursts \citep[e.g.][]{Wolff2005}, reprocessed X-ray bursts \citep[e.g.][]{Paul2012}, and so-called in-eclipse X-ray bursts \citep{Knight2025_ineclipse, Rikame2025}. Further analysis of the X-ray bursts determined that the system resides at a distance of $7.1 \pm 0.1$~kpc \citep{Galloway2008b}. The NS spin frequency is estimated to be within a few Hertz of the $552$ Hz burst oscillation frequency detection by \citet{Galloway2010}.

The X-ray eclipses in EXO 0748 arise due to the occultation of the $\sim 2 M_{\odot}$ NS \citep{Ozel2006, Knight2022a} and accretion disc by the $\sim 0.4 M_{\odot}$ M-dwarf companion \citep{Parmar1986, Parmar1991}. However, the observed eclipse transitions (ingresses and egresses) and totality durations are highly variable on both long (years) and short (single orbit) timescales. The early \textit{EXOSAT} observations revealed ingress and egress durations ranging from less than one second to more than thirty seconds. \citet{Parmar1991} suggested that an X-ray-induced evaporative wind may explain the drastic variability among the eclipse transitions, as it can sufficiently increase the atmospheric scale height of the main-sequence companion and thus produce long eclipse transitions. Subsequent studies of the eclipses observed by \textit{XMM-Newton} and \textit{RXTE} confirmed the eclipse transition variability (see e.g. \citealt{Wolff2009, Knight2023}) and also revealed that the X-ray eclipses are energy-dependent and asymmetric (\citealt{Knight2022a}). The properties of EXO 0748's X-ray eclipses are analogous to the radio eclipses observed in spider pulsars, which led to the discovery that the companion star in EXO 0748 is being strongly irradiated by the incident X-rays, causing the liberation of its surface material; a process known as ablation. This discovery ultimately resulted in EXO 0748 being sub-classified as a false widow \citep{Knight2023}.

Formally, false widows are NS binaries (i.e. X-ray binaries) that ablate their companion star via X-ray irradiation while actively accreting \citep{Knight2023}. They are named in reference to a species of arachnid, to highlight that the observable process of ablation is universal among `spider' systems (false widows and spider pulsars; redbacks and black widows, respectively distinguished by companion masses, M$_{\rm{c}}$ $\approx 0.2 -0.4$ M$_{\odot}$ and M$_{\rm{c}} << 0.1$ M$_{\odot}$; e.g. \citealt{Roberts2012}), but is less progressed in false widows. However, we note that even though ablation is clearly less progressed in false widows than spider pulsars, there is significant evidence of ablation in these sources that is not universal among LMXBs, thus making false widows distinct from the broader population of accreting NS systems. Absorption by the ionised, clumpy ablated material naturally explains X-ray eclipses exhibited by EXO 0748 (and other false widows; see e.g. \citealt{Knight2022b}). Furthermore, false widow systems may represent an intermediate evolutionary stage between NS LXMBs and spider pulsars, as this route extends the timescales over which ablation can occur.

The premise of this hypothesis is that the ablation of the companion star begins while the source is actively accreting and continues throughout the spider pulsar phase, leading to a short-period binary with an undermassive companion like those observed in spider pulsars \citep{Knight2023}. In turn, these systems may evolve into an isolated millisecond pulsar once the companion star is completely consumed \citep{Ruderman1989a}. Ablation driven by pulsar radiation is not an efficient mass-loss mechanism, which makes it difficult to produce an isolated millisecond pulsar on the correct evolutionary timescale \citep{Polzin2018, Polzin2020}. Although if the ablation begins in the earlier false widow phase, the total time over which ablation can occur gets extended, which may make the aforementioned evolutionary pathway feasible. Further support for EXO 0748's false widow classification comes from the detection of a broad C IV emission line by \citet{Parikh2021}, who draw similarities between their quiescent observations of EXO 0748 and the known transitional redback pulsar, PSR J1023$+$0038, in its rotation-powered state, and the reversal of the eclipse asymmetry uncovered by \citet{Knight2023}, which is only otherwise observed in the black widow pulsar PSR J2051--0827 \citep{Polzin2019a}.

EXO 0748 entered its first known X-ray quiescence in 2008, during which follow-up studies revealed the relatively slow cooling of the NS crust \citep[c.f.][]{Degenaar2009, Degenaar2011, DiazTrigo:2011A&A...528A.150D}. Several optical spectroscopic studies were performed in quiescence, resulting in a dynamically constrained NS mass of $\mathrm{M_{NS} > 1.27\,M_\odot}$ (\citealt{Bassa2009}; although the mass of the NS may be closer to $2$ M$_{\odot}$ e.g. \citealt{Knight2022a, Ozel2006}) and the discovery of irradiation-driven emission lines, arising from the irradiated face of the companion star \citep{Ratti:2012MNRAS.420...75R}, modulated on the orbital period \citep{Pearson2006, MD2009, Bassa2009}. \citet{Ratti:2012MNRAS.420...75R} concluded that a likely cause of the extra rotational broadening is an outflow from the companion star, driven by a pulsar wind and/or X-ray heating. Combined with a tentative small phase lag with respect to hydrogen lines, the authors suggest that the emission originates near the trailing face of the donor, where material is being ablated, in agreement with EXO 0748's false widow classification. 

EXO 0748 indicated its return to outburst on 10th June 2024, after a $\sim 16$ year quiescence \citep{Baglio2024ATel16646, Rhodes2024}, with a type I X-ray burst \citep{Kuulkers2024} that triggered the Burst Alert System (BAT) on board the Neil Gherls Swift Observatory (\textit{Swift}), with a peak count rate of $\sim 500$ cts/sec in the 15-350 keV band (GCN \#36653). Roughly two months before the \textit{Swift}-BAT detected X-ray burst, the Asteroid Terrestrial-Impact Last Alert System, ATLAS, survey captured EXO 0748 rising in brightness \citep{Rhodes2024}. As X-ray binaries typically spend more time in quiescence than outburst, EXO 0748's return was somewhat unexpected. Following the initial X-ray burst detection, a global observing effort commenced, including observations by the Neutron Star Interior Composition Explorer, \textit{NICER}, and the Nuclear Spectroscopic Telescope Array, \textit{NuSTAR}, which captured several more type I X-ray bursts \citep{Kuulkers2024, Aoyama2024, Mihara2024, Bhattachrarya2024Paper, Bhattacharya2024Atel, Kashyap2024} and X-ray eclipses \citep{Buisson2024Atel}, the latter of which provided an update to the orbital ephemeris. Complementary optical and X-ray observations by \textit{XMM-Newton} captured X-ray eclipses and one reprocessed X-ray burst \citep{Knight2025}. \textit{AstroSat} also observed several type I X-ray bursts and eclipses \citep{2025arXiv251013556P}.

While its previous two-decade-long outburst could have suggested another prolonged outburst, EXO 0748 started to return to X-ray quiescence in January 2025, as the ATLAS light curve showed a substantial decay after just $\sim 10$ months in outburst \citep{Rhodes2025_Atel}. Subsequently, several \textit{Swift} X-ray Telescope (XRT) observations in May 2025 showed its measured temperatures and fluxes were consistent with quiescent NSs \citep{Degenaar2025_Atel}, thus confirming its return to X-ray quiescence. Here, we summarise the surprisingly short 2024/25 outburst of EXO 0748 utilising many datasets from the global multi-wavelength monitoring effort, to explore the differences between this $\sim 10$ month outburst and its previous $\sim 24$ year long outburst. In Section \ref{sx:obs} we detail the observations used, describe the corresponding data reduction procedures and provide a broad overview of the 2024/25 outburst (Section \ref{subsec:lightcurve}). In Section \ref{sx:x-ray_spec}, we explore the X-ray spectral behaviour of the source during the 2024/25 outburst and compare it to the 1985-2008 outburst. In Sections \ref{sx:lrlx} and \ref{sx:optical} we interpret the multi-wavelength observations utilising the radio/X-ray correlation, colour-magnitude diagram and the system's orbital phase dependence. We discuss our findings in \ref{sx:disc} and conclude in Section \ref{sx:summary}.


\section{Observations, Data Reduction and Preliminary Analysis}
\label{sx:obs}
To provide the first detailed multi-wavelength study of the 2024/25 outburst of EXO 0748, we collate data from nine facilities, covering wavelengths from radio to X-ray. Here, we detail the observations utilised from each facility, their respective data reduction procedures, and perform some preliminary analysis to produce a multi-wavelength light curve (Figure~\ref{fig:lightcurve}), summarising the outburst. 

\subsection{Radio}
\subsubsection{MeerKAT}
\label{subsec:Meerkat}

 \begin{table}
     \centering
     \begin{tabular}{cc}
     \hline
     Start Date $\&$ Time (UTC) & 3~$\sigma$ Upper Limit ($\mu$Jy) \\
\hline
2024-06-10 17:19:21 & 200 \\
2024-06-15 19:16:03 & 120 \\
2024-06-22 18:40:26 & 170 \\
2024-06-29 16:56:35 & 160 \\
2024-07-08 16:04:15 & 180 \\
2024-08-11 14:59:51 & 190 \\       
2024-10-14 11:18:05 & 170 \\
\hline
     \end{tabular}
     \caption{A log of the radio observations of EXO 0748--676 obtained with MeerKAT during the 2024/25 outburst.}
     \label{tab:radio_obs}
 \end{table}
MeerKAT is a 64-dish interferometer based in the Karoo Desert in South Africa. Each dish is $13.5$~m in diameter with a maximum baseline of $8$~km, enabling an angular resolution of $\sim5$~arcsec at $1.28$~GHz \citep{2016mks..confE...1J}. As part of the XKAT large observing program (PI: Fender), the MeerKAT radio telescope observed the field of EXO 0748 seven times between June and October 2024, as detailed in Table \ref{tab:radio_obs}. Each observation lasted 15 minutes and we indicate the times of these observations with black arrows in the top panel of Figure \ref{fig:lightcurve}. 

The data were reduced using \textsc{oxkat}, a series of semi-automated scripts designed for MeerKAT data-reduction \citep{oxkat}. \textsc{oxkat} performs flagging, calibration (including a round of phase-only self-calibration) and imaging of the calibrated data. We did not detect any radio emission at the position of EXO 0748 in any of the seven observations. Instead, we provide 3~$\sigma$ upper limits on the flux density in Table \ref{tab:radio_obs} and utilise these measurements in Section \ref{sx:lrlx}. The root mean square (RMS) noise values in each image were higher than expected for a point-source dominated field \citep[e.g.][]{2026MNRAS.546ag139T} due to the presence of a bright source on the edge of the primary beam. Therefore, we concatenated all seven observations to produce a tighter image. We did not detect a radio counterpart in the deeper image either. However, we obtained an image with an RMS noise of 42\,$\mu$Jy/beam, corresponding to a 3\,$\sigma$ upper limit of $\sim$120\,$\mu$Jy/beam.

\subsection{Optical/UV}

\subsubsection{LCO and Faulkes Telescopes}
\label{sec:LCO}

In June 2024, the Las Cumbres Observatory \citep[LCO;][]{Brown2013} 1-m network began regular optical observations of EXO 0748 \citep{Baglio2024ATel16646}. The LCO network telescopes at the South African Astronomical Observatory (SAAO) in South Africa, Siding Spring Observatory in Australia, and Cerro Tololo Interamerican Observatory in Chile, took images in the SDSS ${g}^{\prime}$ and ${i}^{\prime}$ filters (central wavelengths of 468.4 and 748.0\,nm, respectively). A source consistent with the coordinates of EXO 0748 was detected between 2024-06-10 and 2025-03-21 (MJD 60471--60755), after which the source faded to undetectable flux levels. Monitoring continued with the 1-m network until 2025-04-27 (MJD 60792). On 2025-05-08 (MJD 60803), the 2-m Faulkes Telescope South at Siding Spring Observatory, Australia, observed EXO 0748. The 4-channel imager MuSCAT4 \citep[Multicolor Simultaneous Camera for studying Atmospheres of Transiting exoplanets 4;][]{Narita2020} was used, producing simultaneous images in the SDSS ${g}^{\prime}$, ${r}^{\prime}$ (616.6\,nm), ${i}^{\prime}$ and $z_{\rm s}$ (893.2\,nm) passbands (all magnitudes are in the AB system). The optical monitoring is part of a long-term program observing $\sim 50$ LMXBs coordinated by the Faulkes Telescope Project \citep{Lewis2008, Lewis2018}.

The images were processed using the X-ray Binary New Early Warning System (\textsc{XB-NEWS}), a real-time data analysis pipeline \citep[e.g.][]{Russell2019, Goodwin2020, Saikia2026}. The \textsc{XB-NEWS} pipeline performed standard data reduction using LCO calibration files and several quality control steps to ensure that only good quality images are analysed. Following this, an astrometric solution was computed for each image using positions from the Gaia DR2 catalogue\footnote{\url{https://www.cosmos.esa.int/web/gaia/dr2}}. The coordinates of the source, using all detections, were found to be RA = 07\:48\:33.71, Dec = -67\:45\:07.7 (J2000), with systematic errors of the order of $\lesssim 0.1$''. This is consistent with the known position of EXO 0748. Aperture photometry of all the stars in each image was achieved, solving for zero-point calibrations between epochs \citep{Bramich2012}. Flux calibration of the photometry used an extended version of the ATLAS All-sky Stellar Reference Catalogue \citep[ATLAS-REFCAT2;][]{Tonry2018}. \textsc{XB-NEWS} then performed multi-aperture photometry \citep[azimuthally-averaged PSF profile fitting photometry;][]{Stetson1990} on the target in each reduced image. We considered magnitude errors larger than $\sim 0.25$ mag as marginal detections (or non-detections) and discarded them. The resulting magnitudes were in the range ${g}^{\prime} = 17.56$--$19.35$; ${i}^{\prime} = 17.71$--$19.20$ for the majority of the outburst (with errors in the range 0.01--0.09 mag), fading to ${g}^{\prime} = 20.43 \pm 0.20$ on 2025-03-21 (MJD 60755). We detected emission in $41/51$ ${i}^{\prime}$-band observations and $40/49$ g'-band observations. Only one observation occurred with each of the $z_{\rm s}$ and ${r}^{\prime}$ filters, resulting in upper limits. The deepest $3\sigma$ upper limits during quiescence were ${g}^{\prime} > 21.01$; ${r}^{\prime} > 21.09$; ${i}^{\prime} > 20.33$; $z_{\rm s} > 19.68$. The light curve for these data is shown in the upper panel of Figure \ref{fig:lightcurve}.

\subsubsection{ATLAS}
The Asteroid Terrestrial-impact Last Alert System, ATLAS, consists of four robotic telescopes, one in South Africa, one in Chile and two in Hawaii \citep{2018PASP..130f4505T}. Every night, the ATLAS telescopes scan the sky performing image subtraction to identify both moving near-Earth objects and static Galactic and extragalactic transients \citep{2020PASP..132h5002S}. The telescopes observe using two filters: \textit{cyan} (420--650\,nm) and \textit{orange} (560--820\,nm). ATLAS data is publicly available via a forced photometry server \citep{2021TNSAN...7....1S}.

Following the report by \citet{2024GCN.36653....1D}, we queried the ATLAS forced photometry server at the position of EXO 0748 (RA: 07h:48m:33.71s, DEC-67:45:07.70) and reported the rise of an optical counterpart \citep{Rhodes2024}. We utilised the difference-imaging version of the magnitudes, with a reference image from before the outburst began. EXO 0748 is too faint during quiescence to be detected by ATLAS (the $5\sigma$ limit is 19.7\, magnitudes in both bands) and so will not significantly affect the measurements obtained through difference imaging. We found that ATLAS observed the position of EXO 0748 roughly every two to three days throughout the 2024/25 outburst, which allowed us to report that the source showed signs of returning to quiescence in January 2025 \citep{RhodesKnight2025ATEL}. We show the ATLAS observations in the top panel of Figure \ref{fig:lightcurve}.

\subsubsection{Swift-UVOT}
The Ultra-Violet Optical Telescope \citep[UVOT;][]{roming04} onboard \textit{Swift} can observe using six bands (\textit{v}, \textit{b}, \textit{u}, \textit{uvw1}, \textit{uvm2} and \textit{uvw2}) with central wavelengths spanning 5468 to 1928\r{A}\footnote{\url{https://www.swift.ac.uk/analysis/uvot/filters.php}}. As part of the global monitoring effort, we requested two separate target-of-opportunity (ToO) observations of EXO 0748. A further nine ToO observations occurred at the request of other teams both during and after the 2024/25 outburst (ToO IDs 21971, 21987, 22377, 22483; target IDs 19755 \footnote{\url{https://www.swift.psu.edu/operations/obsSchedule.php?t=19755}} and 31272 \footnote{\url{https://www.swift.psu.edu/operations/obsSchedule.php?t=31272}}). The flux densities were measured from the standard UVOT images using \texttt{uvotsource}. We applied a 5\,arcsec radius circular source region centred on the known coordinates of EXO 0748, and a 20\,arcsec radius circular region on a nearby, starless portion of the sky to model the background. We provide a summary of these observations in Table~\ref{tab:swift}, and we show the measured flux densities in the top panel of Figure~\ref{fig:lightcurve}.

\begin{table}
\setlength{\tabcolsep}{3pt}
    \centering
    \begin{tabular}{cccc}
    \hline
    ObsID & Start Date \& Time & End Date \& Time & UVOT Filters \\
    \hline
    \vspace*{0.1cm}
    1236151000	& 2024-06-10 07:37:35 & 2024-06-10 07:55:59 & v \\
    \vspace*{0.1cm}
    0031272058 & 2024-06-22 00:33:00 & 2024-06-22 00:50:00 & uvw1 \\
    \vspace*{0.1cm}
    0031272059 & 2024-06-29 01:04:00 & 2024-06-29 01:24:00 & uvm2 \\
    \vspace*{0.1cm}
    0031272060 & 2024-07-13 04:55:00 & 2024-07-13 11:26:00 & u \\
    \vspace*{0.1cm}
    0031272061	& 2024-07-20 00:55:00 & 2024-07-20 01:09:00 & uvw1 \\
    \vspace*{0.1cm}
    0031272062 & 2025-02-07 12:27:00 & 2025-02-07 16:02:00 & uvw2 \\
    \vspace*{0.1cm}
    0031272063 & 2025-02-12 09:46:00 & 2025-02-12 09:54:00 & uvw1, uvw2 \\
    \vspace*{0.1cm}
    0031272064 & 2025-04-13 05:16:00 & 2025-04-13 05:35:00 & uvw1, uvw2 \\
    \vspace*{0.1cm}
    0019755001 & 2025-05-08 09:26:00  & 2025-05-08 20:36:00 & uvw1 \\
    \vspace*{0.1cm}
    0019755002 & 2025-05-11 07:39:00 & 2025-05-11 18:52:00	 & uvm2 \\
    \vspace*{0.1cm}
    0019755003 & 2025-05-15 05:02:00 & 2025-05-15 13:14:00 & uvm2 \\
    \vspace*{0.1cm}
    0019755004 & 2025-05-19 04:04:00 & 2025-05-19 15:26:00 & uvm2 \\
    \hline
    \end{tabular}
    \vspace*{-0.25cm}
    \caption{A summary of the \textit{Swift} observations of EXO 0748--676 during the 2024/25 outburst.}
    \label{tab:swift}
\end{table}

\subsubsection{XMM-Newton Optical Monitor}
\label{subsubsec:OM}
The optical monitor (OM; \citealt{2001A&A...365L..36M}) on board \textit{XMM-Newton} observed simultaneously with the European Photon Imaging Camera (EPIC) on 2024-06-30 (ObsID 0935191701), cycling through the \textit{U}, \textit{V} and \textit{B} filters. We follow the data reduction procedure described in section 2 of \citet{Knight2025}. However, we provide a brief overview here for completeness. We reduce the \textit{U}, \textit{V} and \textit{B} band OM exposures using the \textsc{xmm-sas} processing chain \texttt{omfchain}, leaving most parameters to their default values. We utilise circular source regions of 6 arcsec radius around the optical counterpart to extract light curves with 60~s binning and subsequently apply barycentric corrections with \texttt{barycen}. We analysed the reprocessed X-ray burst present in these data in \citet{Knight2025}. In this work, they are used to explore the orbital phase-folded modulation in Section \ref{subsec:phase}.

\subsection{X-ray}
\label{subsec:X-ray}
The pointed X-ray observations presented here (\textit{XMM-Newton}, \textit{Swift-XRT}, \textit{NICER}, \textit{AstroSat} and \textit{NuStar}) are supplemented with public data from the Monitor of All-sky X-ray Image (\textit{MAXI}) Gas Slit Camera (GSC) and the \textit{Swift} Burst Alert Telescope (BAT) Hard X-ray Transient Monitor which observe between 2–20\,keV and  15–50\,keV, respectively \citep{2011PASJ...63S.677H, 2011PASJ...63S.635S, 2013ApJS..209...14K}. The \textit{MAXI} and \textit{Swift}-BAT count rates in the above bands are converted to $0.5-10$ keV fluxes using energy conversion factors (ECFs) detailed in Appendix \ref{sx:x-ray_appendix}, Tables \ref{tb:maxi_conversion} and \ref{tb:swiftbat_conversion} respectively. As noted in Appendix \ref{sx:x-ray_appendix}, the ECFs for \textit{MAXI} are derived using the spectral models of the closest-in-time pointed observation to the \textit{MAXI} time bin. The same procedure is used for the \textit{Swift}-BAT conversions, except we only utilise the spectral models of the closest-in-time pointed observation with a strong hard/high energy component. This method provides a more accurate view of the source flux during the $\sim 10$ month outburst than assuming a constant ECF.

\subsubsection{AstroSat}

\emph{AstroSat} observed EXO 0748 on 2024 July 24 (MJD 60515.33), corresponding to 2024-07-24 08:01:03 UTC. We utilise this observation to obtain a broadband X-ray spectrum of the source during its 2024/25 outburst.

The Large Area X-ray Proportional Counter \citep[LAXPC;][]{Yadav2016, Antia2017, Antia2021} observations were obtained in Event Analysis (EA) mode. Due to the gain instability caused by the gas leakage, \textsc{LAXPC10} data were not used while \textsc{LAXPC30} was switched off during these observations\footnote{LAXPC30 is switched off since 8 March 2018, refer to \url{http://astrosat-ssc.iucaa.in/}}. We utilsied the~\textsc{LaxpcSoft} v3.3\footnote{\url{http://www.tifr.res.in/~astrosat\_laxpc/LaxpcSoft.html}} software package to extract light curves and spectra. Background files are generated using the blank sky observations \citep[see,][for details]{Antia2017}. The appropriate response files were used for spectral analysis.

The Soft X-ray Telescope \citep[SXT;][]{Singh2014, Singh2017} observed EXO 0748 in Fast Window mode. We used the Level-2 cleaned event files generated by the standard \textit{AstroSat} processing pipeline. The cleaned event files from individual orbits were merged using the using \textsc{SXT} Event Merger Tool (Julia Code\footnote{\url{http://www.tifr.res.in/~astrosat\_sxt/dataanalysis.html}}), and analysed using the \texttt{xselect} task within \textsc{heasoft} to extract a source image and spectra. A circular region with radius of 5~arcmin centered on the source was used to extract source events. The standard background spectrum, redistribution matrix file, and ancillary response file supplied by the SXT team were used for subsequent analysis.

The \textsc{SXT} and \textsc{LAXPC} spectra were grouped to have at least 25 counts/bin and were fitted simultaneously. To account for cross-calibration differences between the instruments, we adopted a spectral model of the form \texttt{constant*tbabs*zxipcf*(cutoffpl+bbodyrad)}, with the \textsc{SXT} normalization fixed at unity and the LAXPC20 normalization allowed to vary. The equivalent neutral hydrogen column density was fixed at $N_{\rm H}=0.1\times10^{22}$ cm$^{-2}$, while all other model parameters were allowed to vary freely. From the best-fitting model, we derived an unabsorbed flux of $1.53\times10^{-10}$ erg cm$^{-2}$ s$^{-1}$ in the 0.5--10 keV energy band. This flux measurement is included in the long-term X-ray light curve shown in the lower panel of Figure~\ref{fig:lightcurve}.

\subsubsection{Swift-XRT}
We utilised all \textit{Swift-XRT} exposures between 2024 June 16 and 2025 May 19 (MJD 60477 and 60814) and provide details of these exposures in Table \ref{tab:swift}. We first ran the \textsc{xrtpipeline} on the Level-1 event files, keeping all arguments to their default values and specifying the source position as RA 07h 48m 33.7s and Dec -67$^{\circ}$ 45' 07.7", to obtain the Level-2 cleaned event lists. For each observation, we extracted images using \textsc{xselect} to determine suitable source and background regions. In all cases, these were circles with radii of $30$ and $60$\,arcsec, respectively. Subsequently, we utilised these regions to extract spectra from each observation in the $0.5-10$ keV band. We build Ancillary Response Files (ARFs) using 
\texttt{xrtmkarf} and bin the spectra to have at least one count per bin with \textsc{grppha}, making the spectra suitable for fitting with the $W$-statistic within \textsc{xspec}. We determined the $0.5-10$ keV flux for each observation through spectral fitting with \textsc{xspec} and provide details of the models for each observation in Table \ref{tb:swift-xrt}. We show these fluxes with diamonds in the lower panel of Figure \ref{fig:lightcurve}. Due to the short exposure, ObsID 0031272063 is not included within Figure \ref{fig:lightcurve} as there were insufficient counts to perform spectral fitting. 

To perform X-ray colour analysis, we utilised \textsc{xselect} again to extract light curves in the following four bands: $2-3.5$, $3.5-6$, $6-8$ and $7-10$ keV. We converted the average rates from each observation and each band, with eclipses and bursts removed, to \textit{RXTE} equivalent count rates in the $2-3.5$, $3.5-6$, $6-9.7$ and $9.7-16$ keV bands, assuming the spectral models in Table \ref{tb:swift-xrt} and provide the resulting ECFs in Table~\ref{tb:swiftxrt_conversions}. We note that ObsIDs 0031272063, 0019755001, 0019755002, 0019755003 and 0019755004 were not included in the X-ray colour analysis as the source count rate was too low to extract light curves with sufficient count rates in the required four energy bands.

\subsubsection{NICER}
\textit{NICER} observed EXO 0748 throughout the 2024/25 outburst, with the first of these observations occurring on 2024 June 10, $\sim 10$ hours after the initial type I X-ray burst detection by \textit{Swift}-BAT. Here, we utilise all of the publicly available \textit{NICER} data of EXO 0748, with non-zero exposure (52 observations) taken between June 2024 and April 2025 inclusive; these are ObsIDs beginning with 70500301 or 80500301.

We reduced the data using the \textit{NICER} data reduction software \textsc{nicerdas} version 2025-03-11$\_$V013a, in conjunction with \textsc{heasoft} version 6.35.1 and \textsc{xti-caldb} 20240206. First, we ran the \texttt{nicerl2} data processing pipeline to perform standard calibration, screening, and filtering of the data, and kept most filtering criteria at their default values. We included data taken at low Sun angles by relaxing the undershoot rate limit to allow up to 400 ct~s$^{-1}$ (per focal plane module; FPM). At low Sun angles, optical loading is relatively high, which can deteriorate the response and raise the background at energies $\lesssim 0.4$~keV. However, this did not impact our analysis as we only utilise energies $\geq 0.5$~keV here. Following the online documentation, we ran \texttt{nicerl2} twice, once with the default threshold setting of `night' and once with the threshold setting of `day'. Here, the `day' threshold setting limits the data selection to that taken during the standard orbit day time ranges, thus separating data impacted by the optical light leak in May 2023~\footnote{\url{https://heasarc.gsfc.nasa.gov/docs/nicer/analysis_threads/light-leak-overview/}}. Subsequently, we ran \texttt{nicerl3-lc}, specifying the Space Weather background model, to extract day and night light curves with five-minute time bins in the  $0.5-10$ keV band. At this stage, we noticed several observations appeared heavily background affected, as their light curves displayed exceptionally bright points that were unlikely to originate from the source and did not coincide with an X-ray burst. In these cases, we utilise 12-15 keV light curves to identify and exclude periods of high background. As EXO 0748 is a dipping, eclipsing and bursting source, we carefully inspected the resulting light curves to remove these features, where possible, to reduce their contribution to our time-averaged, per-observation measurements. However, we note that EXO 0748 was in a dipped state during the early part of the 2024/25 outburst, making the complete removal of this variability challenging. In addition, we screened for and removed any achromatic dips arising from occultation of the detector plane by parts of the International Space Station~\footnote{\url{https://heasarc.gsfc.nasa.gov/docs/nicer/analysis_threads/iss_obstruction/}}. We repeated the light curve extraction steps to obtain day and night light curves in the following four bands: $2-3.5$, $3.5-6$, $6-9$ and $9-12$~keV to perform X-ray colour analysis in Section~\ref{sx:x-ray_spec}, and provide the ECFs in Table~\ref{tb:nicer_conversions}.

For each observation, we also extracted a $0.5-10$~keV spectrum using  \texttt{nicerl3-spect} and kept all arguments to their default values. We fit the source spectra with models of the form \texttt{tbabs*zxipcf(diskbb + bbody)} using \textsc{xspec} version 12.13.1, assuming the abundances of \citet{Wilms2000}. Here, we employed the \textit{NICER} SCORPEON background model, which was simultaneously fit with our source model to prevent background over- or under-subtraction and thus return more realistic estimates of the source parameter uncertainties. We maintained the source model form for the majority of the \textit{NICER} spectra to provide a simple way to see how the spectra evolved during the course of the outburst. However, there are a few cases for which this model form did not provide a statistically acceptable fit, leading us to apply a modified model in these instances. The best-fitting spectral parameters for all \textit{NICER} observations are given in Appendix \ref{sx:x-ray_appendix}, Tables \ref{tb:nicer705} - \ref{tb:nicer805_powerlaw} inclusive. We determined the $0.5-10$ keV flux for each observation from the modelled spectra using \texttt{cflux} and show these with circles in the lower panel of Figure \ref{fig:lightcurve}.

\begin{table}
    \centering
    \begin{tabular}{c c c}
    \hline
    ObsID & Start Date \& Time & Day | Night Exposure [s] \\
    \hline
    \vspace*{0.055cm}
    7050030101 & 2024-06-10 \ 17:20:00.00 & \ \ \ \ \ \ \ \  0  ~|~ \ \ 810 \\
    \vspace*{0.055cm}
    7050030102 & 2024-06-11 \ 04:08:48.00 & \ \ \ \ 594 ~|~ 4986 \\
    \vspace*{0.055cm}
    7050030103 & 2024-06-12 \ 01:51:40.00 & \ \ \ \ 227 ~|~ 5597 \\
    \vspace*{0.055cm}
    7050030104 & 2024-06-13 \ 01:05:20.00 & \ \ \ \ 436 ~|~ 3110 \\
    \vspace*{0.055cm}
    7050030105 & 2024-06-14 \ 09:37:40.00 & \ \ \ \ \ \ \ \ 0 ~|~ 2824 \\
    \vspace*{0.055cm}
    7050030106 & 2024-06-15 \ 02:33:50.00 & \ \ \ \ 137 ~|~ 1269 \\
    \vspace*{0.055cm}
    7050030107 & 2024-06-16 \ 02:03:00.00 & \ \ \ \ \ \ \ \ 0 ~|~ \ \ 314 \\
    \vspace*{0.055cm}
    7050030108 & 2024-06-17 \ 01:15:40.00 & \ \ \ \ \ \ \ \  0 ~|~ 2586 \\ 
    \vspace*{0.055cm}
    7050030109 & 2024-06-18 \ 06:41:40.00 & \ \ \ \ \ \ \ \ 0 ~|~ 1766 \\
    \vspace*{0.055cm}
    7050030110 & 2024-06-19 \ 02:50:40.00 & \ \ \ \ \ \ \ \ 0 ~|~ 2353 \\
    \vspace*{0.055cm}
    7050030111 & 2024-06-20 \ 00:25:00.00 & \ \ \ \ \ \ \ \ 0 ~|~ 5237 \\
    \vspace*{0.055cm}
    7050030112 & 2024-06-21 \ 01:12:00.00 & \ \ \ \ \ \ \ \ 0 ~|~ 4268 \\
    \vspace*{0.055cm}
    7050030113 & 2024-06-22 \ 00:27:00.00 & \ \ \ \ \ \ \ \ 0 ~|~ 3134 \\
    \vspace*{0.055cm}
    7050030114 & 2024-06-24 \ 09:21:00.00 & \ \ \ \ 980 ~|~ \ \ \ \ \ \ 0 \\
    \vspace*{0.055cm}
    7050030115 & 2024-06-29 \ 15:11:40.00 & \ \ \ \ \ \ \ \  0  ~|~ \ \ 691 \\
    \vspace*{0.055cm}
    7050030121 & 2024-07-07 \ 03:52:48.00 & \ \ \ \ 301 ~|~ \ \ \ \ \ \ 0 \\
    \vspace*{0.055cm}
    7050030122 & 2024-07-08 \ 03:23:25.00 & \ \ 1461 ~|~ \ \ \ \ \ \ 0 \\
    \vspace*{0.055cm}
    7050030123 & 2024-07-09 \ 05:25:18.00 & \ \ \ \ 382 ~|~ \ \ \ \ \ \ 0  \\
    \vspace*{0.055cm}
    7050030124 & 2024-07-10 \ 00:01:00.00 & \ \ \ \ 985 ~|~ \ \ \ \ \ \ 0  \\
    \vspace*{0.055cm}
    7050030125 & 2024-07-11 \ 03:52:31.00 & \ \ 1266 ~|~ \ \ \ \ \ \ 0  \\
    \vspace*{0.055cm}
    7050030126 & 2024-07-11 \ 23:57:20.00 & \ \ 1009 ~|~ \ \ \ \ \ \ 0 \\
    \vspace*{0.055cm}
    7050030128 & 2024-07-14 \ 00:15:17.00 & \ \ \ \ \ \ 93 ~|~ \ \ \ \ \ \ 0  \\
    \vspace*{0.055cm}
    7050030139 & 2024-07-31 \ 09:41:40.00 & \ \ \ \ \ \ \ \ 0 ~|~ 1571 \\
    \vspace*{0.055cm}
    7050030141 & 2024-10-08 \ 16:14:20.00 & \ \ \ \ \ \ \ \ 0 ~|~ \ \ 335 \\
    \vspace*{0.055cm}
    7050030143 & 2024-10-16 \ 05:27:40.00 & \ \ \ \ \ \ \ \ 0 ~|~ \ \ \ \ 16 \\
    \vspace*{0.055cm}
    7050030144 & 2024-11-13 \ 16:28:00.00 &  \ \ \ \ \ \ \ \ 0 ~|~ \ \ 563 \\
    \vspace*{0.055cm}
    7050030145 & 2025-01-30 \ 07:08:33.00 & \ \ \ \ \ \ \ \ 0 ~|~ 1287 \\
    \vspace*{0.055cm}
    7050030146 & 2025-02-01 \ 01:06:40.00 & \ \ \ \ 350 ~|~ \ \ 963 \\
    \vspace*{0.055cm}
    7050030147 & 2025-02-03 \ 02:50:00.00 & \ \ 1637 ~|~ \ \ 590 \\
    \vspace*{0.055cm}
    7050030148 & 2025-02-14 \ 14:17:02.00 & \ \ \ \ \ \ \ \ 0 ~|~ 1032 \\
    \vspace*{0.055cm}
    7050030149 & 2025-02-20 \ 17:22:00.00 & \ \ \ \ \ \ \ \ 0 ~|~ 4521 \\
    \vspace*{0.055cm}
    7050030150 & 2025-02-21 \ 01:14:00.00 & \ \ \ \ \ \ \ \ 0 ~|~ 1536 \\
    \vspace*{0.055cm}
    7050030151 & 2025-02-22 \ 01:55:25.00 & \ \ \ \ \ \ \ \ 0 ~|~ \ \ 370 \\
    \vspace*{0.055cm}
    7050030152 & 2025-02-23 \ 02:52:20.00 & \ \ \ \ \ \ \ \ 0 ~|~ 3340 \\
    \vspace*{0.055cm}
    7050030153 & 2025-02-24 \ 00:25:18.00 & \ \ \ \ \ \ \ \ 0 ~|~ \ \ 864 \\
    \vspace*{0.055cm}
    8050030101 & 2025-03-01 \ 07:02:03.00 & \ \ \ \ \ \ \ \ 0 ~|~ \ \ \ \ 67 \\
    \vspace*{0.055cm}
    8050030102 & 2025-03-02 \ 00:02:45.00 & \ \ \ \ \ \ \ \ 0 ~|~ \ \ \ \ 17 \\
    \vspace*{0.055cm}
    8050030107 & 2025-03-08 \ 00:07:00.00 & \ \ \ \ 513 ~|~ \ \ \ \ \ \ 0 \\
    \vspace*{0.055cm}
    8050030108 & 2025-03-09 \ 00:46:40.00 & \ \ \ \ 443 ~|~ \ \ \ \ \ \ 0 \\
    \vspace*{0.055cm}
    8050030109 & 2025-03-10 \ 00:07:40.00 & \ \ \ \ 567 ~|~ \ \ \ \ \ \ 0 \\
    \vspace*{0.055cm}
    8050030110 & 2025-03-11 \ 01:06:00.00 & \ \ 4087 ~|~ \ \ \ \ \ \ 0  \\
    \vspace*{0.055cm}
    8050030111 & 2025-03-11 \ 23:59:11.00 & 11756 ~|~ \ \ \ \ \ \ 0  \\
    \vspace*{0.055cm}
    8050030112 & 2025-03-13 \ 00:38:20.00 & \ \ 9148 ~|~ \ \ \ \ \ \ 0 \\
    \vspace*{0.055cm}
    8050030113 & 2025-03-14 \ 00:01:20.00 & \ \ 6112 ~|~ \ \ \ \ \ \ 0 \\
    \vspace*{0.055cm}
    8050030114 & 2025-03-15 \ 00:52:40.00 & \ \ 2704 ~|~ \ \ \ \ \ \ 0 \\
    \vspace*{0.055cm}
    8050030115 & 2025-03-16 \ 00:06:40.00 & \ \ \ \ 193 ~|~ \ \ \ \ \ \ 0 \\
    \vspace*{0.055cm}
    8050030116 & 2025-03-23 \ 15:40:08.00 & \ \ \ \ \ \ \ \ 0 ~|~ 1137 \\
    \vspace*{0.055cm}
    8050030117 & 2025-03-31 \ 14:13:14.00 & \ \ \ \ \ \ \ \ 0 ~|~ 1158 \\
    \vspace*{0.055cm}
    8050030118 & 2025-04-02 \ 09:16:37.00 & \ \ \ \ \ \ \ \ 0 ~|~ 1754 \\
    \vspace*{0.055cm}
    8050030119 & 2025-04-04 \ 07:53:40.00 & \ \ \ \ \ \ \ \ 0 ~|~ \ \ 118 \\
    \hline
    \end{tabular}
    \caption{A summary of the \textit{NICER} observations of EXO 0748--676 during the 2024/25 outburst used in this work. Here, the quoted `day' and `night' exposures define the clean exposure durations, with data gaps ignored, obtained from each threshold run during the reduction stage.}
    \label{tab:nicer}
\end{table}

\subsubsection{XMM-Newton}
\textit{XMM-Newton} observed EXO 0748 utilising all on-board instruments between 14:30:35 on 2024-06-30 and 12:42:15 on 2024-07-01 (ObsID: 0935191701, PI: Dr Norbert Schartel. The data were obtained from the XMM-Newton Science Archive (XSA) and reduced with \textit{XMM-Newton's} Science Analysis Software \textsc{XMM-SAS}, version 21.0 and the latest calibration files. We follow the data reduction procedure described in \citet{Knight2025} up to the point of applying barycentric corrections (to Barycentric Dynamical Time using the default DE200 ephemeris). We extracted light curves with 5-minute time bins in the following five bands: $0.5-10$, $2-3.5$, $3.5-6$, $6-8$ and $7-10$ keV. All light curves are background-subtracted and corrected using \texttt{epiclccorr}. We utilise the latter four bands for the X-ray colour analysis conducted in Section \ref{sx:x-ray_spec} and provide the ECFs to the corresponding \textit{RXTE} bands in Table \ref{tb:xmm_conversions}.

We subsequently extracted source plus background spectra from each of the EPIC detectors in the $0.5-10$ keV band, using the same source regions as \citet{Knight2025}. We utilised singles, doubles, triples and quadruples when extracting MOS1 and MOS2 spectra, as well as singles and doubles when extracting the PN spectra. We used channels 0-11999 for both MOS detectors and 0-20479 for PN, and a spectral bin size of 5. We repeated this step to extract background spectra for subtraction, utilising a 40 arcsecond radius, circular source-free region close to the position of EXO 0748 and on the same chip as the source for MOS1 and MOS2. For PN, which was operating in timing mode, we utilised the background region 4 $\leq$ \texttt{RAWX} $\leq$ 12. We applied \texttt{backscale} to determine the areas of the source and background regions used to extract the spectral files, accounted for any bad pixel locations, and then generated corresponding redistribution matrices and ancillary response files. Finally, we grouped the spectra to have at least 25 counts for each spectral channel and did not oversample the intrinsic energy resolution by a factor higher than 3.

We fit the three 0.5 - 10 keV EPIC spectra simultaneously with \textsc{xspec} version 12.13.1 with the model \texttt{tbabs*zxipcf(diskbb+bbody+nthcomp)*gabs}. Here, the \texttt{diskbb} and \texttt{bbody}, respectively, describe emission from the multi-temperature accretion disc and neutron star surface. Guided by \citet{Knight2022a} and \citet{Knight2023}, we include a partial covering absorber (\texttt{zxipcf}) to describe absorption by the clumpy, highly ionised, ablated material that is known to be present in false widow binaries. We also applied a constant factor to model the differences in detector calibration, finding that they are consistent to within 10~$\%$. From this fit, we determine the observed $0.5 - 10$ keV EPIC flux to be $(5.79 \pm 0.02) \times 10^{-11}$ erg cm$^{-2}$ s$^{-1}$, and show this measurement with a purple pentagon in the lower panel of Figure \ref{fig:lightcurve}. Further details of the spectral fits and best-fitting parameters are in Appendix~\ref{sx:x-ray_appendix}.

\begin{figure*}
    \centering
    \includegraphics[width=\textwidth]{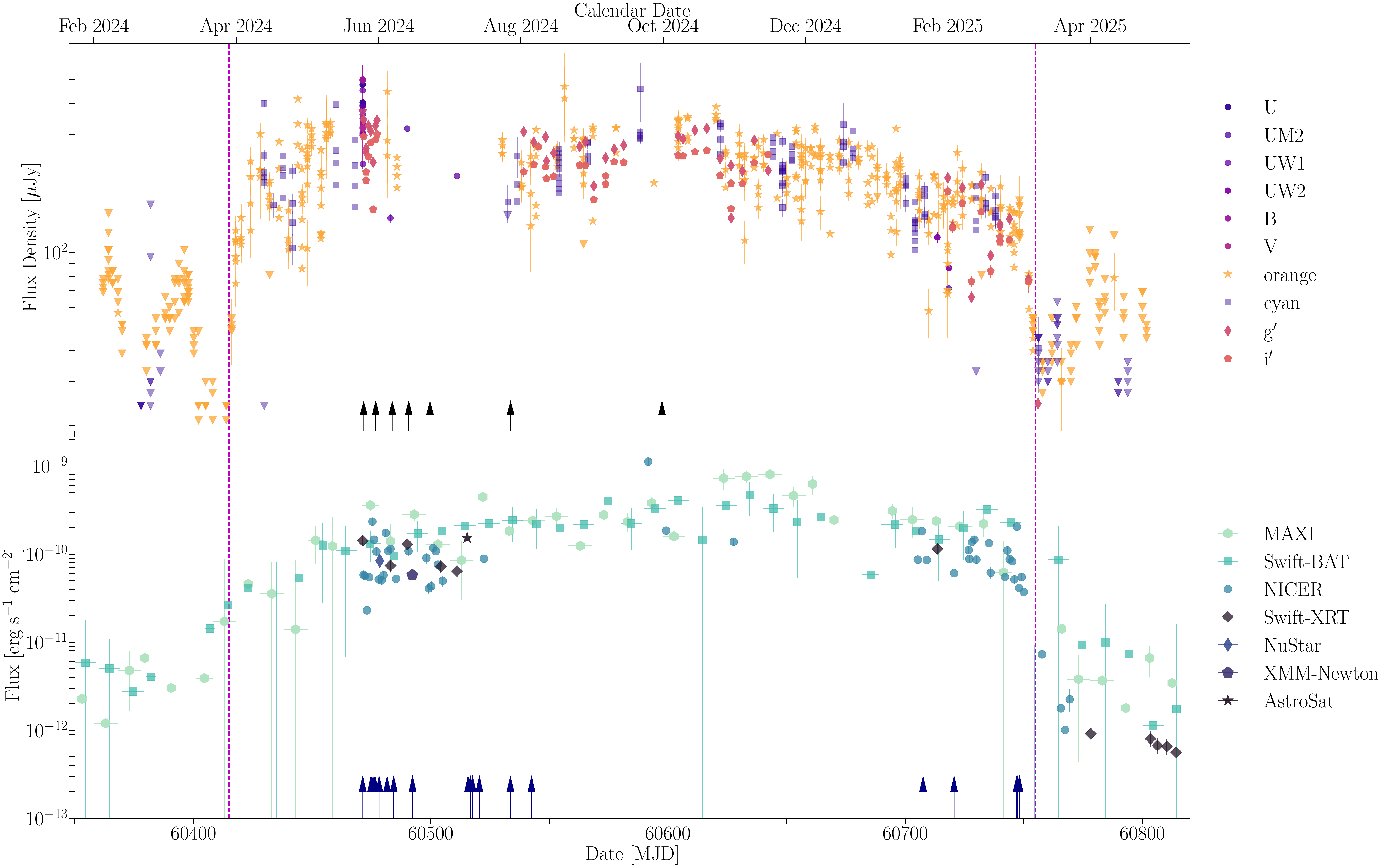}
    \vspace*{-0.5cm}
    \caption{Top: Optical/UV light curve of EXO 0748 during its 2024/25 outburst, comprising monitoring from ATLAS (orange and cyan), and Las Cumbres Observatory (LCO; ${g}^{\prime}$ and ${i}^{\prime}$) and \textit{Swift-UVOT} (U, UM2, UW1 and UW2). Here, inverted triangles represent upper limits. The arrows along the lower axis of the top panel indicate the times of the MeerKAT radio observations described in Section \ref{subsec:Meerkat}. Bottom: X-ray light curve of EXO 0748 during its 2024/25 outburst, consisting of monitoring from \textit{Swift-BAT} (15-50 keV) and \textit{MAXI-GSC} (2-20 keV), and pointed exposures from \textit{AstroSat}, \textit{NICER}, \textit{Swift-XRT} and \textit{XMM-Newton} (0.5 - 10 keV) and \textit{NuStar} (3 - 78 keV). Where necessary, the X-ray fluxes are bandpass corrected to the 0.5 - 10 keV band, using the spectral models and energy conversion factors provided in Appendix \ref{sx:x-ray_appendix}. The arrows along the lower axis of the bottom panel highlight the times of known type I X-ray bursts. In both panels, the vertical lines indicate the start (MJD 60415) and end (MJD 60755) of the outburst, determined by the date of the first and last optical detections, respectively.}
    \label{fig:lightcurve}
\end{figure*}

\subsubsection{NuStar}
\textit{NuSTAR} conducted two observations of EXO 0748 on 2024-06-17. The first of these was a 548 s exposure, starting at 00:51:07 (ObsID 91001324001). The second was a 62648 s exposure, starting at 01:31:09 (ObsID 91001324002). We obtained these data from the High Energy Astrophysics Science Archive Research Center (HEASARC) data archive and reduced them using \textsc{nustardas} version 2.1.5 in conjunction with \textsc{caldb} version 20250224.

We followed the standard data reduction procedures and first performed data calibration and screening of both observations with \texttt{nupipeline}, keeping all arguments to their default values. For both FPMA and FPMB, we select 60 arcsecond radius circular source regions, centred on EXO 0748, and 120 arcsecond radius circular background regions. Our chosen background regions are South-East of EXO 0748, while remaining in the same detector quadrant. We subsequently use \texttt{nuproducts} to extract $3 - 78$~keV spectra and light curves in the following five bands: $3-4$, $4-6$, $6-10$, $10-20$ and $3-78$ keV. 

We fit the extracted FPMA and FPMB spectra simultaneously with \textsc{xspec} version 12.13.1 with the model \texttt{tbabs*zxipcf(bbody+cutoffpl)}. We account for the cross-normalisation difference between the two detectors with a constant factor, finding that they are consistent to within $10~\%$. To achieve a statistically acceptable fit ($\chi^{2} = 2385.13$, $\nu = 2328$, $p = 0.2$; see Appendix \ref{sx:x-ray_appendix} for further details), we ignore energies below $3$~keV and over $50$~keV. From this fit, the $3-50$ keV flux was determined to be $2.56 \times 10^{-10}$ erg cm$^{-2}$ s$^{-1}$. The flux becomes $8.40 \times 10^{-11}$ erg cm$^{-2}$ s$^{-1}$ when bandpass correcting to the $0.5-10$ keV band using the ECF in Table \ref{tb:nustar_conversions}. We show this measurement with a blue diamond in the lower panel of Figure~\ref{fig:lightcurve}.

\subsection{Multi-Wavelength Light Curve of the 2024/25 Outburst}
\label{subsec:lightcurve}

Figure \ref{fig:lightcurve} shows the multi-wavelength light curve compiled from the measurements described throughout Section \ref{sx:obs}. Here, the top panel shows the optical and UV monitoring, while the bottom panel depicts the X-ray observations. We discuss each panel in turn.

The ATLAS survey captured the onset of the outburst, with a sharp rise in flux density around MJD 60415, and observed throughout the $\sim 10$ month outburst with observations every 1--3 days. After MJD $\sim 60670$, the flux density slowly decays back towards quiescent levels, and the optical counterpart became undetectable by MJD 60755. Based on the ATLAS monitoring, we define the start and end of the outburst as MJD 60415 and MJD 60755, respectively. We highlight these dates in both panels of Figure \ref{fig:lightcurve} by vertical dashed lines. However, we note that these are only approximate start and end times as the quiescent levels for EXO 0748 are below the sensitivity level of ATLAS. As such, we were unable to capture the optical rise from and decay into quiescence completely. \textit{NICER} and \textit{Swift}-XRT observations between MJD 60778 and 60810 \citep{Degenaar2025_Atel} confirmed EXO 0748's return to quiescence.

Between MJD $\sim 60450$ and $\sim 60670$, the optical flux density remained relatively stable; we consistently measure a \textit{cyan} band (420--650\,nm) flux density between 200 and 300\,$\mu$Jy until the source begins to fade. However, we do observe a lot of scatter, with no monotonic trends. The large amount of scatter in the optical/UV light curve is not unexpected. The optical counterparts of transient LMXBs can exhibit high levels of variability because multiple emitting regions, such as the accretion disc, the jet and the companion star, contribute to the total observed optical flux. For EXO 0748, the observed scatter likely arises from changes in the accretion rate as well as the strong orbital phase dependence exhibited by the optical counterpart to EXO 0748 \citep[as discussed in Section \ref{subsec:phase},][]{Schmidtke1987A}. There is also likely some contamination from reprocessed type I X-ray bursts (e.g. around MJD 60470), which adds to the observed scatter. 

As there was no long-term optical monitoring of EXO 0748 during the 1985 outburst, there is insufficient data to compare the light curve morphology between the two recorded outbursts. However, we can compare the reported flux densities. The broadband \textit{cyan} filter overlaps, in wavelength space, with the \textit{Swift}-UVOT \textit{v}-band filter, which intermittently monitored EXO 0748 during its 1985 outburst \citep[e.g.][]{1988A&AS...76..185V, Schmidtke1987A,1993ApJ...408..651T}. The reported flux densities from this monitoring range from 230 to 580$\mu$Jy (17.0 -- 18.0\,mag, with type I X-ray bursts removed), depending on the phase of the outburst and the assumed orbital period. From this, we estimate that the optical counterpart (i.e. the optical emission from the outburst) was up to a factor of two brighter (a difference of 0.8\,magnitudes) in the 1985 outburst than the recent 2024/25 outburst. However, we also note that the optical luminosity during outburst can change by four to five orders of magnitude \citep[as shown in figure 1 of][]{2006MNRAS.371.1334R}. Therefore, the differing flux densities between the 1985 and 2024/25 outbursts should not be considered significant in the context of explaining why the 2024/25 outburst was different.

The lower panel of Figure \ref{fig:lightcurve} shows the X-ray light curve for EXO 0748, with all instruments bandpass corrected to $0.5-10$ keV. The rise in the X-rays is much shallower than observed at optical wavelengths, and the general shape of the light curve follows a slow rise and fast decay; this appears to be the opposite of the general shape of the optical light curve. The X-ray flux spans $2-3$ orders of magnitude over the course of the outburst (note the logarithmic y-axis in the lower panel of Figure \ref{fig:lightcurve}), peaking at MJD $\sim 60650$. Variations of this kind are often observed in accreting sources, but are likely due to eclipses, dips, type I X-ray bursts (arrows in the lower panel of Figure \ref{fig:lightcurve}; likely incomplete due to the lower number of pointed observations during the middle of the outburst) and absorption by ablated material in EXO 0748. While we have taken steps to exclude sections of data encompassing type I X-ray bursts, eclipses and periods of strong dipping, absorption by the ablated material, and indeed absorption-induced dips are more challenging to identify and thus remove. Therefore, absorption-induced phenomena are the most likely cause of the flux variations among the pointed observations. For the X-ray monitors, there are likely eclipses and type I X-ray bursts hidden within the data due to the broad observing cadence, causing some of the observed variations.

Overall, the light curve phenomenology is reminiscent of a low-intensity state (see e.g. \citealt{MaitraBailyn:2008ApJ...688..537M}): the optical and X-ray fluxes are relatively stable and there is a lack of a fast rise-exponential decay (FRED) light curve behaviour. We consider the stability of the light curves to be somewhat uncharacteristic of transient LMXB outbursts. Often transient outbursts exhibit `canonical' FRED-type light curves in which the optical and X-ray emissions are strongly correlated \citep[e.g.][]{2006MNRAS.371.1334R, 2018ApJ...867L...9T, 2025MNRAS.541.1851B}. While, it is challenging to fully characterise the shape of the X-ray light curve as there were no pointed observations during the rise into outburst, the light curve morphology is clearly not canonical. Even so, the data from \textit{MAXI} and \textit{Swift}-BAT depict the first observed rise of EXO 0748 into outburst at X-ray wavelengths (because the discovery of EXO 0748 in 1985 occurred when the system was already in outburst; \citealt{Parmar1986}). Thus, we can see for the first time that EXO 0748 rose gradually at X-ray wavelengths and remained reasonably stable in flux while active. 

Perhaps one of the most intriguing properties of the 2024/25 outburst is its apparent X-ray faintness compared to the 1985 outburst, particularly towards the start. This comparative faintness could indicate that the source was either highly absorbed/obscured during the 2024/25 outburst or that the source was intrinsically fainter. To demonstrate this, we consider \textit{XMM-Newton} ObsID 0935191701 (June 2024), which detected EXO 0748 with a variable, combined EPIC out-of-eclipse count rate of $5 - 10$ cts/s in the $0.5 - 10$ keV band (see e.g. figure 1 of \citealt{Knight2025}). This range becomes $6.65 - 13.3$ ct/s when bandpass corrected to $2 - 16$ keV \textit{RXTE} equivalent rates, roughly $1.5 - 3.8$ times lower than measured \textit{RXTE} hard state count rates from the previous outburst ($\approx 20-25$ cts/s/PCU). However, the start of the previous outburst of EXO~0748 was not well covered; the first detection occurred in July 1984 as part of the \textit{EXOSAT} Medium Energy Slew Survey, where it was detected with $1 - 8$~keV count rate of $9.7 \pm 1.9$ cts~s$^{-1}$~half$^{-1}$ \citep{Reynolds1999}. Assuming the spectral model from \citet{Parmar1986} ($\Gamma \sim 2.1$, N$_{\rm{H}} = 3 \times 10^{21}$ atom cm$^{-2}$), the equivalent $0.5 - 10$~keV flux is $2 \times 10^{-10}$ erg cm$^{-2}$ s$^{-1}$, consistent with measurements from the 1985 outburst \citep{sidoli2005, Boirin2007, DiazTrigo2006} and later measurements from the 2024/25 outbursts (from pointed observations; see Figure \ref{fig:lightcurve} and Appendix \ref{sx:x-ray_appendix}). EXO 0748 was serendipitously detected in May 1980 with Einstein \citep{Parmar1986,Garcia1999}, with a flux consistent with quiescence ($0.25 \pm 0.13 \times 10^{-12}$ erg cm$^{-2}$ s$^{-1}$ in the $0.5 - 10$~keV band) and a NS crust temperature (95 eV) lower than that measured in 2013 (110 eV, roughly 5 years after EXO 0748 entered its first quiescence; \citealt{Degenaar2014}). This leaves an unobserved, four year window in which the source would have entered outburst. Therefore, a faint phase similar to that seen early on in the 2024/25 outburst could also have occurred at the beginning of the 1985 outburst, but there are no observations available to assess this possibility.


\section{X-ray Spectral Evolution}
\label{sx:x-ray_spec}

\begin{figure}
    \centering
    \includegraphics[width=\columnwidth]{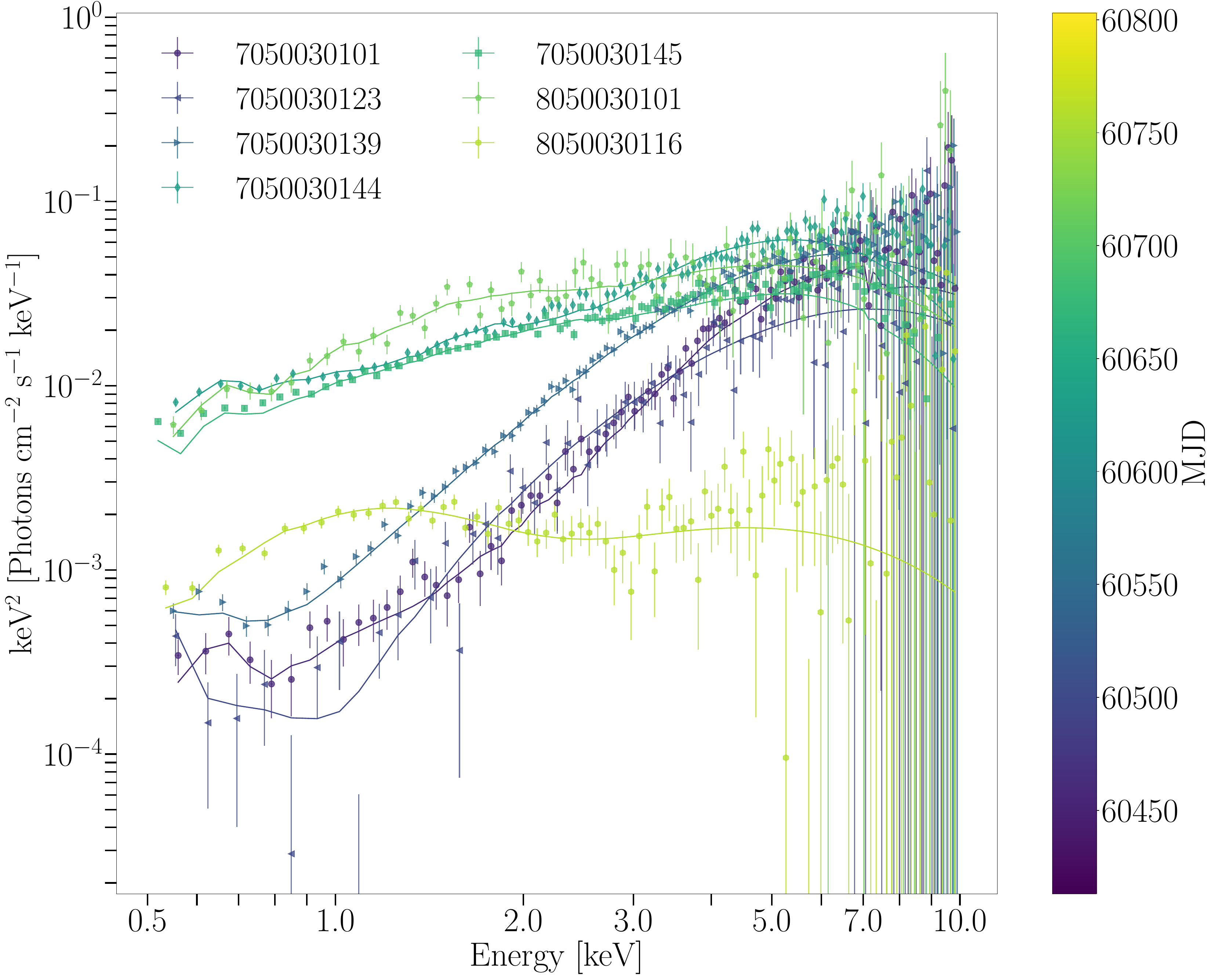}
    \vspace*{-0.5cm}
    \caption{Seven representative $0.5 -10$ keV \textit{NICER} spectra of EXO 0748--676, extracted from observations taken during the 2024/25 outburst, as a function of MJD. Each spectrum was fit in \textsc{xspec}; the best-fitting models are provided in Tables \ref{tb:nicer705} and \ref{tb:nicer805}. The spectra are shown unfolded around their respective best fitting models. Clearly seen is the strong absorption/obscuration at the beginning of the outburst, which dampens the soft X-ray contribution.}
    \label{fig:Spectra}
\end{figure}

As discussed in Section \ref{subsec:lightcurve}, EXO 0748 was detected by multiple X-ray instruments at lower count rates than expected when compared to its 1985 outburst, indicating that either the X-ray emitting region was strongly absorbed or obscured, particularly towards the beginning of the outburst, or that the source was intrinsically fainter. To determine the cause of EXO 0748's apparent faintness, we begin by looking at seven representative $0.5 - 10$ keV \textit{NICER} spectra (Figure~\ref{fig:Spectra}), selected to be distributed at $\sim 30-40$ day intervals across the outburst. The best-fitting X-ray spectral parameters for the seven representative spectra (and all pointed observations) are in Appendix \ref{sx:x-ray_appendix}, where we demonstrate that the majority of the pointed spectra can be well fit with a model of the form \texttt{tbabs*zxipcf(diskbb+bbody)}. Maintaining this model form throughout was a deliberate choice that permits a straightforward comparison of the spectral properties across the outburst. For all spectral fits, \texttt{tbabs} describes the Galactic absorption and was fixed to a value of $0.116 \times 10^{22}$ atom cm$^{-2}$, as determined by the Colden Galactic Neutral Hydrogen Density Calculator~\footnote{\url{https://cxc.harvard.edu/toolkit/colden.jsp}} using the NRAO dataset \citep{Dickey1990}. 

Looking at the representative spectra in Figure~\ref{fig:Spectra}, we see a clear difference between spectra from observations taken during the first and second halves of the outburst. The early spectra have fewer counts and are strongly absorbed, most notably at soft X-ray wavelengths (energies $\lesssim 5$~keV). This absorption manifests as a low flux accretion disc component in our spectral modelling (i.e. the ratio of the flux of the multi-temperature accretion disc component to the flux of the blackbody component is $< 1.0$; see e.g. Table \ref{tb:nicer705}). The soft X-ray absorption disappears as the outburst progresses, in agreement with the flux variations seen in the lower panel of Figure \ref{fig:lightcurve} and our spectral modelling, which shows the accretion disc component becoming the dominant of the two thermal components towards the end of the outburst (see e.g. Table \ref{tb:nicer805}). This spectral behaviour suggests that EXO 0748's apparent faintness is due to absorption/obscuration of the central X-ray source by a medium that moves or disperses as the outburst progresses. Further support for this interpretation is provided in Figure \ref{fig:specvar} where we see that the temperature inferred from the multi-temperature accretion disc component (circles) is lower during the early stages of the 2024/25 outburst (and is the less dominant thermal component). As the outburst progresses, our spectral fits favour higher temperatures for this accretion disc component, and it becomes the dominant thermal component (ratios $> 1$; Tables \ref{tb:nicer705} - \ref{tb:swift-xrt}). However, we note that some of the later spectral fits return a higher accretion disc temperature than blackbody temperature, contrary to expectations if the blackbody is the inner boundary layer. For spectra where the blackbody contribution is small and the disc temperature is high, it may simply be that the true boundary layer is too hot to be distinguished with soft X-ray instruments, which forces the blackbody component to fit to other features at soft energies. In any case, these spectra still show a greater soft X-ray contribution than in the early observations, and the specific components contributing may be hard to distinguish given the variable absorption.

Despite the absorption being more prominent at the start of the outburst, we find no discernible trends in the best-fitting parameters of the \texttt{zxipcf} model component (column density, $N_{\rm{H}}$, ionisation, $\log(\xi)$, and covering fraction, f$_{\rm{cov}}$; see Figure \ref{fig:specvar}). A likely explanation for the lack of clear trends is that in EXO 0748, the variations in the properties of the local absorber are known to occur as a function of orbital phase (see e.g. \citealt{Knight2023}), so the time-averaged spectral fitting performed here will not detect such variations. Nonetheless, we do find a slight (although not statistically significant) drop in the covering fraction toward the latter stages of the outburst, which is consistent with the idea that the absorbing medium dispersed as the outburst progressed. However, the clearest evidence we have of the local absorber moving/dispersing during the latter stages of the outburst is that our latter spectral fits often did not require the \texttt{zxipcf} component to achieve a statistically acceptable fit (see e.g. Table \ref{tb:nicer805}). Although it is important to note that the latter spectra (when the source was fading) also have lower signal-to-noise, thus less complex spectral models are more likely to be statistically acceptable than in the middle of the outburst, when the spectra have a higher signal-to-noise ratio. 

\begin{figure*}
    \centering
    \includegraphics[width=\textwidth]{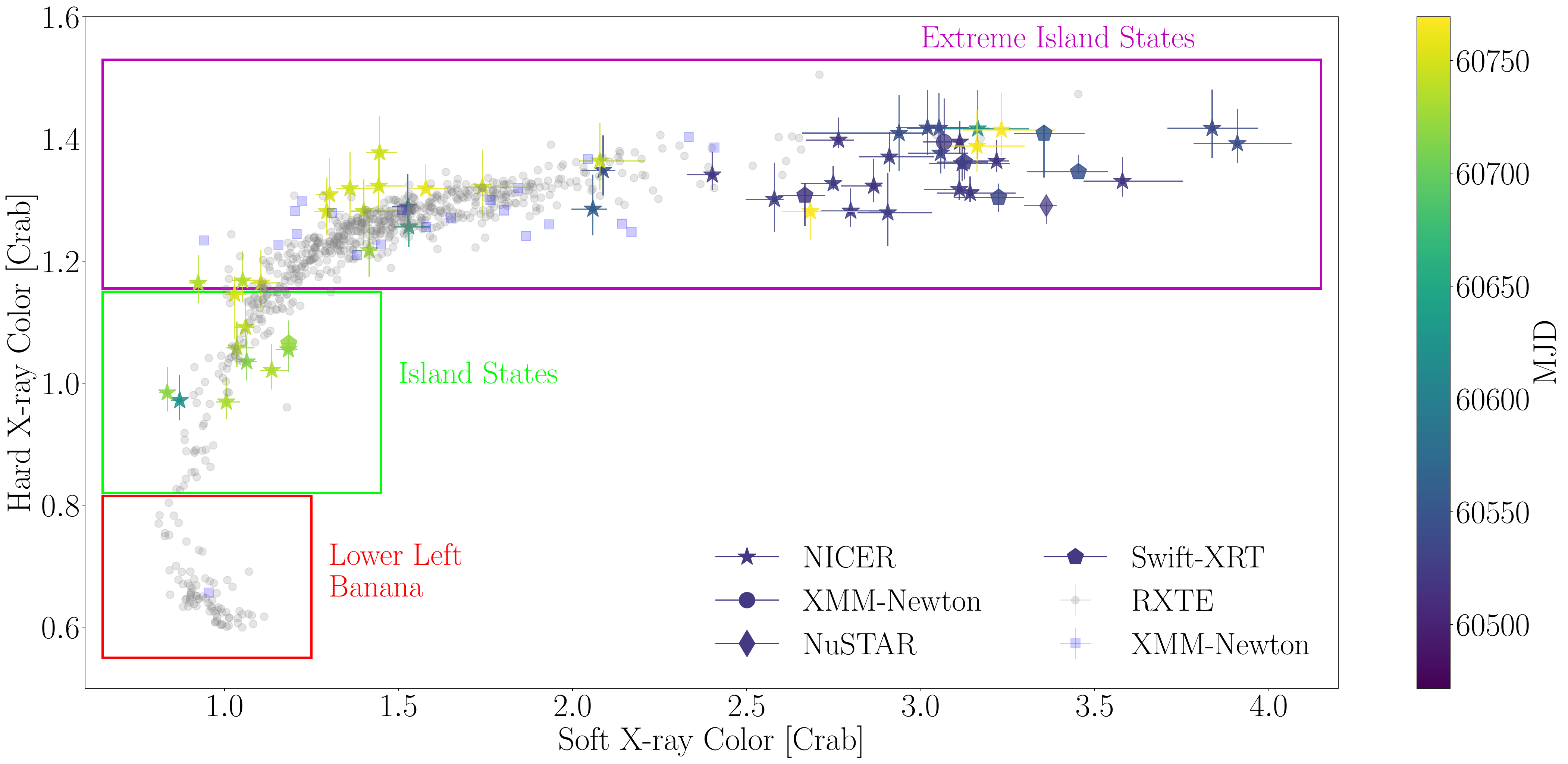}
    \vspace*{-0.5cm}
    \caption{A colour-colour diagram of EXO 0748--676, comprised of data from both the 1985-2008 and 2024/25 outbursts. Data collected during its 2024/25 outburst are shown as a function of MJD and consist of pointed exposures from \textit{NICER} (solid stars), \textit{Swift-XRT} (solid pentagons), \textit{XMM-Newton} (solid circle) and \textit{NuStar} (solid diamond). In all cases, the soft and hard X-ray colours are defined as $3.5 - 6 / 2 - 3.5$ keV and $9.7 -16 / 6-9.7$ keV, respectively, to enable comparison to the \textit{RXTE} colours measured in the previous outburst. We provide details the bandpass corrections used to compile this Figure for each instrument in Appendix \ref{sx:x-ray_appendix}. The transparent data points consist of \textit{RXTE} (transparent circles; reproduced here from \citealt{Knight2025_ineclipse}. See also \citealt{Mancuso2019}) and \textit{XMM-Newton} (transparent squares) observations from the 1985-2008 outburst for comparison.}
    \label{fig:CCD}
\end{figure*}

\begin{figure*}
    \centering
    \includegraphics[width=\textwidth]{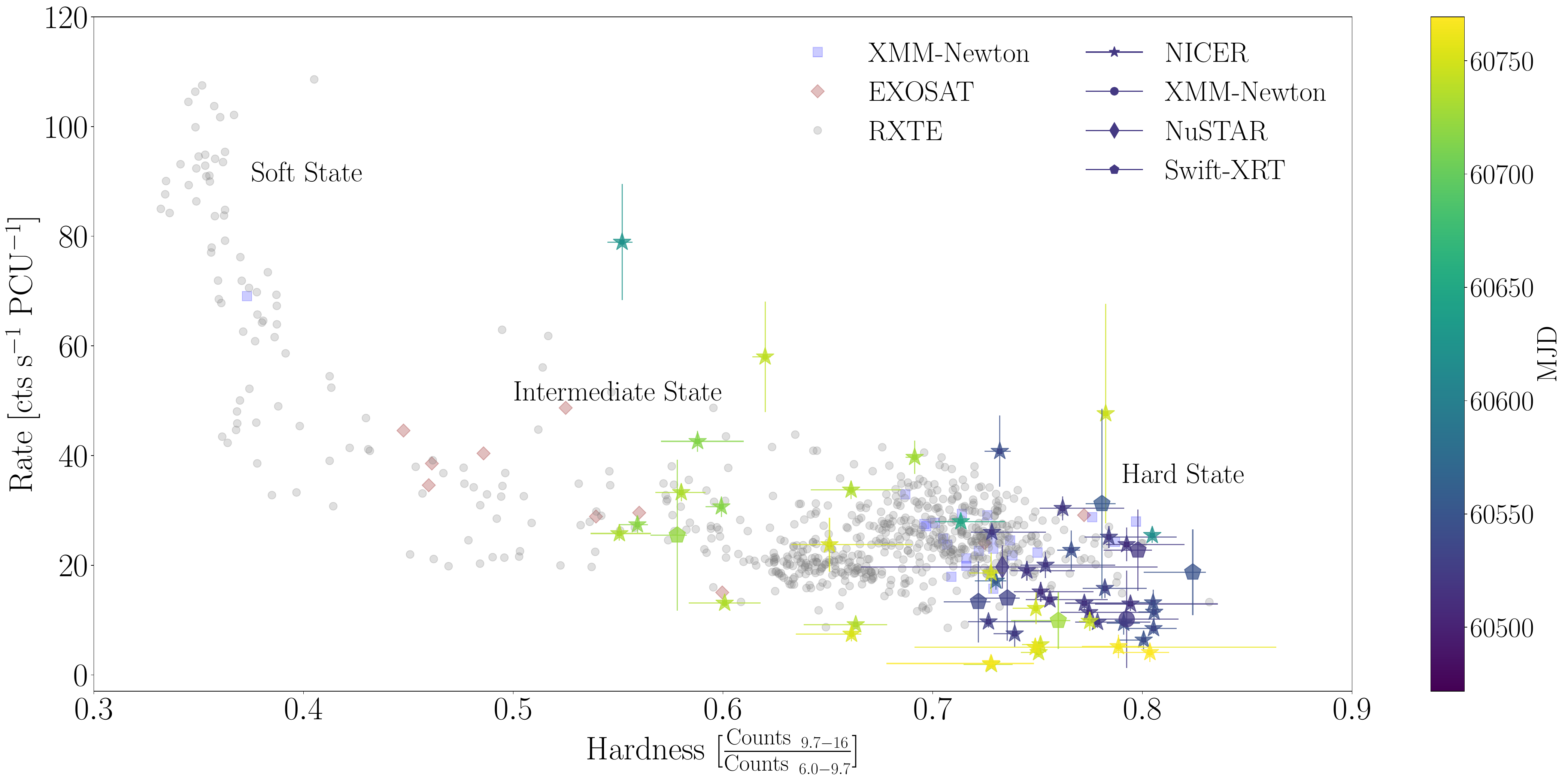}
    \vspace*{-0.5cm}
    \caption{A hardness-intensity diagram of EXO 0748--676, comprised of data from both the 1985-2008 and 2024/25 outbursts. Data collected during its 2024/25 outburst are shown as a function of MJD and consist of pointed exposures from \textit{NICER} (solid stars), \textit{Swift-XRT} (solid pentagons), \textit{XMM-Newton} (solid circle) and \textit{NuStar} (solid diamond). In all cases, the count rates are converted to \textit{RXTE} $2 - 16$ keV band equivalent rates. The hardness is the counts ratio in the $9.7 - 16 / 6 - 9.7$ keV bands, as in \citet{Ponti2014}. We provide the bandpass corrections used to compile this figure in Appendix \ref{sx:x-ray_appendix}. The transparent data points consist of \textit{RXTE} (transparent circles), \textit{XMM-Newton} (transparent squares) and \textit{EXOSAT} (transparent diamonds) observations from the 1985-2008 outburst for comparison.}
    \label{fig:HID}
\end{figure*}

We also explore the spectral evolution of EXO 0748 through the colour-colour diagram (CCD) and hardness-intensity diagram (HID) presented in Figures \ref{fig:CCD} and \ref{fig:HID}, respectively. In both Figures, we utilise \textit{RXTE} and \textit{XMM-Newton} observations from the 1985 outburst to compare to the 2024/25 outburst. The HID also includes measurements from the Medium Energy Instrument on board \textit{EXOSAT} to highlight the differences between the 2024/25 outburst and the earliest records of EXO 0748's 1985 outburst. As EXO 0748 is an atoll source~\footnote{Atoll sources are LMXB systems with weak magnetic fields, characterised by a spectral evolution comprised of distinct states, (e.g. island and banana; see \citealt{vanderKlis1989}).}, monitored regularly with \textit{RXTE}, we converted all 2024/25 data presented in these Figures to \textit{RXTE} equivalent rates specified in their respective captions and, in the case of the CCD, normalised to Crab, following the normalisation procedure of \citet{Altamirano2008}. We perform the conversions and normalisations to maintain consistency with the class definitions of \citet{vanderKlis1989} (see also \citealt{Lewin2006}) and enable comparison to the large quantity of data taken during the 1985 outburst. In the CCD, the extreme island and island states correspond to hard and hard-intermediate spectral states, respectively. The lower left banana state is a soft-intermediate state, which is the softest spectral state exhibited by EXO 0748 (it has never been observed in the banana state, which is the softest part of the CCD and would roughly coincide with the position of the legend in Figure \ref{fig:CCD}).

Much like its 1985 outburst, EXO 0748 spent the majority of the 2024/25 outburst in a low-intensity, hard state in the HID and the extreme island state in the CCD. As the outburst progressed and the soft X-ray contribution increased, EXO 0748 transitioned into a slightly softer and brighter state (island state in the CCD; intermediate state in the HID). However, unlike its 1985 outburst, there is no transition to the soft (banana) state in the 2024/25 outburst. Instead, after entering the island state, EXO 0748 promptly hardened and dropped in intensity as it returned to X-ray quiescence. This spectral evolution suggests that the 2024/25 outburst is a failed-transition outburst, having not developed into a canonical soft state and instead remaining hard and partially obscured for the entire outburst \citep{2021MNRAS.507.5507A}.

Regardless, it is crucial to note that both the CCD and HID show never-before-seen spectral behaviour of the source during the earliest stages of its outburst. In Figure \ref{fig:HID}, we see the source in a fainter and slightly harder position on the HID than previously captured with \textit{RXTE} and \textit{EXOSAT}. However, there is some overlap between the faintest and hardest \textit{RXTE} measurements, which correspond to when EXO 0748 was entering its first quiescence in 2008, and the 2024/25 measurements. This is mimicked in Figure \ref{fig:CCD}; the extreme island state extends to higher soft X-ray colours (corresponding to a harder spectral state) than were found with \textit{RXTE} during the 1985 outburst. As noted by \citet{Homan2015}, EXO 0748 is known to have a broader extreme island state than other atoll sources, which we speculate may be connected to its lack of a soft/banana state. However, exploring this is beyond the scope of this paper. 


\section{Radio/X-ray Correlation}
\label{sx:lrlx}

\begin{figure}
    \centering
    \includegraphics[width=\columnwidth]{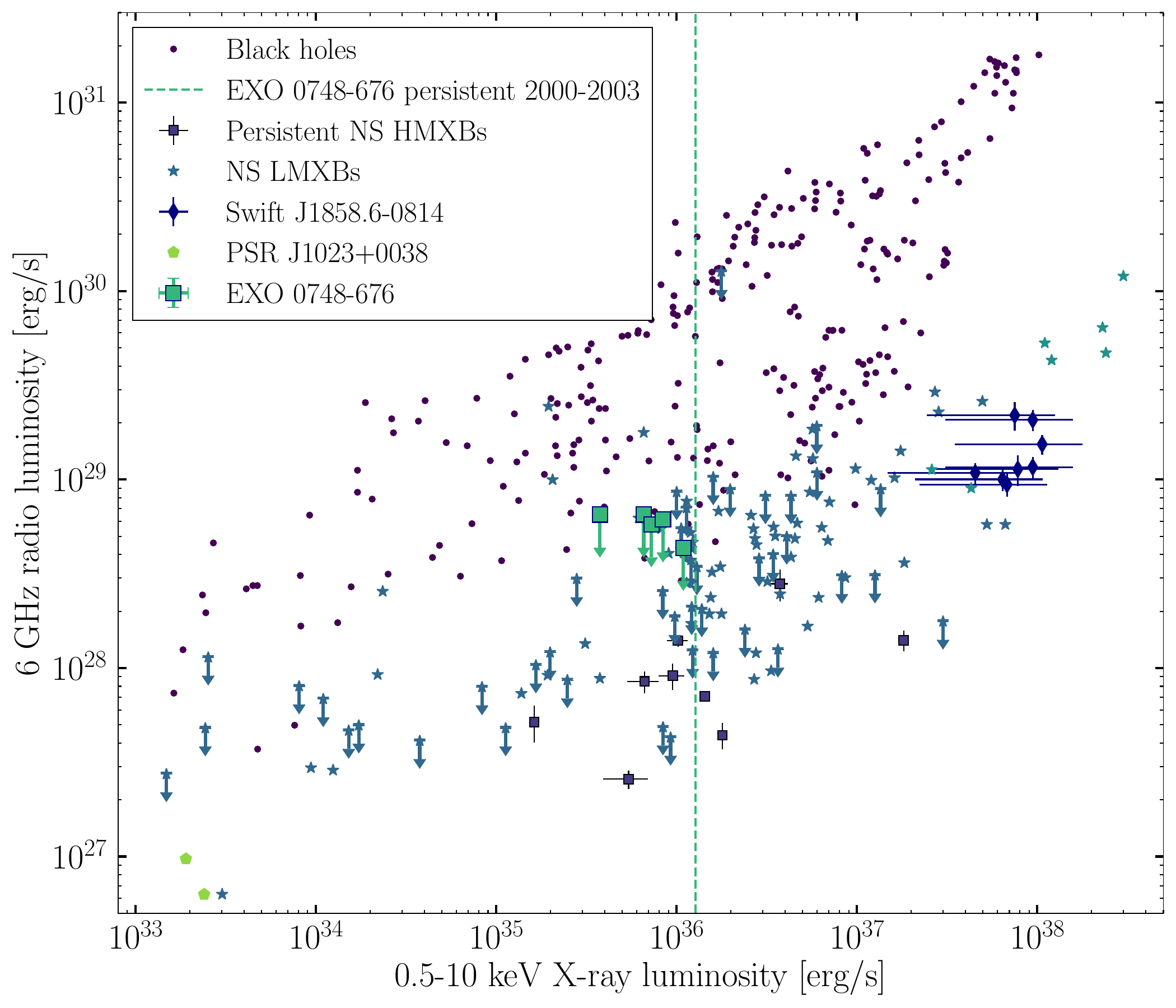}
    \vspace*{-0.5cm}
    \caption{The radio/X-ray correlation for both black hole and neutron star LMXBs \citep{vandenEijnden2021}. The green downwards facing arrows correspond to the five epochs where we have contemporaneous MeerKAT and \textit{NICER} observations (Table \ref{tab:radio_obs} and \ref{tab:nicer}). We also show an average persistent emission X-ray luminosity from the previous outburst with the vertical dashed line \citep{sidoli2005, Boirin2007}. The transitional millisecond pulsar PSR J1023+0038 (pentagons; \citealt{2015ApJ...809...13D}) and false widow candidate \textit{Swift} J1858.6-0814 (diamonds; \citealt{Rhodes2022}) are shown for comparision}.
    \label{fig:LrLx}
\end{figure}

In Figure \ref{fig:LrLx}, we contextualise the 2024/25 outburst of EXO 0748 by placing it on the radio/X-ray correlation for the first time. Here, the dark purple circles and blue stars correspond to hard state (or equivalent) black hole and NS LMXBs, respectively \citep{vandenEijnden2021}. Also shown are a small group of persistent NS high-mass X-ray binaries (purple squares), which sit in a similar region of the parameter space to the NS LMXBs. The green squares correspond to the five epochs where we have contemporaneous (within 24\,hours) radio (MeerKAT) and X-ray (\textit{NICER}) observations. For each epoch, we convert the modelled X-ray fluxes (presented in Table \ref{tb:nicer705}) to luminosities, assuming a a distance of 7.1\,kpc \citep{Galloway2008b, Hynes2006}. As mentioned in Section \ref{subsec:Meerkat}, we do not detect a radio counterpart. Therefore, the data points are upper limits in the y-direction. Based on the position EXO 0748 in Figure \ref{fig:LrLx}, if any radio emission is present, it would likely be consistent with the rest of the NS LMXB population. We find that our X-ray luminosities span nearly an order of magnitude, ranging from $3\times10^{35}$ to $1.5\times10^{36}$\,erg/s. Only our most luminous X-ray epoch (MJD 60476) matches the persistent X-ray luminosities from 2000 - 2003 (dashed line in Figure \ref{fig:LrLx}). This is interesting given the apparent faintness and early-time absorption/obscuration of EXO 0748 during the 2024/25 outburst, as discussed in Sections \ref{sx:obs} and \ref{sx:x-ray_spec}. 

The difference in luminosity between the two outbursts is likely due to the dampened soft X-ray contribution during the first part of the 2024/25 outburst. While we have shown the absorbing medium dispersed during the latter stages of the 2024/25 outburst, the contemporaneous MeerKAT and \textit{NICER} observations used here all occurred before MJD 60600, meaning the absorption was still present and thus, a reasonable cause of the differing luminosity measurements. Indeed, strong local absorption was employed to explain the X-ray properties of the false widow candidate \textit{Swift} J1858.6--0814 (diamonds in Figure \ref{fig:LrLx}; see also \citealt{Hare2020, Rhodes2022}). However, we note that EXO 0748 appeared with lower luminosities at both optical and X-ray wavelengths this time, suggesting that there may also be an intrinsic difference (i.e. the source was less powerful in the recent outburst) between the 1985 and 2024/25 outbursts.

As EXO 0748 is a false widow, we also compare our measurements to the related subclass of transitional redback pulsars using the archetypal system PSR J1023$+$0038. As shown with light green pentagons in Figure \ref{fig:LrLx}, PSR J1023$+$0038 sits in a less luminous region of the correlation space. \citet{2025MNRAS.536...99G} compared the radio and X-ray luminosities of known and candidate transitional millisecond pulsars (TMSPs), including  PSR J1023$+$0038, and found that their X-ray luminosities all sit at or below $10^{35}$\,erg/s, making them far less luminous than other NS systems. The authors also showed that, compared to NS LMXBs, TMSPs tend to be more radio luminous for a given X-ray luminosity, occasionally more so than black hole X-ray binaries. Therefore, even though we have proposed similarities between TMSPs and false widows, the relation between the radio and X-ray counterparts across the two classes appears quite different and implies a gap in our understanding of false widow evolution.


\section{Optical/UV Analysis}
\label{sx:optical}
\subsection{Colour-Magnitude Diagram}
\label{subsec:cmd}

\begin{figure}
    \centering
    \includegraphics[width=\columnwidth]{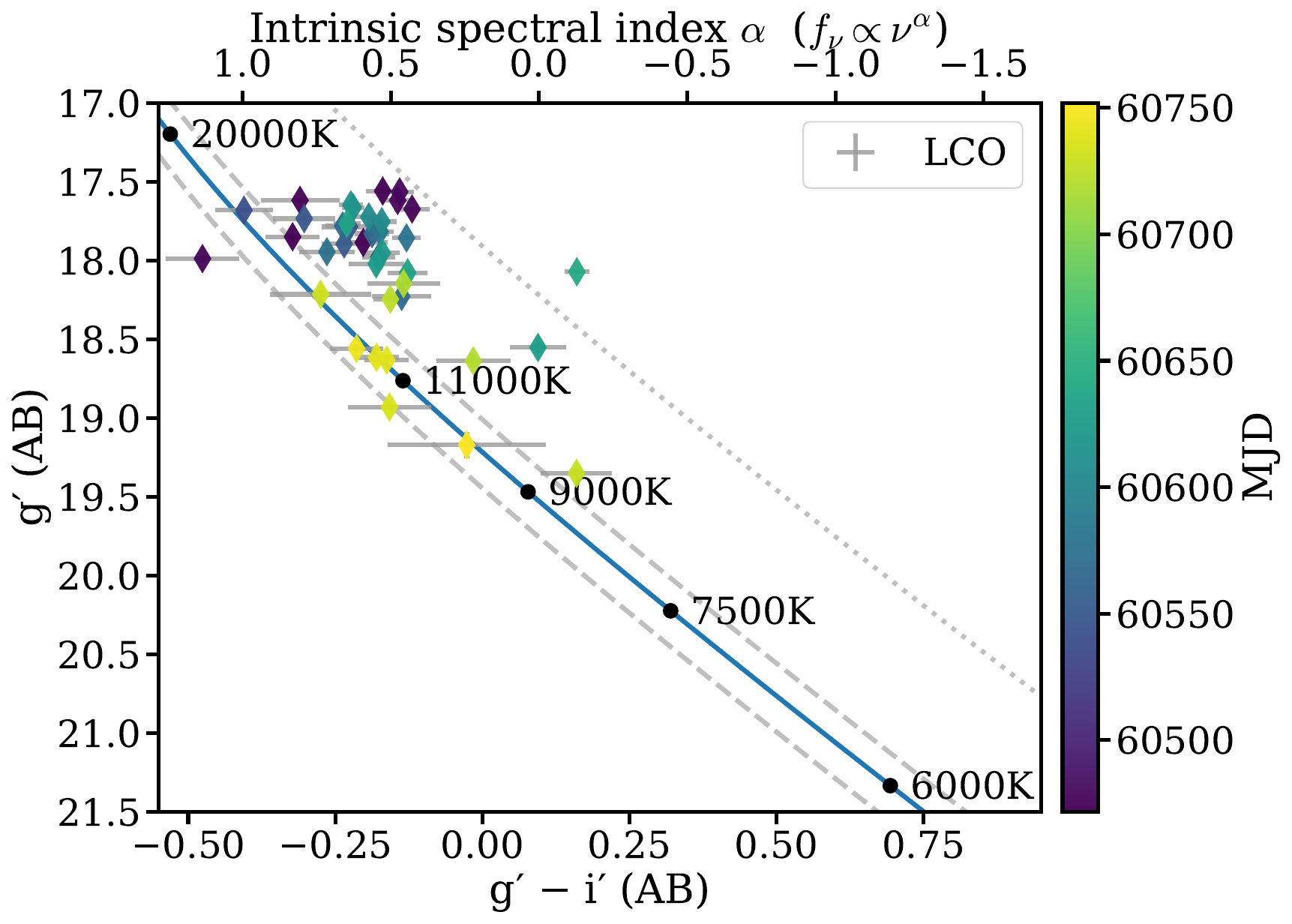}
    \vspace*{-0.5cm}
    \caption{Colour–magnitude diagram of EXO 0748-676 as observed by LCO. The solid line represents the expected colours of blackbodies at different temperatures with a fixed circular emitting area with a radius of $0.42\,\mathrm{R_\odot}$ at a distance of $7.1 $~kpc \citep{Galloway2008b}, showing the evolution with temperature. Circles along the solid line mark the corresponding positions in the diagram for the different labelled temperatures. The dashed lines illustrate the same relation but with surface areas varying by $\pm10\%$. The dotted line represents the tidal radius of a binary system using the parameters derived from \citep{Knight2022b}. An extinction of $E(B-V)=0.06$ was applied to the models \citep{Hynes2006}. Diamonds indicate the individual LCO measurements, colour-coded according to the corresponding MJD.
    }
    \label{fig:color_mag}
\end{figure}

Figure \ref{fig:color_mag} shows the colour-magnitude diagram \citep[CMD; e.g.][]{MaitraBailyn:2008ApJ...688..537M,Russell_CMD:2011MNRAS.416.2311R} for the 2024/25 outburst of EXO 0748 using quasi-simultaneous (i.e. back-to-back) observations from LCO. The CMD is used to study outburst evolution of LMXBs as it can detect changes in the dominant emission mechanism at optical wavelengths. When the optical emission is dominated by the accretion disc, the CMD offers a way to infer how the disc temperature structure evolves, especially during the rise and decline of outbursts \citep[e.g. ][]{Zhang:2019ApJ...876....5Z,Baglio:2020ApJ...905...87B,2022ApJ...932...38S, Baglio_cig_x-4:2022ApJ...930...20B, Rout_MAXI1807:2025ApJ...988..153R}. We note that, in LMXBs, the possible contribution of non-thermal emission from a jet in the optical regime makes these inferences challenging in most cases \citep[e.g.][]{2006MNRAS.371.1334R}. 

In addition to the quasi-simultaneous LCO observations, to guide the eye, we have overlaid the colours and magnitudes expected from circular blackbodies at different temperatures with radius $R=0.42\,\mathrm{R_\odot}$, as a representative value for the donor radius at a distance of 7.1$\pm$0.1\,kpc extinguished by $E(B-V)=0.06$ \citep{Galloway2008b, Hynes2006}. 
The area of these blackbodies is not intended as a physical estimate of the disc but as a fiducial reference: it represents the smallest radius expected if the donor dominated the emission \citep[$T_{\mathrm{donor}}\gtrsim5000\,\mathrm{K}$, e.g.][]{Bassa2009}. The accretion disc is expected to extend over a larger physical scale, while the effective reprocessing area depends on poorly constrained geometric factors, including the disc inclination, vertical structure, warping, and the fraction of the donor surface visible and irradiated. Therefore, the blackbody tracks should be interpreted as illustrative temperature guides rather than direct measurements of physical emitting radii.

Unfortunately, the rise component of the outburst was not covered by the LCO monitoring, which was used in this analysis. During most of the outburst (up to approximately 250 days after onset), the majority of the data points cluster around $g'\simeq18$ mag ($\sim580\,\mu$Jy) and $(g'-r')\simeq-0.2$, with a few outliers near $(g'-r')\simeq0.1$. The outlying values at $\rm MJD\simeq 60626\,\rm{d}$ and $60642\,\rm{d}$ are obtained within a few percent of the mid-eclipse in orbital phase ($\phi_{\rm orb}\sim0$), we therefore attribute these outlying data points to contamination from dips or eclipses, where the colour change originates from the non-simultaneity of the different bands. In addition to the colour $(g'-r')$, we also add the spectral index between the two LCO bands on the top axis of Figure \ref{fig:color_mag}. The spectral index for most epochs sits in the range of  $0.5\lesssim\alpha\lesssim1$, for a spectral shape $f_\nu\propto\nu^\alpha$. 
The spectral index alone is not sufficient to determine the origin of the optical emission, we therefore computed the relation between the optical and X-ray variation ($L_{\rm opt}\propto L_{\rm X}^{\beta}$, see Fig.~\ref{fig:Lopt-Lx}), obtaining $\beta=0.72\pm0.16$ (for \textit{Swift}-XRT and ATLAS $o$-band, since they are the only simultaneous X-ray and optical observations with sufficient dynamic range). 
While not conclusive, the combination of the colour/spectral index and the $L_{\rm opt}$–$L_{\rm X}$ relation favour a scenario in which the optical emission is dominated by X-ray reprocessing, most likely in the irradiated disc, with additional contributions from the donor star and/or extended material in the system \cite[c.f.][]{2006MNRAS.371.1334R, Russell_CMD:2011MNRAS.416.2311R}. 

Without invoking a detailed geometric model, the blackbody tracks in Figure \ref{fig:color_mag} provide a useful framework for interpreting the outburst evolution in the context of the disc instability model (DIM; see \citealt{Lasota_DIM_review:2001NewAR..45..449L} and \citealt{Hameury_DIM_review:2020AdSpR..66.1004H} for reviews). Despite the lack of quasi-simultaneous observations during the rise of the outburst, the later stages of the outburst are well sampled and trace a distinct evolution. During the main outburst phase, most measurements are clustered in a relatively small region of the diagram, suggesting that the source maintains a high, approximately constant effective area. The brightest points are also slightly above and/or redder than the simple blackbody guide tracks. Together with the orbital modulation observed during the bright phase (see Section \ref{subsec:phase}), this suggests that the optical light cannot be produced by the irradiated disc alone. A natural additional contributor to the scatter is reprocessed emission from the donor star and/or ablated material, which would be cooler and redder than the disc since it is located at larger radii.

The points during the main outburst represents the disc near peak expansion. The evolution from MJD $\sim$60650 onwards shows a drop in flux at approximately constant temperature before evolving along a roughly constant area track in which the temperature decreases steadily (from MJD\,$\sim60700$). A declining temperature at a constant effective area is characteristic of a cooling irradiated disc, as seen during the decay of several transients \citep[e.g.][]{MaitraBailyn:2008ApJ...688..537M,Zhang:2019ApJ...876....5Z,2022ApJ...932...38S}. We therefore interpret this phase as the cooling of the dominant irradiated optical component, most likely associated with the outer accretion disc. In the absence of ultraviolet data, the CMD alone does not allow us to determine whether the thermal evolution proceeds from the inside out or outside in, since the optical bands primarily trace the outermost part of the disc.  
Instead, a key observable appears after the first drop in flux, at MJD\,$\sim60750$, where the effective temperature steadily drops while the effective emitting area remains approximately constant when it falls below $\sim11000\,\mathrm{K}$. This behaviour was observed in the black hole transient Swift~J$1753.5-0127$ \citep[][]{Zhang:2019ApJ...876....5Z} and interpreted as the onset of hydrogen recombination in the disc, marking the start of the DIM cooling front once the ionized--neutral transition
($\sim11000-6000\,\mathrm{K}$) is reached. This interpretation accounts well for the behaviour in EXO~0748, and to our knowledge represent the first identification of such transition in a NS LMXB. 

\subsection{Optical Orbital Phase-Folded Light Curves}
\label{subsec:phase}

\begin{figure}
    \centering
    \includegraphics[width=\columnwidth]{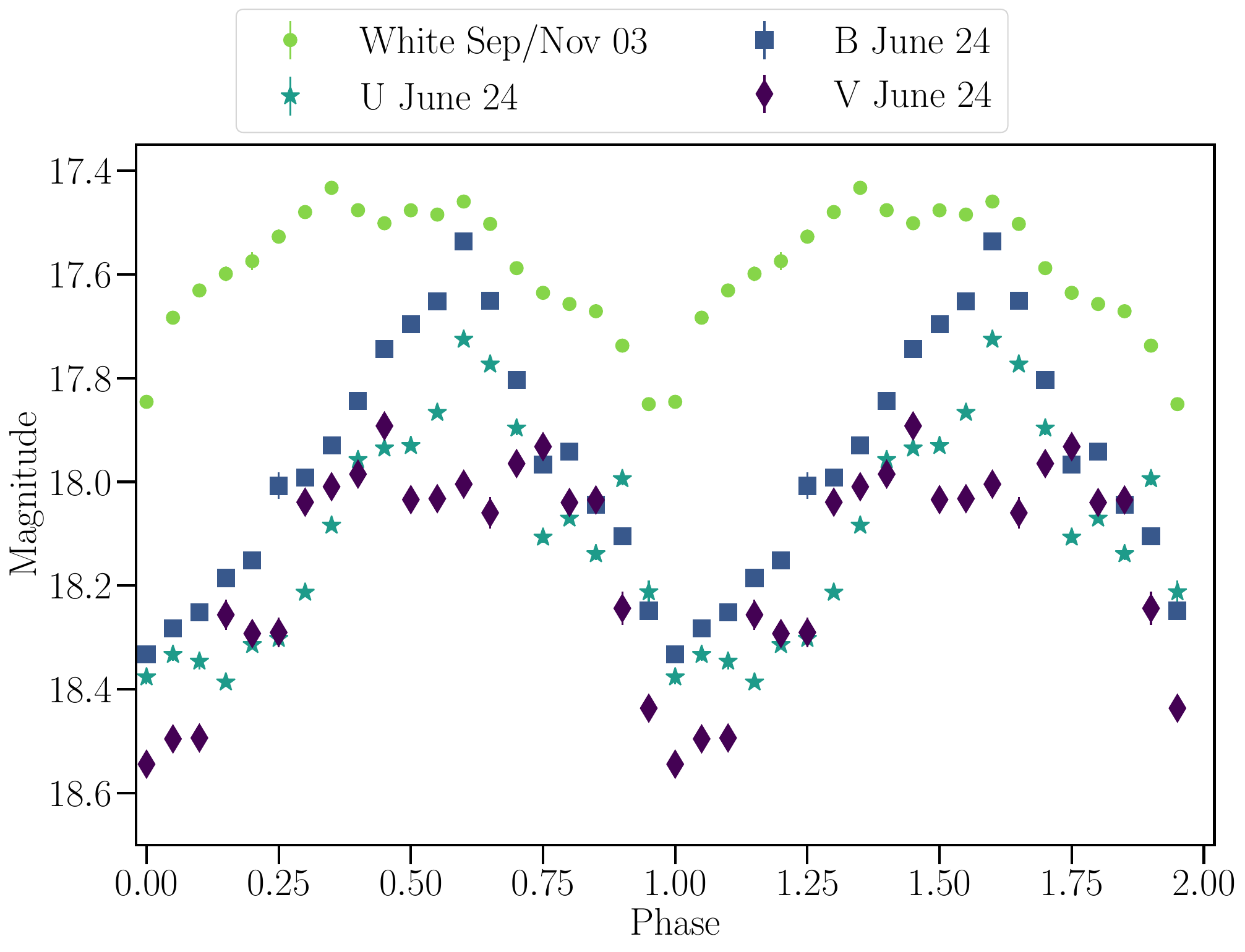}
    \caption{The optical phase curve for EXO 0748 using the \textit{XMM-Newton} fast photometry data  \citep{Knight2025}. The \textit{U}- and \textit{B}-band data covers two full orbits and the \textit{V}-band data covers just over one orbit. The uncertainties on each data point are the standard error for that phase bin. We also show the \textit{White} broadband phase curve (circles) constructed using nearly five days of observing time between September and November 2003 also using \textit{XMM-Newton}.}
    \label{fig:phase_curve}
\end{figure}

Figure \ref{fig:phase_curve} shows the optical orbital-phase folded light curve for EXO 0748, in the \textit{U}, \textit{B} and \textit{V} bands, constructed from \textit{XMM-Newton} ObsID 0935191701 (June 2024; OM fast; see Section \ref{subsubsec:OM}). While it is possible to produce phase curves from the ATLAS and LCO data, which cover the majority of the outburst, we choose not to include these data here as it is challenging to confidently separate any phase-dependent changes from other outburst variations, e.g. accretion rate variations. Therefore, we only use the \textit{XMM-Newton} OM data for this analysis. We utilise the most up-to-date orbital ephemeris to perform the orbital-phase folding (a complete derivation and details of the updated ephemeris will be presented in Buisson et al. \textit{in prep}) and define phase zero as the point of superior conjunction (centre of the NS eclipse). We also exclude the times ($\pm 10$~minutes) of known type I X-ray bursts.

As seen in Figure \ref{fig:phase_curve}, EXO 0748 is brightest between phases $0.6 - 0.7$, and faintest at the point of superior conjunction (phase zero). This is clearly seen in the \textit{B} (squares) and \textit{U} (stars) bands. The \textit{U} and \textit{V} (diamonds) bands appear similar in magnitude, but are both consistently fainter than the \textit{B} band. The exception to this is orbital phases $0.5 - 0.7$ where there is a notable drop in the \textit{V} band curve, which we speculate could be due to extinction within the binary system. We further find that the decay portion of the phase curve is steeper than the rise, causing it to appear as a skewed sinusoid. Asymmetric optical orbital modulation of this kind is atypical during outburst, but has also been observed in the false widow candidate \textit{Swift} J1858.6--0814 \citep{Rhodes2022}. This similarity suggests that optical modulation of this kind may be unique to false widows and we discuss this possibility further in Section \ref{sx:disc}.

Also shown in Figure \ref{fig:phase_curve} is the optical phase-folded light curve, constructed from the \textit{XMM-Newton} \textit{White} band data taken during the latter stages of the 1985 outburst (circles; September and November 2003). As shown, these data exhibit a higher flux density and display different orbital phase dependencies. Compared to the 2024/25 outburst, the 2003 phase curve appears more symmetrical about phase $\sim 0.5$, albeit with some scatter, and shows a double-peaked shape. Although we acknowledge that the lack of colour information hinders a more precise comparison between the 1985 and 2024/25 outbursts. Nonetheless, this indicates an evolution in the phase-resolved behaviour in EXO 0748. We will explore this evolution fully in a forthcoming study (Knight and Rhodes, \textit{in prep}), but discuss briefly in Section \ref{sx:disc}. 

\section{Discussion}
\label{sx:disc}
After sixteen years in quiescence, EXO 0748 unexpectedly returned to outburst in 2024 and remained active for $\sim 10$ months; a stark difference to its only other recorded outburst, which lasted more than two decades. Here, we have utilised observations from nine facilities, covering radio, optical/UV and X-ray wavelengths, to explore the differences between the two outbursts and improve our understanding of the evolution and outburst mechanisms of accreting sources.

\subsection{Outburst Duration and Light Curve Morphology}
\label{sx:disc_lc}
The most obvious difference between the 2024/25 outburst of EXO 0748  and the 1985 - 2008 outburst is the duration. Having been active for over two decades, EXO 0748 was previously assumed to be a persistent NS LXMB (e.g. \citealt{Hynes2006}), until 2008, when it was first observed to enter quiescence. In contrast, the 2024/25 outburst lasted only $\sim 10$ months. It is for this reason we refer to the 2024/25 outburst as `short'. However, we note that for other classes of X-ray transients, a 10-month outburst is not necessarily short and would be considered long for Very Faint X-ray Transients (VFXTs; see \citealt{Stel2026} for an example). Regardless, the short 2024/25 outburst and its first recorded X-ray quiescence (2008 - 2024; see e.g. \citealt{Hynes2009}, \citealt{Degenaar2009} and \citealt{Degenaar2011} for discussions of EXO 0748's first quiescence) demonstrates that EXO 0748 should now be considered a transient source.

Based on its correlated X-ray timing and spectral behaviour, EXO 0748 is considered an atoll source, having clearly displayed three of the four main spectral states during its 1985 - 2008 outburst (see \citealt{Mancuso2019} and references therein). As such, we investigate how the transient behaviour of EXO 0748 aligns with the broader population of atoll sources. Atoll sources are usually persistently accreting sources (e.g. GX9$+$1 and GX3$+$1; \citealt{vandenBerg2017, Molkov1999, Lutovinov2003}), but can be transient sources (e.g. 1A 1744-361 \citealt{2024ApJ...966..232N, Wang2025}), with outbursts typically lasting less than a year. The duration of the 2024/25 outburst is, therefore, well-aligned with the transient atoll population, while its previous long-duration outburst would have comfortably placed it within the persistent group (before it entered its first observed quiescence). Notably, transient LMXBs, with outbursts of comparable duration to that seen here, can exhibit canonical (FRED) light curves. However, the stable X-ray flux and optical flux density of EXO 0748 during the majority of the 2024/25 outburst (see Figure \ref{fig:lightcurve}) is, morphologically speaking, more characteristic of persistent sources or low-intensity states such as that observed in Aquila X-1 \citep{MaitraBailyn:2008ApJ...688..537M}. The reasonably stable X-ray flux seen here is also reminiscent of the stable X-ray flux observed in EXO 0748's previous outburst, before it entered a period of more variable flux in early 2004 (see e.g. figure 1 of \citealt{Homan2015}). Nonetheless, EXO 0748's dramatic change in outburst duration clearly demonstrates that an individual source can exhibit behaviours of both the persistent and transient atoll groups, or even that atoll sources can change from persistent to transient. 

Within the framework of the DIM, whether a source appears as persistent or transient depends on whether the mass transfer rate is, respectively, higher or lower than the critical value required to keep the accretion disc thermally stable (fully ionised). Interestingly, the upper limit and new estimations of EXO 0784's mass transfer rate lie in the interface between transient and persistent sources \citep[][Buisson et al in prep.]{Coriat:2012MNRAS.424.1991C, Dubus:2019A&A...632A..40D}. Therefore, variations in the long-term mass transfer rate could, in principle, move a system between these regimes (e.g. \citealt{Lasota_DIM_review:2001NewAR..45..449L, Hameury_DIM_review:2020AdSpR..66.1004H}). Causes of such variation may be geometric (e.g. a change in the orbital period due to mass and angular momentum transfer) or evolutionary (e.g. the onset of a transitional stage). Here, the latter is an interesting prospect. False widows may evolve through a transitional pulsar-like stage on their way to becoming spider pulsars, during which periods of accretion-powered activity are separated by periods of rotation-powered activity (see e.g. \citealt{Linares2014} for discussion of transitional redback pulsars). Therefore, the short-lived 2024/25 outburst and newfound X-ray transience of EXO 0748 may indicate the onset of such a stage. Although, it is important to note that without definitive observations of a false widow transitioning into a spider pulsar, a transitional false widow stage remains hypothetical. Regardless, we note that the DIM predicts FRED-light curve behaviour and soft X-ray spectra, which is at odds with some black hole X-ray binaries and VFXTs \citep{2004NewA....9..249B, 2013MNRAS.434.1586A}, and the 2024/25 outburst of EXO 0748. 

Another possible interpretation of the differing outburst lengths is that EXO 0748's first observed outburst was a full (major) outburst, while the 2024/25 outburst was a short (mini) outburst. Mini outbursts often occur some time after a major outburst and have been observed in various accreting systems, albeit typically with a faster outburst timescale \citep{MD2017, Zhang:2019ApJ...876....5Z, Ozbey2022, 2023ApJ...949..104S}. However, the characteristics of mini outbursts are wide-ranging. For example, \citet{2023ApJ...949..104S} and \citet{Ozbey2022}, respectively, discuss mini outbursts in \textit{Swift} J1910.2-0546 and \textit{MAXI} J1820$+$070. Unlike the 2024/25 outburst of EXO 0748, both sources exhibit FRED-like light curves during their mini outbursts. Also, as discussed by \citet{MD2017}, the 2015 mini outburst of V404 Cygni included intense flares and a fast optical wind similar to those seen during its major outburst. However, the mini-outbursts of \textit{Swift} J1753.5-0127 \citep{Zhang:2019ApJ...876....5Z} and \textit{MAXI} J1348-630 \citep{2026ApJ...997....7F} both exhibit relatively flat optical and X-ray light curves, which are morphologically similar to the light curves presented in Figure \ref{fig:lightcurve}. We also note that \textit{Swift} J1753.5-0127 exhibited a `persistent' style outburst, lasting more than a decade, prior to its mini outburst.
These similarities tentatively permit a mini-outburst interpretation for the 2024/25 outburst of EXO 0748.

To assess this possibility further, we apply the classification system of \citet{Zhang:2019ApJ...876....5Z}. Within this classification system, the 2024/25 outburst of EXO 0748 can be classified as a mini outburst if $\Delta~\rm{t_{i}}/\rm{t_{out}} \leq 1$ and $\rm{F_{i}}/\rm{F_{out}} < 0.7$. Alternatively, it would classify as a new outburst if $\Delta~\rm{t_{i}}/\rm{t_{out}} > 1$ or if $\Delta~\rm{t_{i}}/\rm{t_{out}} \leq 1$ and $\rm{F_{i}}/\rm{F_{out}} > 0.7$. Here, $\Delta~\rm{t_{i}}$ is the time separating the start of the quiescent period from the start of the rebrightening, $\rm{t_{out}}$ is the duration of the main outburst and $\rm{F_{i}}/\rm{F_{out}}$ is the ratio of the rebrightening peak flux (in a given waveband) to the outburst peak flux. Taking the duration of the 1985 outburst as $\sim 24$ years and the duration of the quiescence as $\sim 16$ years, we find that $\Delta~\rm{t_{i}}/\rm{t_{out}} = 0.667$ (3sf). Given the long duration of the 1985 outburst, we cannot reliably locate a single peak flux measurement. Instead we employ an average $0.5 - 10$ keV value of $2 \times 10^{-10}$ erg s$^{-1}$ cm$^{-2}$ in line with measurements from \citet{sidoli2005}, \citet{Boirin2007} and \citet{DiazTrigo2006}. For consistency, we compare it to the average $0.5 - 10$ keV flux measured from this outburst, $1.13 \times 10^{-10}$ erg s$^{-1}$ cm$^{-2}$. With these values we determine $\rm{F_{i}}/\rm{F_{out}} = 0.565$ (3sf). Therefore, the 2024/25 outburst does classify as a mini outburst within this framework. However, as noted by \citet{2021MNRAS.507.5507A} mini outbursts might sometimes be confused with failed transition outbursts (a possibility we discuss in Section \ref{sx:disc_spec}) because both events reach lower X-ray fluxes compared to a main outburst, thus questioning this interpretation.

Nonetheless, if we assume the mini outburst interpretation, the outburst timescale of EXO 0748 would fall between that of typical transient and persistent sources. Given the relative normalcy of its binary properties, it is unclear why its outbursts would be longer than typical transients, and fitting it into the short end of `persistent' outbursts is also challenging. One reason this is challenging is due to the uncertainty in outburst durations in sources considered persistent, as data are only available since the beginning of human monitoring ($\sim50$\, years for X-rays). Extending this by a significant factor will be slow; instead, we must either find a record of outbursts `frozen' into natural phenomena (such as echo mapping re-emission from nearby gas clouds; \citealt{Peterson2004}), using The Harvard Plate Stacks~\footnote{\url{https://platestacks.cfa.harvard.edu/}} to assess the duration of historic outbursts, or leave this investigation to future generations. For completeness, we searched for outburst signatures in the optical light curves from the Harvard Plate Stacks but find no reliable detections. Their automated pipeline reported two detections at about 10\textsuperscript{th}\ mag and a handful at around 17\textsuperscript{th}\ mag in 1941 (similar to measurements from the previous active phase). On inspection of the digitised photographic plate images, we determine that the 10\textsuperscript{th}~mag detections are tracking artefacts rather than real detections. We further inspected the plate images corresponding to each of the 17\textsuperscript{th} detections manually. However, EXO 0748 was not present in any of the plate images. Nearby known sources which are significantly brighter than EXO 0748 (13\textsuperscript{th} and 15\textsuperscript{th}\ mag) are also not present in the images, so we do not consider the 17\textsuperscript{th}~mag pipeline detections as reliable. Therefore, we conclude that the sensitivity of the digitised Harvard Plates is not sufficient to determine whether or not EXO 0748 was active in the 1940s. 

In any case, determining the true outburst durations of persistent sources, as with other population-related properties, is most easily achieved with monitoring of a wide variety of sources. Continued wide-field X-ray monitors are, therefore, vital to monitor both persistent and transient LMXBs to improve our understanding of the true distribution of outburst timescales. 

\subsection{Absorption and Spectral Evolution}
\label{sx:disc_spec}
In Sections \ref{sx:obs} - \ref{sx:lrlx}, we have demonstrated that during the first part of the 2024/25 outburst, EXO 0748 was partially obscured by a local absorbing medium that preferentially absorbs soft X-rays. We suggest that this medium disperses as the outburst progresses, based on the slight rise in X-ray flux during the second half of the outburst and that many of the later spectra did not require a local absorber to achieve a statistically good fit. However, this should not be considered conclusive. 

The cause of the early outburst absorption remains unclear. From all spectral fits including a partial covering absorber, we characterise the absorbing medium as a neutral/low ionisation medium with an average column density of $N_{\rm{H}} \sim 5 \times 10^{21}$\,cm$^{-2}$ and varying covering fractions in the range 0.6 - 0.9. There are several possible causes of the absorption in EXO 0748. The first is the material is ablated from the companion star, which is consistent with the inferred density and covering fraction of the local absorber. However, ablated material is known to be highly ionised and thus is unlikely to be the primary cause of the absorption observed towards the start of the outburst. We also note that, while the energy deposited through X-ray heating allows the liberated material to escape the surface of the companion star, it remains gravitationally bound to the system, and so we would not expect it to vanish on short timescales. However, analysis of \textit{NuSTAR} eclipses in the same observations used here by \citet{2026JHEAp..5300595S} reveals longer ingress/egress at softer energies that sharpen with increasing X-ray energy. This is a characteristic of ablation-induced eclipse energy dependence \citep{Knight2022a, Knight2022b, Knight2023} and suggests that some ablation-related absorption is present at near-eclipse phases. If so, this would imply that the companion star in EXO 0748 continued to be ablated during X-ray quiescence, suggesting a radio-pulsar phase. This would align with the false widow evolution and support the interpretation that this short outburst marks the start of the transitional phase for EXO 0748. 

An alternative scenario is that the absorber is a disc wind or structure that is thickest around where the accretion stream would meet the disc edge, similar to that observed in \textit{Swift} J1858.6--0814 \citep{2021MNRAS.503.5600B} or an accretion disc corona (ADC) as discussed by \citet{Church2004}. However, it is also unclear whether such a structure would disperse over the course of a few months. Due to the lack of observations during the rise of EXO 0748 into its 1985 outburst, it is unclear whether the early absorption is a repeated feature. This question can only be answered with detailed X-ray observations of a future rise into outburst. 

A second notable difference in the spectral behaviour of EXO 0748 between its two recorded outbursts is the lack of a clear soft spectral state during the 2024/25 outburst, which we demonstrate in Figures \ref{fig:CCD} and \ref{fig:HID}. Much like its 1985 outburst, EXO 0748 spent the majority of its time in the extreme island (hard) state. As the local absorbing medium disperses, we see EXO 0748 venturing into the island (hard/intermediate) state, but quickly returns to the extreme island states as it fades towards quiescence. In this regard, the CCD of the 2024/25 is somewhat unremarkable; EXO 0748 occupied the same area of the CCD as in its 1985 outburst. However, as shown in Figure \ref{fig:CCD}, the soft colours span a much larger range than typically observed in other atoll sources (see e.g. \citealt{Fred2011} for a comparison of the CCDs of other atoll sources). This wide range of soft colours was also noted by \citet{Homan2015} during an analysis of EXO 0748's previous outburst. The authors attributed this to incomplete dip removal, which predominantly impacts the low-energy bands and therefore softer colours. As noted in Section \ref{sx:obs}, we have taken every effort to remove sources of variability (e.g. bursts, dips) for the colour analysis. However, the complete removal of dips was challenging, so the removal may be incomplete here as well. There are additional effects which may play a role in widening the range of soft colours, such as the bandpass corrections for all instruments to the \textit{RXTE} bandpass. Even so, there is some overlap between 2024/25 measurements and the hardest \textit{RXTE} points from the previous outburst, suggesting a gap in our understanding of the spectral evolution of atoll sources. The extended range of soft colours may also be connected to the lack of a banana state in EXO 0748, but exploring this is beyond the scope of this paper. 

In its 1985 outburst, EXO 0748 appeared in three clear intensity states within the HID: low/hard, intermediate and high/soft states (see e.g. \citealt{Homan2015}, \citealt{Sch1999}, \citealt{Ponti2014} and \citealt{MD2014}). The majority of its time was spent in the low-intensity hard spectral state, but did transition into high-soft states on several occasions, during which the fractional RMS/fast variability would drop to around 5 per cent \citep{Ponti2014} in a manner consistent with other atoll sources and black hole systems \citep{MD2014}. However, EXO 0748 did not transition to a high-intensity, soft state during the 2024/25 outburst. As the local absorbing medium dispersed, EXO 0748 migrated a towards the intermediate state (same observations captured in the island state), but this should not be considered as a genuine state transition.

The previously observed transitions of EXO 0748 to the high-intensity soft state on the HID are diagonal, which, as discussed by \citet{2007MNRAS.378...13G}, is not uncommon for atoll sources. \citet{MD2014} explored these transitions on the HID, finding that the count-rate increases during the hard-to-soft transition and decreases during the reverse transition. This means that the state transitions are never parallel, like those observed for many black hole systems. Interestingly, \citet{MD2014} also found EXO 0748 exhibited hysterisis loops of 60-90 day durations during the 1985 outburst, which feature diagonal transitions between the high-intensity soft state and low-intensity hard state. No such behaviour is found during the 2024/25 outburst, even though the outburst duration was long enough to include a 60-90 day hysterisis loop. However, the brief transitions of EXO 0748 into the intermediate region of the HID and back, during the 2025/25 outburst, are diagonal, as expected. One explanation for the lack of full hysterisis loop and transition into a bright, soft state could be that EXO 0748 \textit{behaves} like a quasi-persistent LMXB and thus the source simply wasn't in outburst long enough to exhibit a full state change. However, we favour the interpretation that the 2024/25 outburst is a failed transition outburst, having not developed into a canonical soft state and instead remaining hard and partially obscured for the entire outburst, similar to the sources discussed by \citet{2021MNRAS.507.5507A}. However, the authors also note that during a mini outburst (which could be an explanation for the 2024/25 outburst of EXO 0748; see Section \ref{sx:disc_lc}) a source can either stay in a faint, hard state or it can undergo a transition to a brighter soft state. Thus the lack of an observed state transition does not permit a definitive explanation.

While outbursts without a state transition can be dubbed as \textit{failed} for black hole transients, it is not as uncommon for NS systems. Notably, the majority of the accreting millisecond X-ray pulsars do not transition between the hard and soft states during their outbursts (see e.g. \citealt{DiSalvo2022}), typically resulting in their classification as hard X-ray transients. As shown by \citet{2007MNRAS.378...13G}, accreting millisecond X-ray pulsars often spend their entire outburst in the island (hard) states and are generally observed at lower luminosities than atoll sources. As such, one can draw similarities between the outburst properties of accreting millisecond X-ray pulsars and the faint, hard 2024/25 outburst of EXO 0748. These similarities provide support the interpretation that this short outburst marks the start of the transitional pulsar-like phase for EXO 0748. 

\subsection{False Widow Ablation and Optical Phase Curves}
Perhaps the most intriguing finding in this work is the non-standard, asymmetric optical orbital phase curve with a peak at $\phi \sim 0.7$ (Figure \ref{fig:phase_curve}), observed during an outburst. In contrast, optical orbital modulation is often observed in irradiated systems during quiescence, where the phase curve is brightest at $\phi \sim 0.5$, when the face of the irradiated companion star is pointing at the observer. However, EXO 0748 is not the only system to exhibit this atypical behaviour. The false widow candidate, \textit{Swift} J1858.6--0814, displayed a very similar optical phase curve (an asymmetric, skewed sinusoid) while in outburst, also with a peak at $\phi \approx 0.7$ \citep{2025MNRAS.536.3421R}. In \citet{2025MNRAS.536.3421R}, it was suggested that the atypical phase curve was a result of excess ablated material trailing behind the companion star as it orbits, thereby increasing the reflected/reprocessed radiation from the accretion disk at phases around $\phi = 0.7$ and/or increased blocking of light from the disk at phases around $\phi = 0.25$. As EXO 0748 is also evidenced to be undergoing ablation, we suggest that similar reflection/obscuring processes are likely occurring here. However, it is prudent to note that EXO 0748 shows recurring pre-eclipse dips between orbital phases $0.6 - 0.9$ (see e.g. \citealt{Ramam2017} and references therein). These dips indicate that the accretion stream impact point (hotspot) in the outer disk is suitably positioned to contribute to the observed, asymmetric optical modulation by increasing thermal emission observed between phases $0.6 - 0.9$. We will explore these scenarios further in a forthcoming study.

Figure \ref{fig:phase_curve} also shows evidence that the phase curve has evolved over the last two decades. Between 2003 and 2024, the peak of the phase curve shifts from 0.5 to $\sim$0.7. Also, \citet{Schmidtke1987A} published an optical phase curve using data collected in 1985 that peaks around phases $0.3 - 0.4$.

We propose that the ablated material is not constrained to the orbital path of the binary. As found by \citet{Knight2023} and \citet{2025MNRAS.536.3421R}, the ablated material can extend far around the orbital path of the binary. Over time, the material will gain angular momentum, forcing it to move out to larger orbits. The ablated material is still gravitationally bound to the system but any material trailing behind the companion star will slowly moves outwards, forming a material spiral or perhaps feeding a circumbinary disc. The extended, ablated material may be complemented by material ejected from the accretion disk in the form of winds. Over densities in the extended material at larger radii or perhaps even where the ablated material meets a circumbinary disk could change the phase curve shape on decades-long timescales. We will test our proposed scenario in future works. 

Arguably more interesting than the phase curve itself is the discovery of a second false widow system exhibiting asymmetric optical orbital modulation while in outburst. This discovery suggests that asymmetric optical orbital modulation may be a defining feature of a false widow system or simply any \textit{accreting} system undergoing ablation. Ablation is currently only found in `spider' systems: false widows, redback pulsars and black widow pulsars. While false widows ablate their companion stars through X-ray heating, pulsar wind radiation drives ablation in redback and black widow pulsars. Because of this, a direct comparison of their optical emission finds opposing results. The transitional millisecond redback pulsar PSR J1023+0038 shows asymmetric orbital phase curves during its \textit{pulsar} mode but not in its accreting mode \citep{2014MNRAS.444.1783C, 2015MNRAS.453.3461S, 2021MNRAS.507.2174S}; the opposite to that observed in false widows. This suggests that asymmetric phase curves are observable when the companion star heating mechanism is active. This is during the X-ray bright phase of false widows and the radio bright phase of TMSPs \citep{2021MNRAS.507.2174S}. Also, the redback pulsar PSR J2055$+$1545 shows evidence of asymmetry at optical wavelengths, where the asymmetry is attributed to asymmetric heating of the companion by the pulsar wind \citep{2025MNRAS.538..380T}; the same interpretation as for PSR J1023+0038. Interestingly, PSR J2055$+$1545 has a higher orbital inclination \citep[76--79$^{\circ}$;][]{2025MNRAS.538..380T}, than the PSR J1023+0038, similar to that of EXO 0748. This may indicate that a high orbital inclination is required to observe phase curve asymmetry and that the ablated material does not extend high above the orbital plane.

In addition, we find evidence of ongoing ablation in EXO 0748 in the CMD (Figure~\ref{fig:color_mag}): the blue spectral index and the optical--X-ray correlation ($\beta = 0.72 \pm 0.16$) indicate that the irradiated disc dominates the optical emission throughout the outburst. The brightest points lying above and redward of the single-temperature tracks point to an additional, cooler contribution from reprocessing on the face of the donor and by the ablated material. As the X-ray luminosity fades, these outer components are expected to weaken, leaving the constant-area irradiated disc as the dominant optical source, although the present data do not allow a quantitative decomposition. The CMD also traces the onset of the cooling front, seen as the rapid light-curve decline around MJD\,$\sim\,$60750. This transition occurs when the (colour) temperature drops below $\sim11 000\,\mathrm{K}$, consistent with the temperature range over which hydrogen recombines in an irradiated disc and the cooling front of the disc instability model is triggered. The same coincidence was reported for the black hole transient Swift~J$1753.5-0127$ \citep{Zhang:2019ApJ...876....5Z}, and to our knowledge this is the first such identification in a NS LMXB.

\section{Conclusion}
\label{sx:summary}
In this paper, we have utilised observations from the MeerKAT radio telescope, the LCO and Faulkes Telescopes, ATLAS, \textit{Swift} XRT, UVOT and BAT, \textit{XMM-Newton's} OM and EPIC detectors, \textit{AstroSat}, \textit{NICER}, \textit{NuSTAR} and \textit{MAXI} to present a multi-wavelength view of the 2024/25 outburst of EXO 0748--676 and the first-ever multi-wavelength coverage of a full outburst from this source, including both the rise and decay.

Compared to the previous outburst (1985 -2008), the 2024/25 outburst was notably shorter, lasting $\sim 10$ months, and barely matched the typical persistent X-ray luminosity achieved in the previous outburst (Figure \ref{fig:LrLx}). The morphology of the 2024/25 outburst light curve fails to show the hallmarks of a canonical transient LMXB outburst and instead remains stable for the majority, as if it were a persistent source. Therefore, we speculate that the recent outburst could be connected to evolutionary changes within the system. However, exploring this fully is beyond the scope of this paper.

The X-ray spectra reveal strong early-time obscuration, which suppresses the soft X-rays (Figure \ref{fig:Spectra}). Corresponding X-ray colour analysis reveals that EXO 0748--676 remained in a hard, partially obscured state for the majority of the 2024/25 outburst, appearing both fainter and harder than previously observed. The obscuration subsides during the latter stages of the outburst, allowing the source to soften slightly and move into the hard-intermediate regions of the CCD and HID (Figures \ref{fig:CCD} and \ref{fig:HID}). However, the source did not enter a true soft state during the 2024/25 outburst. This spectral behaviour, combined with the low observed luminosity \citep{2016ApJS..222...15T,2021MNRAS.507.5507A}, leads us to suggest that the 2024/25 outburst of EXO 0748--676 was a failed-transition outburst. 

Our optical analysis shows the 2024/25 outburst of EXO 0748--676 was consistent with an irradiation-dominated outburst, with the CMD (Figure \ref{fig:color_mag}) tracing the cooling of a large, approximately constant effective emitting area in the outer disc. The rapid fade at the end of the outburst occurs when the inferred temperature reaches $\sim11000,\mathrm{K}$, suggesting that the final decline is related to the onset of hydrogen recombination and the cooling-front transition in the irradiated disc. To our knowledge, this provides the first evidence for the onset of a cooling front at optical wavelengths in a NS LMXB. In addition, we find that the optical orbital phase-folded light curve (Figure \ref{fig:phase_curve}) is asymmetric, peaking at $\sim0.7$ rather than the expected position of 0.5. This behaviour appears characteristic of false widows, or more simply, accreting systems with ongoing ablation. The asymmetric phase curve is likely due to an asymmetric layer of material liberated from the companion star's surface that is gravitationally bound to the system, which can either block or reprocess light from the X-ray-emitting regions. 

Overall, our results suggest that EXO 0748--676 lies in a regime between persistent and transient LMXBs, where obscuration, irradiation, and ablation all play a role in shaping the observed outburst behaviour. Studying future outbursts of EXO 0748--676 or other false widow systems is thus essential to improve our understanding of these effects on the outbursts and evolution of NS LMXBs. 

\section*{Acknowledgements}
\subsection*{Communications}
LR thanks J. Hessels for insightful conversations that led to some of the interpretations in this work. AHK gratefully acknowledges valuable conversations with A. Ingram that aided the structure and presentation of the results in this paper. AHK thanks G. Mancuso for the provision of the \textit{RXTE} colour measurements (included with permission; originally from \citealt{Mancuso2019}. See also \citealt{Knight2025_ineclipse}) that appear in Figure \ref{fig:CCD} and their assistance with the Crab normalisation required to overlay the 2024/25 outburst data onto the \textit{RXTE} colour-colour diagram. DMR and AHK acknowledge communications with T. Mu\~{n}oz-Darias, whose perception and feedback prompted further consideration of the contributions of the accretion disc hotspot to the optical phase curve.

\subsection*{Funding}
AHK acknowledges support from the Science and Technology Facilities Council (STFC) as part of the consolidated grant award ST/X001075/1. 
LR acknowledges funding from the  Trottier Space Institute Fellowship, the Natural Sciences and Engineering Research Council of Canada (NSERC) Arthur B. McDonald Fellowship and Discovery Grant programs, the Canada Research Chairs (CRC) program, the Fondes de Recherche Nature et Technologies (FRQNT), the Centre de recherche en astrophysique du Québec (un regroupement stratégique du FRQNT), the Trottier Space Institute at McGill and the AstroFlash research group. The AstroFlash research group at McGill University, University of Amsterdam, ASTRON, and JIVE is supported by: a Canada Excellence Research Chair in Transient Astrophysics (CERC-2022-00009); an Advanced Grant from the European Research Council (ERC) under the European Union’s Horizon 2020 research and innovation programme (`EuroFlash’; Grant agreement No. 101098079); an NWO-Vici grant (`AstroFlash’; VI.C.192.045); an NSERC Discovery Grant (RGPIN-2025-06681); an ERC Starting Grant (`EnviroFlash’; Grant agreement No. 101223057); and an NWO-Veni grant (VI.Veni.222.295).
NCS acknowledge support from the Science and Technology Facilities Council (STFC) grant ST/X001121/1 and the Ministry of Science, Innovation and Universities project PID2023-151905OB-I00. 
DMR is supported by Tamkeen under the NYU Abu Dhabi Research Institute grant CASS. 
M.C.B. acknowledges support from the INAF Fundamental Research Grant XBOOM. 
JvdE was supported by funding from the European Union's Horizon Europe research and innovation programme under the Marie Skłodowska-Curie grant agreement No 101148693 (MeerSHOCKS).
MM acknowledges support by the Science and Technology Facilities Council (STFC) as part of the small grant award ST/Y001699/1.
CD acknowledges support from the Science and Technology Facilities Council (STFC) through the grant ST/T000244/1.

\subsection*{Instruments, Facilities and Software}
The Asteroid Terrestrial-impact Last Alert System (ATLAS) project is primarily funded to search for near earth asteroids through NASA grants NN12AR55G, 80NSSC18K0284, and 80NSSC18K1575; byproducts of the NEO search include images and catalogs from the survey area. ATLAS is partially funded by Kepler/K2 grant J1944/80NSSC19K0112 and HST GO-15889, and STFC grants ST/T000198/1 and ST/S006109/1. The ATLAS science products have been made possible through the contributions of the University of Hawaii Institute for Astronomy, the Queen’s University Belfast, the Space Telescope Science Institute, the South African Astronomical Observatory, and The Millennium Institute of Astrophysics (MAS), Chile. This work makes use of observations from the Las Cumbres Observatory (LCO). The MeerKAT telescope is operated by the South African Radio Astronomy Observatory, which is a facility of the National Research Foundation, an agency of the Department of Science and Innovation. 

This research has made use of data from the Nuclear Spectroscopic Telescope Array (\textit{NuSTAR}), a project led by the California Institute of Technology, managed by the Jet Propulsion Laboratory, and funded by the National Aeronautics and Space Administration (NASA). Data analysis was performed using the NuSTAR Data Analysis Software (\textsc{nustardas}), jointly developed by the ASI Science Data Center (SSDC, Italy) and the California Institute of Technology (USA). This research has made use of data and software provided by the High Energy Astrophysics Science Archive Research Center (HEASARC), which is a service of the Astrophysics Science Division at NASA/GSFC. This research has made use of data from the Neutron star Interior Composition Explorer (\textit{NICER}), a 0.2-12 keV X-ray telescope operating on the International Space Station. The \textit{NICER} mission and portions of the \textit{NICER} science team activities are funded by NASA. This research has made use of data provided by the \textit{XMM-Newton} Science Archive and the \textit{XMM-Newton} Science Analysis Software (SAS) made publicly available by the European Space Agency (ESA). This research made use of data supplied by the UK \textit{Swift} Science Data Centre at the University of Leicester, the XRT Data Analysis Software (\textsc{xrtdas}) developed under the responsibility of the ASI Science Data Center (ASDC), Italy, and data obtained by the \textit{Swift-BAT} transient monitor \citep{Krimm2013}. This research has made use of MAXI data provided by RIKEN, JAXA, and the MAXI team. 

\section*{Data Availability}
The \textit{NICER} and \textit{NuSTAR} data are readily available from the High Energy Astrophysics Science Archive Research Center (HEASARC) website. The \textit{XMM-Newton} data are publicly available through HEASARC or from the \textit{XMM-Newton} Science Archive provided by the European Space Agency. The \textit{Swift} data used in this work is publicly available from either the UK \textit{Swift} Science Data Centre, HEASARC, or \textit{Swift}-BAT transient monitor websites. The MAXI data are publicly available through the MAXI monitoring website provided by RIKEN, JAXA, and the MAXI team. The ATLAS data is publicly available from \href{https://fallingstar-data.com/forcedphot/}{the ATLAS forced photometry server}. The LCO data published in this work are available upon request to the authors. The MeerKAT upper limits are provided in Table \ref{tab:radio_obs}.



\bibliographystyle{mnras}
\bibliography{EXO0748_2025} 



\appendix
\section{X-ray Spectral Fits and Energy Conversion Factors}
\label{sx:x-ray_appendix}
Here we provide a series of tables summarising all X-ray spectral fits performed in this paper and the energy conversion factors (ECFs) derived from them that required to reproduce Figures \ref{fig:lightcurve}, \ref{fig:CCD} and \ref{fig:HID}. To determine the ECFs, we supply PIMMS with the values of Galactic and local absorption and the values of the best-fitting spectral components (T$_{\rm{in}}$, kT and/or $\Gamma$, depending on the spectral model) for each observation. The resulting ECFs are the values needed to convert 1 ct/s in the original energy band to 1 erg/s/cm to the new energy band (for Figure \ref{fig:lightcurve}) or 1 ct/s in the original energy band to 1 ct/s in the specified \textit{RXTE} energy band (for Figures \ref{fig:CCD} and \ref{fig:HID}). These are computed on a per observation basis to provide a more accurate view of the temporal and colour evolution of the source over the course of the outburst.

\begin{table}
    \setlength{\tabcolsep}{4pt}
    \begin{tabular}{c c c c}
    Component & Parameter & Best-Fitting Value & Unit \\
    \hline
    \vspace*{+0.5cm}
    & EPIC Flux & $0.579 \pm 0.002$ & $\times 10^{-10}$ erg cm$^{-2}$ s$^{-1}$\\
    \vspace*{+0.1cm}
    \texttt{zxipcf} & Nh & $7.486 \pm ^{0.505}_{0.468}$ & $\times 10^{22}$ atom cm$^{-2}$\\
    \vspace*{+0.1cm}
    & $\log({\xi})$ & $0.123 \pm ^{0.117}_{0.116}$ \\
    \vspace*{+0.5cm}
    & f$_{\rm{cov}}$ & $0.758 \pm ^{0.020}_{0.025}$ \\
    \vspace*{+0.1cm}
    \texttt{diskbb} & T$_{\rm{in}}$ & $0.217 \pm ^{0.014}_{0.013}$ & keV \\
    \vspace*{+0.5cm}
    & Flux & $0.034 \pm 0.0004$ & $\times 10^{-10}$ erg cm$^{-2}$ s$^{-1}$ \\
    \vspace*{+0.1cm}
    \texttt{bbody} & kT & $1.693 \pm _{0.077}^{0.053}$ & keV \\
    \vspace*{+0.5cm}
    & Flux & $0.692 \pm 0.002 $ & $\times 10^{-10}$ erg cm$^{-2}$ s$^{-1}$ \\
    \vspace*{+0.1cm}
    \texttt{nthComp} & $\Gamma$ & $\geq 1.004$ & \\
    \vspace*{+0.1cm}
    & kT$_{\rm{e}}$ & $17.24 \pm_{16.22}^{165.3}$ & keV  \\
    \vspace*{+0.5cm}
    & Flux & $0.158 \pm 0.003$ & $\times 10^{-10}$ erg cm$^{-2}$ s$^{-1}$ \\
    \vspace*{+0.1cm}
    \texttt{gabs} & Line Energy & $0.755 \pm ^{0.007}_{0.008}$ &  keV \\
    \vspace*{+0.1cm}
    & $\sigma$ & $ 0.068 \pm _{0.009}^{0.010}$ & keV  \\
    \vspace*{+0.5cm}
    & Strength & $0.117 \pm ^{0.016}_{0.015} $ & keV \\ 
    \vspace*{+0.1cm}
    \texttt{gabs} & Line Energy & $1.026 \pm ^{0.015}_{0.018}$ &  keV \\
    \vspace*{+0.1cm}
    & $\sigma$ & $ 0.083 \pm _{0.026}^{0.022}$ & keV  \\
    \vspace*{+0.1cm}
    & Strength & $0.062 \pm 0.005 $ & keV \\ 
    \hline
    \end{tabular}
    \caption{\label{tb:xmm}Best-fitting spectral parameters and corresponding uncertainties obtained when fitting the three 0.5 -10 keV EPIC spectra from \textit{XMM-Newton} ObsID 0935191701 simultaneously within \textsc{xspec}. The spectra were fit with the model \texttt{tbabs*zxipcf(diskbb+bbody+nthcomp)*gabs*gabs}. The \texttt{tbabs} component is fixed to a value of $0.116 \times 10^{22}$ atom cm$^{-2}$, as determined by the Colden Galactic Neutral Hydrogen Density Calculator. We also use a constant factor to model the differences in detector calibration, finding that they are consistent to within $\sim 20~\%$. The first row shows the measured 0.5 - 10 keV combined EPIC flux, determined from the simultaneous spectral fit. The individual component fluxes, determined using \texttt{cflux}, are given with the relevant model component. This fit achieves $\chi = 566.11$ for $\nu = 465$ and a null-hypothesis probability of $8.94 \times 10^{-4}$.}
\end{table}

    \begin{table}
    \centering
        \setlength{\tabcolsep}{10pt}
        \begin{tabular}{c c c}
        \textit{XMM-Newton} & \textit{RXTE} & Conversion \\
        \vspace*{+0.1cm}
        Energy Band & Energy Band & Factor \\
        \hline
        \vspace*{+0.1cm}
        2 - 3.5 keV & 2 - 3.5 keV & 0.355 \\
        \vspace*{+0.1cm}
        3.5 - 6 keV & 3.5 - 6 keV & 1.041 \\
        \vspace*{+0.1cm}
        6 - 8 keV & 6 - 9.7 keV  & 2.494 \\
        \vspace*{+0.1cm}
        7 - 10 keV & 9.7 - 16 keV & 0.719 \\
        \vspace*{+0.1cm}
        0.5 - 10 keV & 2 - 16 keV & 1.330 \\
        \end{tabular}
        \caption{\label{tb:xmm_conversions} Energy conversion factors required to convert \textit{XMM-Newton} combined EPIC count rates in the original energy band to \textit{RXTE} count rates in specified energy band. The conversions are performed assuming the spectral model in Table \ref{tb:xmm}.}
    \end{table}

\begin{table}
    \setlength{\tabcolsep}{2pt}
    \begin{tabular}{c c c c}
    Component & Parameter & Best-Fitting Value & Unit \\
    \hline
    \vspace*{+0.5cm}
    & $0.5 - 10$ keV Flux & $0.840 \pm 0.002$ & $\times 10^{-10}$ erg cm$^{-2}$ s$^{-1}$\\
    \vspace*{+0.1cm}
    \texttt{zxipcf} & Nh & $5.402 \pm ^{1.010}_{0.151}$ & $\times 10^{22}$ atom cm$^{-2}$\\
    \vspace*{+0.1cm}
    & $\log({\xi})$ & $1.971 \pm ^{0.070}_{0.079}$ \\
    \vspace*{+0.5cm}
    & f$_{\rm{cov}}$ & $\geq 0.899$ \\
    \vspace*{+0.1cm}
    \texttt{bbody} & kT & $1.741 \pm _{0.024}^{0.017}$ & keV \\
    \vspace*{+0.5cm}
    & Flux & $0.296 \pm 0.004 $ & $\times 10^{-10}$ erg cm$^{-2}$ s$^{-1}$ \\
    \vspace*{+0.1cm}
    \texttt{cutoffpl} & $\Gamma$ & $0.594 \pm ^{0.046}_{0.034}$ & \\
    \vspace*{+0.1cm}
    & HighECut & $18.28 \pm_{0.26}^{0.23}$ & keV  \\
    \vspace*{+0.5cm}
    & Flux & $0.847 \pm 0.003$ & $\times 10^{-10}$ erg cm$^{-2}$ s$^{-1}$ \\
    \hline
    \end{tabular}
    \caption{\label{tb:nustar}Best-fitting spectral parameters and corresponding uncertainties obtained when fitting the 3 - 78 keV FPMA and FPMB spectra from \textit{NuSTAR} ObsID 91001324001 simultaneously within \textsc{xspec}. The spectra were fit with the model \texttt{constant*tbabs*zxipcf(bbody+cutoffpl)}. The \texttt{tbabs} component is fixed to a value of $0.116 \times 10^{22}$ atom cm$^{-2}$, as determined by the Colden Galactic Neutral Hydrogen Density Calculator. The constant factor models the differences in detector calibration, which are consistent to within $\sim 20~\%$. The first row shows the measured 0.5 - 10 keV flux, determined from the spectral model. The individual component fluxes, determined using \texttt{cflux}, are given with the relevant model component. This fit achieves $\chi = 2386.38$ for $\nu = 2339$ and a null-hypothesis probability of $0.243$.}
\end{table}

    \begin{table}
    \centering
        \setlength{\tabcolsep}{10pt}
        \begin{tabular}{c c c}
        \textit{NuStar} & \textit{RXTE} & Conversion \\
        \vspace*{+0.1cm}
        Energy Band & Energy Band & Factor \\
        \hline
        \vspace*{+0.1cm}
        3 - 4 keV & 2 - 3.5 keV & 3.485 \\
        \vspace*{+0.1cm}
        4 - 6 keV & 3.5 - 6 keV & 4.408 \\
        \vspace*{+0.1cm}
        6 - 10 keV & 6 - 9.7 keV  & 3.083 \\
        \vspace*{+0.1cm}
        10 - 20 keV & 9.7 - 16 keV & 2.912 \\
        \vspace*{+0.1cm}
        3 - 78 keV & 2 - 16 keV & 2.315 \\
        \end{tabular}
        \caption{\label{tb:nustar_conversions} Energy conversion factors required to convert \textit{NuSTAR} count rates in the original energy band to \textit{RXTE} count rates in specified energy band. The conversions are performed assuming the spectral model in Table \ref{tb:nustar}.}
    \end{table}

\begin{table}
    \setlength{\tabcolsep}{10pt}
    \begin{tabular}{c c c}
    \textit{MAXI} Time  & Spectral Model & Conversion \\
    \vspace*{+0.1cm}
    Range $[\rm{MJD}]$ & ObsID, Table & Factor \\
    \hline
    \vspace*{+0.1cm}
    60304.0 - 60407.9 & 8050030119, \ref{tb:nicer805} & $4.920 \times 10^{-10}$ \\
    \vspace*{+0.1cm}
    60408.0 - 60457.9 & 7050030101, \ref{tb:nicer705} & $1.051 \times 10^{-8}$ \\
    \vspace*{+0.1cm}
    60470.0 - 60477.9 & 7050030104, \ref{tb:nicer705} & $9.605 \times 10^{-9}$ \\
    \vspace*{+0.1cm}
    60478.0 - 60487.9 & 7050030113, \ref{tb:nicer705} & $1.040 \times 10^{-8}$ \\
    \vspace*{0.1cm}
    60488.0 - 60497.9 & 0935191701, \ref{tb:xmm} & $1.044 \times 10^{-8}$ \\
    \vspace*{0.1cm}
    60498.0 --	60507.9 & 7050030126, \ref{tb:nicer705} & $9.531 \times 10^{-9}$ \\
    \vspace*{0.1cm}
    60513.0	-- 60507.9 & 7050030128, \ref{tb:nicer705} & $1.031 \times 10^{-8}$ \\
    \vspace*{0.1cm}
    60518.0 --	60567.9 & 7050030139, \ref{tb:nicer705} & $1.047 \times 10^{-8}$ \\
    \vspace*{+0.1cm}
    60568.0	-- 60597.9 & 7050030141, \ref{tb:nicer705} & $1.067 \times 10^{-8}$ \\
    \vspace*{+0.1cm}
    60598.0	-- 60607.9  & 7050030143, \ref{tb:nicer705} & $9.652 \times 10^{-9}$ \\
    \vspace*{+0.1cm}
    60619.0	-- 60667.9 & 7050030144, \ref{tb:nicer705} & $2.237 \times 10^{-8}$ \\
    \vspace*{+0.1cm}
    60668.0	-- 60707.9 & 7050030145, \ref{tb:nicer705} & $1.061 \times 10^{-8}$ \\
    \vspace*{+0.1cm}
    60708.0	-- 60717.9 & 00031272062, \ref{tb:swift-xrt} & $9.674 \times 10^{-9}$ \\
    \vspace*{+0.1cm}
    60718.0	-- 60727.9 & 7050030148, \ref{tb:nicer705} & $1.117 \times 10^{-8}$ \\
    \vspace*{+0.1cm}
    60728.0	-- 60737.9 & 8050030101, \ref{tb:nicer805} & $1.033 \times 10^{-8}$ \\
    \vspace*{+0.1cm}
    60738.0	-- 60747.9 & 8050030107, \ref{tb:nicer805} & $1.207 \times 10^{-8}$ \\
    \vspace*{+0.1cm}
    60758.0	-- 60767.9 & 805003017, \ref{tb:nicer805} & $9.614 \times 10^{-9}$ \\
    \vspace*{+0.1cm}
    60768.0	-- 60817.9  & 8050030119, \ref{tb:nicer805} & $4.920 \times 10^{-10}$ \\
    \hline
    \end{tabular}
    \caption{\label{tb:maxi_conversion} Energy conversion factors required to translate \textit{MAXI} GSC $2 - 20$ keV count rates in the original energy band into a $0.5 - 10$ keV flux. The conversions are performed assuming the spectral model of the closest in time pointed observation, as detailed in column three. \textit{NICER} ObsID 8050030119 is utilised for all quiescent time bins.}
    \end{table}

    \begin{table}
    \centering
        \setlength{\tabcolsep}{10pt}
        \begin{tabular}{c c c}
        \textit{Swift}-BAT Time & Spectral Model & Conversion \\
        \vspace*{+0.1cm}
        Range $[\rm{MJD}]$ & ObsID, Table & Factor \\
        \hline
        \vspace*{+0.1cm}
        60300 -- 60419.9 & 8050030119, \ref{tb:nicer805} & $5.570 \times 10^{-9}$ \\
        \vspace*{+0.1cm}
        60420 -- 60499.9 & 91001324001, \ref{tb:nustar} & $3.879 \times 10^{-8}$ \\
        \vspace*{+0.1cm}
        60500 -- 60609.9 & 0935191701, \ref{tb:xmm} & $6.965 \times 10^{-8}$ \\
        \vspace*{+0.1cm}
        60610 -- 60689.9 & 7050030146, \ref{tb:nicer705} & $9.591 \times 10^{-8}$ \\
        \vspace*{+0.1cm}
        60690 -- 60769.9 & 8050030112, \ref{tb:nicer805_powerlaw} & $7.106 \times 10^{-8}$ \\
        \vspace*{+0.1cm}
        60730 -- 60769.9 & 8050030110, \ref{tb:nicer705} & $ 5.256 \times 10^{-8}$ \\
        \vspace*{+0.1cm}
        60770 -- 60819.9 & 8050030119, \ref{tb:nicer805} & $5.570 \times 10^{-9}$ \\
        \hline
        \end{tabular}
        \caption{\label{tb:swiftbat_conversion} Energy conversion factors required to convert $15-50$ keV \textit{Swift-BAT} count rates into a $0.5 - 10$ keV flux. The conversions are performed assuming the spectral model of the closest in time pointed observation in which the high-energy/power law component dominates, as detailed in the table. \textit{NICER} ObsID 8050030119 is utilised for all quiescent time bins. }
    \end{table}
    
\begin{figure}
    \centering
    \includegraphics[width=\columnwidth]{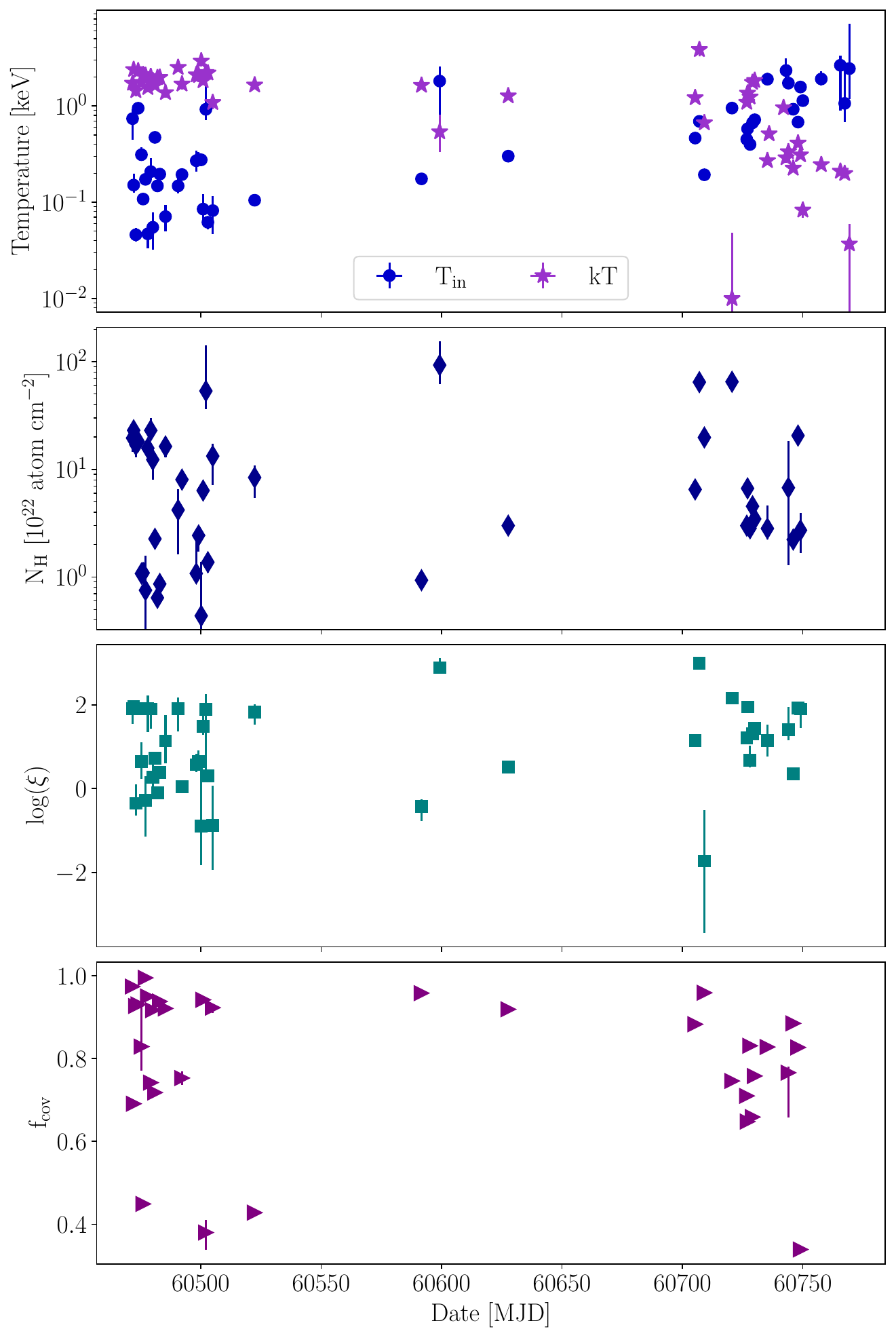}
    \vspace*{-0.5cm}
    \caption{Parameter variations from the \texttt{zxipcf}, \texttt{diskbb} and \texttt{bbody} components used to fit the 0.5 - 10 keV \textit{NICER} spectra in this paper. The corresponding fits appear in Tables \ref{tb:nicer705} - \ref{tb:nicer805_powerlaw}.}
    \label{fig:specvar}
\end{figure}

\begin{figure}
    \centering
    \includegraphics[width=\columnwidth]{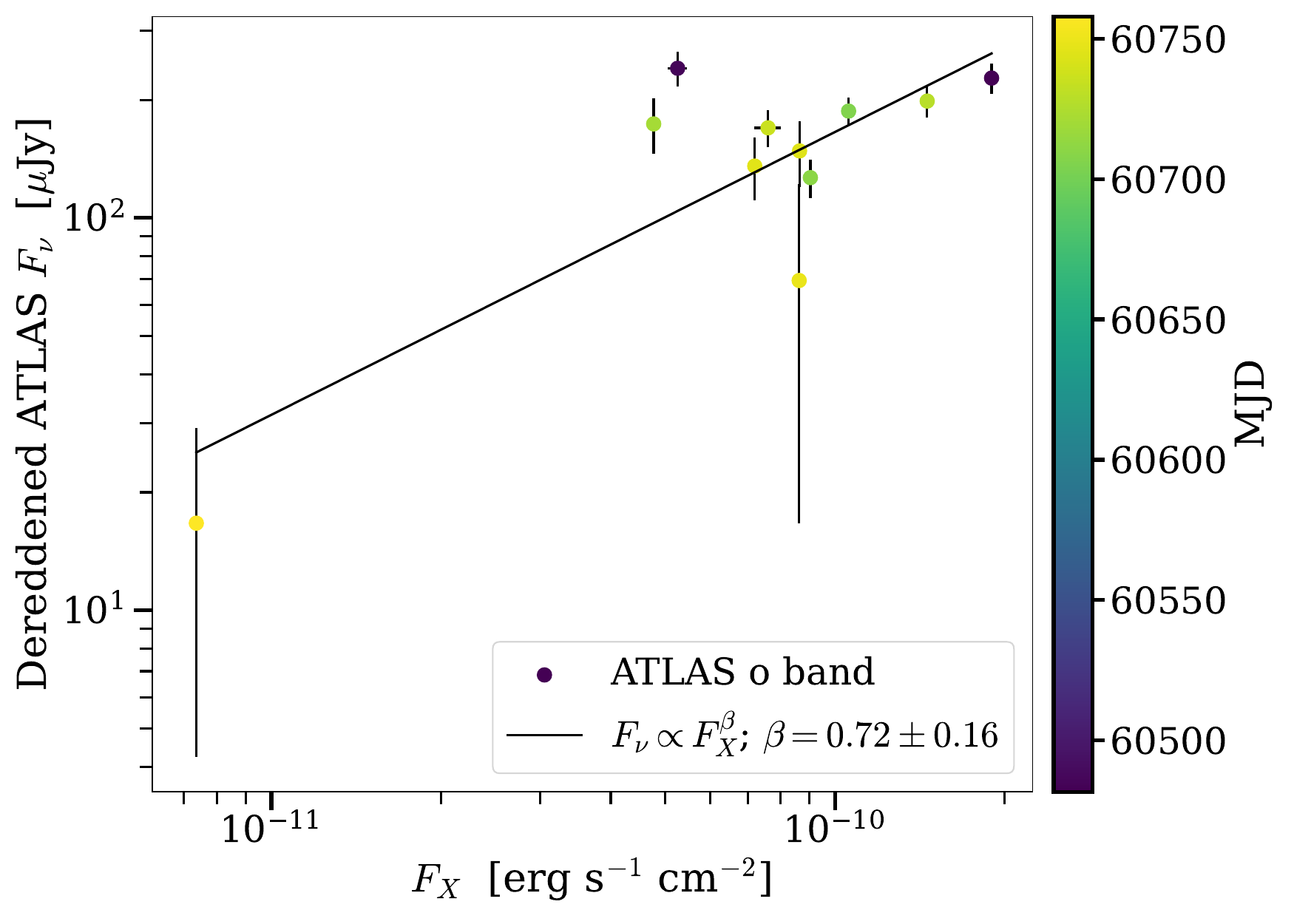}
    \vspace*{-0.5cm}
    \caption{Optical-X-ray correlation of EXO 0748--676 compiled using the modelled X-ray fluxes from \textit{Swift}-XRT and optical flux densities from ATLAS $o$-band during quasi-simultaneous observations (within one day). The fitted relation between the optical and X-ray is shown as a solid line.}
    \label{fig:Lopt-Lx}
\end{figure}

\begin{table*}
\setlength{\tabcolsep}{3.5pt}
\begin{tabular}{c c c c c c c c c c c c}
ObsID & Flux & Nh & $\log({\xi})$ & f$_{\rm{cov}}$ & T$_{\rm{in}}$ & kT & Ratio & $\chi$ & $\nu$ & $p$ \\
 & [$\times 10^{-10}$ erg cm$^{-2}$ s$^{-1}$] & [10$^{22}$ atom cm$^{-2}$] & & & [keV] & [keV]  \\
\hline
\vspace*{+0.05cm}
7050030101 &$0.580 \pm 0.013$ & $19.53 \pm^{1.23}_{5.02}$ & $1.91 \pm^{0.02}_{0.35}$ & $0.974 \pm^{0.001}_{0.058}$ & $0.739 \pm^{0.189}_{0.291}$ & $1.723 \pm^{0.359}_{0.181}$ & 0.599 & 88.68 & 100 & 0.784 \\
\vspace*{+0.1cm}
7050030102 & $0.567 \pm 0.007$ & $23.05 \pm^{3.01}_{2.61}$ & $1.96 \pm^{0.08}_{0.14}$ & $0.691 \pm^{0.038}_{0.039}$ & $0.151 \pm^{0.047}_{0.026}$ & $2.391 \pm^{0.186}_{0.157}$ & 0.008 & 133.36 & 117 & 0.143 \\
\vspace*{+0.1cm}
7050030103 & $0.230 \pm 0.006$ & $16.95 \pm^{3.85}_{3.99}$ & $-0.35 \pm ^{0.46}_{0.30}$ & $0.927 \pm ^{0.028}_{0.052}$ & $0.046 \pm ^{0.008}_{0.007}$ & $1.443 \pm^{0.258}_{0.184}$ & 0.055 & 152.46 & 121 & 0.028 \\
\vspace*{+0.1cm} 
7050030104 & $0.547 \pm 0.006$ & $18.16 \pm^{0.36}_{1.71}$ & $1.91 \pm^{0.16}_{0.01}$ & $0.931 \pm^{0.016}_{0.025}$ & $0.948 \pm ^{0.129}_{0.065}$ & $2.310 \pm^{0.102}_{0.207}$ & 0.464 & 131.73 & 109 & 0.068 \\
\vspace*{+0.1cm} 
7050030105 & $2.343 \pm 0.011$ & $1.077 \pm^{0.24}_{0.11}$ & $0.64 \pm^{0.47}_{0.41}$ & $0.819 \pm^{0.327}_{0.241}$ & $0.312 \pm^{0.061}_{0.035}$ & $1.758 \pm^{0.013}_{0.029}$ & 0.007 & 124.75 & 125 & 0.489 \\
\vspace*{+0.1cm} 
7050030106 & $1.460 \pm 0.016$ & $1.091 \pm^{0.28}_{0.15}$ & $\geq -2.22$ & $0.449 \pm^{0.069}_{0.028}$ & $0.108 \pm^{0.012}_{0.010}$ & $2.121 \pm^{0.088}_{0.120}$ & 0.010 & 134.54 & 109 & 0.049 \\
\vspace*{+0.1cm} 
7050030107 & $1.071 \pm 0.022$ & $0.753 \pm^{0.83}_{1.24}$ & $-0.28 \pm ^{0.58}_{0.87}$ & $0.995 \pm^{0.073}_{0.072}$ & $0.173 \pm^{0.031}_{0.016}$ & $2.077 \pm ^{0.189}_{0.150}$ & 0.122 & 87.98 & 96 & 0.708 \\
\vspace*{+0.1cm} 
7050030108 & $0.515 \pm 0.011$ & $15.74 \pm^{0.47}_{0.24}$ & $1.91 \pm^{0.31}_{0.56}$ & $0.950 \pm ^{0.047}_{0.049}$ & $0.047 \pm^{0.010}_{0.014}$ & $1.554 \pm^{0.200}_{0.170}$ & 0.021 & 134.24 & 121 & 0.194 \\
\vspace*{+0.1cm} 
7050030109 & $0.503 \pm 0.010$ & $23.01 \pm^{7.14}_{3.48}$ & $1.90 \pm^{0.09}_{0.47}$ & $0.742 \pm^{0.079}_{0.046}$ & $0.208 \pm^{0.078}_{0.035}$ & $2.020 \pm^{0.203}_{0.244}$ & 0.016 & 108.57 & 109 & 0.494 \\
\vspace*{+0.1cm} 
7050030110 & $0.577 \pm 0.013$ & $12.30 \pm^{2.71}_{4.24}$ & $0.27 \pm^{0.46}_{0.28}$ & $0.917 \pm^{0.025}_{0.046}$ & $0.055 \pm^{0.024}_{0.023}$ & $1.709 \pm^{0.294}_{0.191}$ & 0.009 & 136.51 & 118 & 0.117 \\
\vspace*{+0.1cm} 
7050030111 & $1.738 \pm 0.007$ & $2.262 \pm^{0.11}_{0.14}$ & $0.72 \pm^{0.10}_{0.08}$ & $0.718 \pm^{0.016}_{0.020}$ & $0.472 \pm^{0.020}_{0.019}$ & $1.637 \pm^{0.013}_{0.024}$ & 0.207 & 119.71 & 129 & 0.709 \\
\vspace*{+0.1cm}
7050030112 & $1.086 \pm 0.006$ & $0.641 \pm ^{0.04}_{0.04}$ & $-0.10 \pm ^{0.09}_{0.06}$ & $\geq 0.993$ & $0.148 \pm ^{0.003}_{0.005}$ & $1.965 \pm ^{0.022}_{0.023}$ & 0.065 & 134.46 & 119 & 0.158 \\
\vspace*{+0.1cm} 
7050030113 & $1.149 \pm 0.007$ & $0.862 \pm ^{0.10}_{0.09}$ & $0.39 \pm ^{0.07}_{0.06}$ & $0.938 \pm ^{0.017}_{0.022}$ & $0.196 \pm ^{0.031}_{0.029}$ & $1.980 \pm ^{0.042}_{0.039}$ & 0.057 & 139.53 & 116 & 0.068 \\
\vspace*{+0.1cm} 
7050030114 & $0.524 \pm 0.021$ & $16.33 \pm ^{2.98}_{3.37}$ & $1.14 \pm ^{0.62}_{0.53}$ & $0.921 \pm ^{0.026}_{0.042}$ & $0.071 \pm ^{0.023}_{0.021}$ & $1.378 \pm ^{0.205}_{0.152}$ & 0.044 & 87.04 & 76 & 0.182 \\
\vspace*{+0.1cm} 
7050030115 & $1.084 \pm 0.016$ & $4.192 \pm ^{2.36}_{2.57}$ & $1.91 \pm ^{0.27}_{0.54}$ & $\leq 0.474$ & $0.148 \pm ^{0.048}_{0.025}$ & $2.513 \pm ^{0.173}_{0.157}$ & 0.006 & 99.45 & 99 & 0.468 \\
\vspace*{+0.1cm}
7050030121 & $0.902 \pm 0.031$ & $1.079 \pm ^{1.34}_{0.08}$ & $0.57 \pm^{0.26}_{0.17}$ & $ \geq 0.999$ & $0.270 \pm ^{0.074}_{0.062}$ & $2.069 \pm ^{0.151}_{0.313}$ & 0.143 & 75.17 & 78 & 0.570 \\
\vspace*{+0.1cm} 
7050030122 & $0.408 \pm 0.013$ & $2.434 \pm ^{0.43}_{0.71}$ & $0.64 \pm ^{0.27}_{0.17}$ & $\geq 0.961$ & $0.276 \pm^{0.030}_{0.039}$ & $2.929 \pm^{0.480}_{0.340}$ & 0.325 & 126.99 & 87 & 0.003 \\
\vspace*{+0.1cm} 
7050030123 & $0.432 \pm 0.019$ & $0.436 \pm ^{0.94}_{0.28}$ & $-0.89 \pm ^{1.68}_{0.94}$ & $\geq 0.627$ & $0.085 \pm^{0.037}_{0.026}$ & $1.830 \pm^{0.189}_{0.161}$ & 0.151 & 91.41 & 74 & 0.083 \\
\vspace*{+0.1cm} 
7050030124 & $1.168 \pm 0.019$ & $6.359 \pm ^{0.67}_{1.15}$ & $1.49 \pm ^{0.51}_{0.21}$ & $0.942 \pm ^{0.020}_{0.032}$ & $0.923 \pm ^{0.869}_{0.207}$ & $2.232 \pm^{0.448}_{0.268}$ & 0.471 & 101.12 & 90 & 0.199 \\
\vspace*{+0.1cm} 
7050030125 & $1.081 \pm 0.017$ & $53.46 \pm _{17.2}^{89.3}$ & $1.89 \pm ^{0.37}_{1.68}$ & $0.380 \pm ^{0.175}_{0.205}$ & $0.062 \pm ^{0.009}_{0.010}$ & $2.201 \pm ^{0.252}_{0.235}$ & 0.012 & 105.85 & 91 & 0.137 \\
\vspace*{+0.1cm} 
7050030126 & $0.759 \pm 0.019$ & $1.373 \pm ^{0.30}_{0.28}$ & $ 0.30 \pm ^{0.03}_{0.02}$ & $\geq 0.994$ & $0.116 \pm ^{0.013}_{0.009}$ & $2.584 \pm^{0.207}_{0.176}$ & 0.396 & 83.19 & 84 & 0.504 \\
\vspace*{+0.1cm} 
7050030128 & $0.498 \pm 0.021$ & $13.28 \pm^{4.02}_{6.11}$ & $ -0.88 \pm^{0.95}_{1.05}$ & $0.923 \pm ^{0.042}_{0.117}$ & $0.082 \pm^{0.034}_{0.035}$ & $1.086 \pm^{0.236}_{0.158}$ & 0.040 & 103.69 & 74 & 0.013 \\
\vspace*{+0.1cm} 
7050030139 & $0.888 \pm 0.008$ & $8.399 \pm^{2.47}_{3.00}$ & $1.83 \pm^{0.19}_{0.30}$ & $0.428 \pm ^{0.079}_{0.052}$ & $0.105 \pm^{0.009}_{0.010}$ & $1.650 \pm^{0.061}_{0.081}$ & 0.012 & 1114.18 & 111 & 0.399 \\
\vspace*{+0.1cm} 
7050030141$^{a}$ & $11.17 \pm 0.091$ & $0.934 \pm^{0.08}_{0.13}$ & $-0.43 \pm ^{0.17}_{0.34}$ & $0.958 \pm ^{0.020}_{0.035}$ & $0.175 \pm^{0.021}_{0.022}$ & $1.693 \pm^{0.029}_{0.027}$ & 0.440 & 127.35 & 112 & 0.152 \\
\vspace*{+0.1cm} 
7050030143 & $1.853 \pm 0.069$ & $93.01 \pm^{62.2}_{31.2}$ & $2.89 \pm ^{0.22}_{0.11}$ & $ \geq 0.959$ & $1.813 \pm^{0.739}_{1.080}$ & $0.539 \pm^{0.267}_{0.208}$ & 0.297 & 72.08 & 75 & 0.574 \\
\vspace*{+0.1cm} 
7050030144 & $1.377 \pm 0.010$ & $3.002 \pm^{0.13}_{0.02}$ & $0.51 \pm ^{0.04}_{0.04}$ & $0.919 \pm ^{0.010}_{0.008}$ & $0.301 \pm^{0.007}_{0.007}$ & $1.273 \pm^{0.024}_{0.024}$ & 0.416 & 137.08 & 103 & 0.014 \\
\vspace*{+0.1cm} 
7050030145 & $0.865 \pm 0.005$ & $6.504 \pm^{0.33}_{0.44}$ & $1.15 \pm ^{0.08}_{0.04}$ & $0.883 \pm ^{0.022}_{0.035}$ & $0.464 \pm^{0.032}_{0.024}$ & $1.218 \pm^{0.059}_{0.047}$ & 0.332 & 114.04 & 106 & 0.280 \\
\vspace*{+0.1cm} 
7050030146 & $1.823 \pm 0.024$ & $64.61 \pm^{13.5}_{1.07}$ & $2.99 \pm ^{0.07}_{0.12}$ & $\geq 0.779$ & $0.692 \pm^{0.063}_{0.058}$ & $3.846 \pm^{0.794}_{0.571}$ & 0.184 & 151.87 & 127 & 0.066 \\
\vspace*{+0.1cm} 
7050030147 & $0.858 \pm 0.007$ & $19.79 \pm^{1.42}_{1.50}$ & $-1.73 \pm ^{1.22}_{1.72}$ & $0.959 \pm ^{0.009}_{0.013}$ & $0.193 \pm^{0.007}_{0.008}$ & $0.669 \pm^{0.020}_{0.027}$ & 0.343 & 112.04 & 103 & 0.255 \\
\vspace*{+0.1cm} 
7050030148 & $0.608 \pm 0.007$ & $65.11 \pm^{2.12}_{2.69}$ & $2.16 \pm ^{0.08}_{0.07}$ & $0.746 \pm ^{0.059}_{0.072}$ & $0.951 \pm^{0.093}_{0.074}$ & $0.010 \pm^{0.038}_{0.033}$ & 8.640 & 144.53 & 117 & 0.043 \\
\vspace*{+0.1cm} 
7050030149 & $1.111 \pm 0.004$ & $2.999 \pm^{0.50}_{0.55}$ & $1.21 \pm^{0.07}_{0.03}$ & $0.710 \pm ^{0.023}_{0.020}$ & $0.451 \pm^{0.015}_{0.007}$ & $1.091 \pm^{0.013}_{0.013}$ & 0.689 & 100.87 & 124 & 0.937 \\
\vspace*{+0.1cm} 
7050030150 & $0.882 \pm 0.005$ & $6.660 \pm^{0.91}_{1.00}$ & $1.95 \pm^{0.03}_{0.04}$ & $0.648 \pm ^{0.019}_{0.022}$ & $0.579 \pm^{0.016}_{0.018}$ & $1.360 \pm^{0.060}_{0.054}$ & 0.700 & 132.26 & 108 & 0.056 \\
\vspace*{+0.1cm} 
7050030151 & $1.371 \pm 0.011$ & $2.834 \pm^{0.41}_{0.14}$ & $0.68 \pm^{0.35}_{0.17}$ & $0.831 \pm ^{0.040}_{0.061}$ & $0.399 \pm^{0.021}_{0.024}$ & $1.240 \pm^{0.055}_{0.054}$ & 0.469 & 113.39 & 101 & 0.188 \\
\vspace*{+0.1cm} 
7050030152 & $1.464 \pm 0.005$ & $4.541 \pm^{0.49}_{0.62}$ & $1.31 \pm^{0.10}_{0.08}$ & $0.659 \pm ^{0.044}_{0.043}$ & $0.669 \pm^{0.028}_{0.028}$ & $1.759 \pm^{0.066}_{0.063}$ & 1.533 & 115.64 & 119 & 0.570 \\
\vspace*{+0.1cm} 
7050030153 & $0.866 \pm 0.007$ & $3.463 \pm^{0.49}_{0.62}$ & $1.44 \pm^{0.10}_{0.08}$ & $0.758 \pm ^{0.044}_{0.043}$ & $0.719 \pm^{0.028}_{0.028}$ & $1.843 \pm^{0.066}_{0.063}$ & 1.345 & 121.09 & 103 & 0.108 \\
\hline
\end{tabular}
\caption{\label{tb:nicer705}Best-fitting spectral parameters and corresponding fit statistics when fitting the 0.5 - 10 keV \textit{NICER} spectra in \textsc{xspec} with a model of the form \texttt{tbabs*zxipcf(diskbb+bbody)}. In all cases, \texttt{tbabs} is fixed to a value of $0.116 \times 10^{22}$ atom cm$^{-2}$, as determined by the Colden Galactic Neutral Hydrogen Density Calculator. The ratio reported is that of the flux of the \texttt{diskbb} component over the flux of the \texttt{bbody} component. The presented X-ray flux is the 0.5 - 10 keV flux determined from the corresponding spectral model for each ObsID. Here, $\chi$, $\nu$ and $p$ are the chi-squared, number of degrees of freedom and null-hypothesis probability, respectively. \\
$^{a}$ This spectrum required additional components to achieve a statistically acceptable fit. These extra values are provided in Table \ref{tb:nicer705_diskline}.}
\end{table*}

\begin{table*}
    \setlength{\tabcolsep}{5pt}
    \begin{tabular}{c c c c c c}
    ObsID & Line & Line & $\sigma$ & Line  & $\sigma$ \\
     & [keV] & & [keV] & [keV] & [keV] \\
    \hline
    \vspace*{+0.1cm} 
    7050030141 & $2.373 \pm^{0.029}_{0.027}$ & $1.488 \pm^{0.013}_{0.014}$ & $0.052 \pm^{0.028}_{0.025}$ & $7.837 \pm^{0.181}_{0.165}$ & $1.734 \pm^{0.311}_{0.262}$ \\
    \end{tabular}
    \caption{\label{tb:nicer705_diskline} Additional best-fitting spectral parameters required to fit the $0.5 -10$ keV spectra of \textit{NICER} 7050030141. The first line energy results from a \texttt{diskline} component. The second and third are both modelled as Gaussian emission lines. The final \textsc{xspec} model for this observation is \texttt{tbabs(zxipcf(diskbb+diskline+bbody+gaussian+gaussian))}}
\end{table*}

\begin{table*}
\setlength{\tabcolsep}{4pt}
\begin{tabular}{c c c c c c c c c c c c}
ObsID & Flux & Nh & $\log({\xi})$ & f$_{\rm{cov}}$ & T$_{\rm{in}}$ & kT & Ratio & $\chi$ & $\nu$ & $p$ \\
 & [$\times 10^{-10}$ erg cm$^{-2}$ s$^{-1}$] & [10$^{22}$ atom cm$^{-2}$] & & & [keV] & [keV]  \\
\hline
\vspace*{+0.1cm} 
8050030101 & $1.326 \pm 0.027$ & $2.822 \pm^{1.81}_{0.59}$ & $1.15 \pm^{0.38}_{0.38}$ & $0.828 \pm^{0.054}_{0.071}$ & $1.896 \pm^{0.316}_{0.233}$ & $0.271 \pm^{0.032}_{0.018}$ & 1.003 & 71.99 & 82 & 0.777 \\
\vspace*{+0.1cm} 
8050030102 & $0.615 \pm 0.036$ & -- & -- & -- & -- & $0.514 \pm^{0.033}_{0.031}$ & -- & 41.83 & 75 & 0.999 \\
\vspace*{+0.1cm} 
8050030107 & $0.549 \pm 0.007$ & -- & -- & -- & -- & $0.960 \pm^{0.026}_{0.025}$ & -- & 103.06 & 102 & 0.452 \\
\vspace*{+0.1cm} 
8050030108 & $1.010 \pm 0.025$ & -- & -- & -- & $2.335 \pm^{0.767}_{0.449}$ & $0.290 \pm^{0.029}_{0.034}$ & 5.847 & 108.95 & 87 & 0.056 \\
\vspace*{+0.1cm} 
8050030109 & $0.875 \pm 0.012$ & $6.749 \pm^{11.5}_{5.46}$ & $1.41 \pm^{0.54}_{0.26}$ & $0.766 \pm^{0.123}_{0.329}$ & $1.728 \pm ^{0.357}_{0.246}$ & $0.335 \pm^{0.026}_{0.056}$ & 0.842 & 89.98 & 92 & 0.540 \\
\vspace*{+0.1cm} 
8050030111 & $0.516 \pm 0.002$ & $2.228 \pm^{0.08}_{0.11}$ & $0.35 \pm^{0.04}_{0.03}$ & $0.885 \pm^{0.019}_{0.024}$ & $0.925 \pm ^{0.025}_{0.024}$ & $0.226 \pm^{0.007}_{0.006}$ & 1.733 & 168.95 & 134 & 0.022 \\
\vspace*{+0.1cm} 
8050030113 & $0.413 \pm 0.002$ & $20.60 \pm^{0.61}_{0.95}$ & $1.92 \pm^{0.01}_{0.01}$ & $0.827 \pm^{0.030}_{0.057}$ & $0.680 \pm ^{0.041}_{0.032}$ & $0.412 \pm^{0.044}_{0.030}$ & 2.838 & 141.30 & 136 & 0.360 \\
\vspace*{+0.1cm} 
8050030114 & $0.548 \pm 0.003$ & $2.720 \pm^{1.21}_{1.04}$ & $1.90 \pm^{0.13}_{0.45}$ & $0.399 \pm^{0.099}_{0.044}$ & $1.573 \pm ^{0.058}_{0.057}$ & $0.311 \pm^{0.005}_{0.008}$ & 2.003 & 171.50 & 111 & 0.002 \\
\vspace*{+0.1cm} 
8050030115 & $0.371 \pm 0.007$ & -- & -- & -- & $1.135 \pm ^{0.046}_{0.043}$ & $0.083 \pm^{0.016}_{0.014}$ & 25.68 & 90.55 & 85 & 0.320 \\
\vspace*{+0.1cm} 
8050030116 & $0.073 \pm 0.001$ & -- & -- & -- & $1.906 \pm ^{0.392}_{0.285}$ & $0.246 \pm^{0.007}_{0.007}$ & 1.291 & 105.00 & 89 & 0.118 \\
\vspace*{+0.1cm} 
8050030117 & $0.018 \pm 0.001$ & -- & -- & -- & $2.645 \pm ^{0.677}_{1.750}$ & $0.210 \pm^{0.016}_{0.022}$ & 1.289 & 87.79 & 96 & 0.713 \\
\vspace*{+0.1cm} 
8050030118 & $0.010 \pm 0.001$ & -- & -- & -- & $1.064 \pm ^{1.280}_{0.380}$ & $0.199 \pm^{0.013}_{0.014}$ & 0.674 & 93.61 & 94 & 0.492 \\
\vspace*{+0.1cm} 
8050030119 & $0.023 \pm 0.004$ & -- & -- & -- & $2.442 \pm ^{4.670}_{1.261}$ & $0.037 \pm^{0.023}_{0.033}$ & 7.060 & 82.77 & 70 & 0.141 \\
\hline
\end{tabular}
\caption{\label{tb:nicer805}Best-fitting spectral parameters and corresponding fit statistics when fitting the 0.5 - 10 keV spectra extracted from \textit{NICER} ObsIDs starting $805$ in \textsc{xspec} with a model of the form \texttt{tbabs*zxipcf(diskbb+bbody)}. Two ObsIDs statistically preferring an alterative model are shown in Table \ref{tb:nicer805_powerlaw}. In all cases, \texttt{tbabs} is fixed to a value of $0.0116 \times 10^{22}$ atom cm$^{-2}$, as determined by the Colden Galactic Neutral Hydrogen Density Calculator. The ratio reported is that of the flux of the \texttt{diskbb} component over the flux of the \texttt{bbody} component, determined with \texttt{cflux} within \textsc{xspec} where possible. Dashes indicate model parameters that were not required to achieve a statistically acceptable fit. The presented X-ray flux is the 0.5 - 10 keV flux determined from the corresponding spectral model for each ObsID. Here, $\chi$, $\nu$ and $p$ are the chi-squared, number of degrees of freedom and null-hypothesis probability, respectively.}
\end{table*}

\begin{table*}
    \setlength{\tabcolsep}{3.5pt}
    \begin{tabular}{c c c c c c c c c c c }
    ObsID & Flux & T$_{\rm{in}}$ & $\Gamma$ & Line & $\sigma$ & Ratio & $\chi$ & $\nu$ & $p$ \\
     & [$\times 10^{-10}$ erg cm$^{-2}$ s$^{-1}$] & [keV] & & [keV] & [keV] \\
    \hline
    \vspace*{+0.1cm} 
    8050030110 & $0.830 \pm 0.039$ & $0.903 \pm^{0.071}_{0.064}$ & $1.782 \pm ^{0.034}_{0.032}$ & $1.022 \pm^{0.007}_{0.007}$ & $0.128 \pm^{0.007}_{0.006}$ & 0.298 & 140.46 & 111 & 0.031 \\
    \vspace*{+0.1cm} 
    8050030112 & $2.048 \pm 0.014$ & $0.469 \pm^{0.042}_{0.041}$ & $0.797 \pm^{0.107}_{0.108}$ & $1.732 \pm^{0.009}_{0.009}$ & $0.074 \pm^{0.016}_{0.015}$ & 0.062 & 155.99 & 153 & 0.418 \\
    \end{tabular}
    \caption{\label{tb:nicer805_powerlaw} Best-fitting spectral parameters and corresponding fit statistics for the $0.5 -10$ keV spectral fit to \textit{NICER} ObsID 8050030110. The spectrum is fit in \textsc{xspec} with the model \texttt{tbabs(diskbb+powerlaw+gaussian)}. Here, \texttt{tbabs} is fixed to a value of $0.116 \times 10^{22}$ atom cm$^{-2}$, as determined by the Colden Galactic Neutral Hydrogen Density Calculator. The ratio reported is that of the flux of the \texttt{diskbb} component over the flux of the \texttt{bbody} component, determined with \texttt{cflux}. The presented X-ray flux is the 0.5 - 10 keV flux determined from the corresponding spectral model for each ObsID. Here, $\chi$, $\nu$ and $p$ are the chi-squared, number of degrees of freedom and null-hypothesis probability, respectively.}
    \end{table*}

\begin{table*}
    \setlength{\tabcolsep}{4pt}
    \begin{tabular}{c c c c c c c c c c c c}
    ObsID & Flux & Nh & $\log({\xi})$ & f$_{\rm{cov}}$ & T$_{\rm{in}}$ & kT & Ratio & $W$ & $\nu$ & $p$ \\
     & [$\times 10^{-10}$ erg cm$^{-2}$ s$^{-1}$] & [10$^{22}$ atom cm$^{-2}$] & & & [keV] & [keV]  \\
    \hline
    \vspace*{+0.1cm}
    00019755001 & $0.008~ \pm$ 0.001 & -- & -- & -- & -- & $0.257 \pm^{0.020}_{0.018}$ & -- & 50.25 & 678 & 1.000 \\
    \vspace*{+0.1cm}
    00019755002 & $0.007~ \pm$ 0.001 & -- & -- & -- & -- & $0.259 \pm^{0.019}_{0.018}$ & -- & 52.65 & 594 & 1.000 \\
    \vspace*{+0.1cm}
    00019755003 & $0.007~ \pm$ 0.001 & -- & -- & -- & -- & $0.244 \pm^{0.022}_{0.019}$ & -- & 74.27 & 398 & 1.000 \\
    \vspace*{+0.1cm} 
    00019755004 & $0.006~ \pm$ 0.001 & -- & -- & -- & -- & $0.244 \pm^{0.026}_{0.022}$ & -- & 49.95 & 705 & 1.000 \\
    \vspace*{+0.1cm} 
    00031272058 & $0.742 \pm 0.041$ & -- & -- & -- & $0.131 \pm^{0.119}_{0.072}$ & $3.064 \pm^{0.521}_{0.383}$ & 0.008 & 236.76 & 284 & 0.194 \\
    \vspace*{+0.1cm} 
    00031272059 & $1.297 \pm 0.049$ & -- & -- & -- & $0.250 \pm^{0.369}_{0.134}$ & $2.992 \pm^{0.295}_{0.250}$ & 0.004 & 368.56 & 452 & 0.055 \\
    \vspace*{+0.1cm} 
    00031272060 & $0.574 \pm 0.037$ & $0.958 \pm ^{1.48}_{0.57}$ & $-0.544 \pm^{2.31}_{0.55}$ & $\geq 0.764$ & $0.064 \pm^{0.079}_{0.055}$ & $2.932 \pm ^{0.648}_{0.479}$ & 0.688 & 194.00 & 299 & 1.000 \\
    \vspace*{+0.1cm} 
    00031272061 & $0.642 \pm 0.075$ & -- & -- & -- & -- & $\leq 11.66$ & -- & 91.50 & 151 & 1.000 \\
    \vspace*{+0.1cm} 
    00031272062 & $1.136 \pm 0.032$ & $8.348 \pm^{12.4}_{0.96}$ & $1.93 \pm^{0.12}_{0.56}$ & $0.792 \pm^{0.108}_{0.046}$ & $0.588 \pm^{0.059}_{0.039}$ & $2.361 \pm^{1.117}_{0.376}$ & 1.522 & 423.49 & 467 & 0.015 \\
    \vspace*{+0.1cm} 
    00031272064 & $0.009 \pm 0.002$ & -- & -- & -- & $0.366 \pm^{0.070}_{0.054}$ & -- & -- & 33.59 & 729 & 1.000 \\
    \vspace*{+0.1cm}
    01236151000 & $1.416 \pm 0.038$ & -- & -- & -- & $0.079 \pm ^{0.038}_{0.030}$ & $2.876 \pm ^{0.196}_{0.162}$ & 0.002 & 569.43& 583 & 0.106 \\
    \end{tabular}
    \caption{\label{tb:swift-xrt}Best-fitting spectral parameters and corresponding fit statistics for the $0.5 -10$ keV spectra extracted from all \textit{Swift-XRT} observations during the 2024/25 outburst. All spectra are fit in \textsc{xspec} with the model of the form \texttt{tbabs*zxipcf(diskbb+bbody)}. In all cases, \texttt{tbabs} is fixed to a value of $0.0116 \times 10^{22}$ atom cm$^{-2}$, as determined by the Colden Galactic Neutral Hydrogen Density Calculator. Dashes indicate model parameters that were not required to achieve a statistically acceptable fit. The ratio reported is that of the flux of the \texttt{diskbb} component over the flux of the \texttt{bbody} component, determined with \texttt{cflux} where applicable. The presented X-ray flux is the 0.5 - 10 keV flux determined from the corresponding spectral model for each ObsID. The W-Statistic (from c-stat fitting), number of degrees of freedom and null-hypothesis probability are $W$, $\nu$ and $p$, respectively.}
    \end{table*}

    \begin{table}
    \centering
        \setlength{\tabcolsep}{10pt}
        \begin{tabular}{c c c c}
        ObsID & \textit{Swift} XRT & \textit{RXTE} & Conversion \\
        \vspace*{+0.1cm}
        & Energy Band & Energy Band & Factor \\
        \hline
        & 2 - 3.5 keV & 2 - 3.5 keV & 2.611 \\
        & 3.5 - 6 keV & 3.5 - 6 keV & 10.28 \\
        00031272058 & 6 - 8 keV & 6 - 9.7 keV & 43.67 \\
        & 7 - 10 keV & 9.7 - 16 keV & 56.31 \\
        \vspace*{+0.5cm}
        & 0.5 - 10 keV & 2 - 16 keV & 19.16 \\

        & 2 - 3.5 keV & 2 - 3.5 keV & 2.606 \\
        & 3.5 - 6 keV & 3.5 - 6 keV & 10.27 \\
        00031272059 & 6 - 8 keV & 6 - 9.7 keV  & 36.95 \\
        & 7 - 10 keV & 9.7 - 16 keV & 64.60 \\
        \vspace*{+0.5cm}
        & 0.5 - 10 keV & 2 - 16 keV & 18.68 \\

        & 2 - 3.5 keV & 2 - 3.5 keV & 2.712 \\
        & 3.5 - 6 keV & 3.5 - 6 keV & 10.27 \\
        00031272060 & 6 - 8 keV & 6 - 9.7 keV  & 45.70 \\
        & 7 - 10 keV & 9.7 - 16 keV & 64.70 \\
        \vspace*{+0.5cm}
        & 0.5 - 10 keV & 2 - 16 keV & 19.08 \\

        & 2 - 3.5 keV & 2 - 3.5 keV & 4.517 \\
        & 3.5 - 6 keV & 3.5 - 6 keV & 10.54 \\
        00031272061 & 6 - 8 keV & 6 - 9.7 keV  & 56.80 \\
        & 7 - 10 keV & 9.7 - 16 keV & 155.8 \\
        \vspace*{+0.5cm}
        & 0.5 - 10 keV & 2 - 16 keV & 44.02 \\

        & 2 - 3.5 keV & 2 - 3.5 keV & 2.530 \\
        & 3.5 - 6 keV & 3.5 - 6 keV & 8.800 \\
        00031272062 & 6 - 8 keV & 6 - 9.7 keV  & 22.80 \\
        & 7 - 10 keV & 9.7 - 16 keV & 25.10 \\
        \vspace*{+0.5cm}
        & 0.5 - 10 keV & 2 - 16 keV & 46.66 \\

        & 2 - 3.5 keV & 2 - 3.5 keV & 2.599 \\
        & 3.5 - 6 keV & 3.5 - 6 keV & 10.26 \\
        01236151000 & 6 - 8 keV & 6 - 9.7 keV  & 46.22 \\
        & 7 - 10 keV & 9.7 - 16 keV & 73.40 \\
        \vspace*{+0.5cm}
        & 0.5 - 10 keV & 2 - 16 keV & 17.91 \\
        
        \end{tabular}
        \caption{\label{tb:swiftxrt_conversions} Energy conversion factors required to convert \textit{Swift}-XRT count rates in the original energy band to \textit{RXTE} count rates in specified energy band. The conversions are performed assuming the spectral model in Table \ref{tb:swift-xrt}.}
    \end{table}
    
    \begin{table}
    \centering
        \setlength{\tabcolsep}{10pt}
        \begin{tabular}{c c c c}
        ObsID & \textit{NICER} & \textit{RXTE} & Conversion \\
        \vspace*{+0.1cm}
        & Energy Band & Energy Band & Factor \\
        \hline
        & 2 - 3.5 keV & 2 - 3.5 keV & 0.415 \\
        & 3.5 - 6 keV & 3.5 - 6 keV & 1.559 \\
        7050030101 & 6 - 9.7 keV & 6 - 9.7 keV & 5.953 \\
        & 9.7 - 12 keV & 9.7 - 16 keV & 27.23 \\
        \vspace*{+0.5cm}
        & 0.5 - 10 keV & 2 - 16 keV & 2.766 \\

        & 2 - 3.5 keV & 2 - 3.5 keV & 0.441 \\
        & 3.5 - 6 keV & 3.5 - 6 keV & 1.633 \\
        7050030102 & 6 - 9.7 keV & 6 - 9.7 keV  & 6.561 \\
        & 9.7 - 12 keV & 9.7 - 16 keV & 37.30 \\
        \vspace*{+0.5cm}
        & 0.5 - 10 keV & 2 - 16 keV & 4.274 \\
        \end{tabular}
        \caption{\label{tb:nicer_conversions} Energy conversion factors required to convert \textit{NICER} count rates in the original energy band to \textit{RXTE} count rates in specified energy band. The conversions are performed assuming the spectral models in Tables \ref{tb:nicer705} - \ref{tb:nicer805_powerlaw}}
    \end{table}

    \begin{table}
    \ContinuedFloat
    \centering
        \setlength{\tabcolsep}{10pt}
        \begin{tabular}{c c c c}
        ObsID & \textit{NICER} & \textit{RXTE} & Conversion \\
        \vspace*{+0.1cm}
        & Energy Band & Energy Band & Factor \\
        \hline
        & 2 - 3.5 keV & 2 - 3.5 keV & 0.392 \\
        & 3.5 - 6 keV & 3.5 - 6 keV & 1.505 \\
        7050030103 & 6 - 9 keV & 6 - 9.7 keV & 5.493 \\
        & 9 - 12 keV & 9.7 - 16 keV & 22.46 \\
        \vspace*{+0.5cm}
        & 0.5 - 10 keV & 2 - 16 keV & 2.131 \\

        & 2 - 3.5 keV & 2 - 3.5 keV & 0.413 \\
        & 3.5 - 6 keV & 3.5 - 6 keV & 1.595 \\
        7050030104 & 6 - 9 keV & 6 - 9.7 keV  & 6.483 \\
        & 9 - 12 keV & 9.7 - 16 keV & 35.72 \\
        \vspace*{+0.5cm}
        & 0.5 - 10 keV & 2 - 16 keV & 3.766 \\

        & 2 - 3.5 keV & 2 - 3.5 keV & 0.242 \\
        & 3.5 - 6 keV & 3.5 - 6 keV & 1.430 \\
        7050030105 & 6 - 9 keV & 6 - 9.7 keV  & 5.938 \\
        & 9 - 12 keV & 9.7 - 16 keV & 26.48 \\
        \vspace*{+0.5cm}
        & 0.5 - 10 keV & 2 - 16 keV & 1.234 \\
        
        & 2 - 3.5 keV & 2 - 3.5 keV & 0.249 \\
        & 3.5 - 6 keV & 3.5 - 6 keV & 1.460 \\
        7050030106 & 6 - 9 keV & 6 - 9.7 keV  & 6.273 \\
        & 9 - 12 keV & 9.7 - 16 keV & 31.59 \\
        \vspace*{+0.5cm}
        & 0.5 - 10 keV & 2 - 16 keV & 1.627 \\

        & 2 - 3.5 keV & 2 - 3.5 keV & 0.244 \\
        & 3.5 - 6 keV & 3.5 - 6 keV & 1.455 \\
        7050030107 & 6 - 9.7 keV & 6 - 9.7 keV  & 6.236 \\
        & 9.7 - 12 keV & 9.7 - 16 keV & 30.98 \\
        \vspace*{+0.5cm}
        & 0.5 - 10 keV & 2 - 16 keV & 1.500 \\

        & 2 - 3.5 keV & 2 - 3.5 keV & 0.385 \\
        & 3.5 - 6 keV & 3.5 - 6 keV & 1.512 \\
        7050030108 & 6 - 9.7 keV & 6 - 9.7 keV  & 5.739 \\
        & 9.7 - 12 keV & 9.7 - 16 keV & 24.24 \\
        \vspace*{+0.5cm}
        & 0.5 - 10 keV & 2 - 16 keV & 2.260 \\

        & 2 - 3.5 keV & 2 - 3.5 keV & 0.438 \\
        & 3.5 - 6 keV & 3.5 - 6 keV & 1.610 \\
        7050030109 & 6 - 9.7 keV & 6 - 9.7 keV  & 6.266 \\
        & 9.7 - 12 keV & 9.7 - 16 keV & 32.10 \\
        \vspace*{+0.5cm}
        & 0.5 - 10 keV & 2 - 16 keV & 3.537 \\

        & 2 - 3.5 keV & 2 - 3.5 keV & 0.361 \\
        & 3.5 - 6 keV & 3.5 - 6 keV & 1.506 \\
        7050030110 & 6 - 9.7 keV & 6 - 9.7 keV  & 5.916 \\
        & 9.7 - 12 keV & 9.7 - 16 keV & 26.51 \\
        \vspace*{+0.5cm}
        & 0.5 - 10 keV & 2 - 16 keV & 2.313 \\

        & 2 - 3.5 keV & 2 - 3.5 keV & 0.254 \\
        & 3.5 - 6 keV & 3.5 - 6 keV & 1.426 \\
        7050030111 & 6 - 9.7 keV & 6 - 9.7 keV  & 5.806 \\
        & 9.7 - 12 keV & 9.7 - 16 keV & 24.73 \\
        \vspace*{+0.5cm}
        & 0.5 - 10 keV & 2 - 16 keV & 1.298 \\

        & 2 - 3.5 keV & 2 - 3.5 keV & 0.241 \\
        & 3.5 - 6 keV & 3.5 - 6 keV & 1.445 \\
        7050030112 & 6 - 9.7 keV & 6 - 9.7 keV  & 6.139\\  
        & 9.7 - 12 keV & 9.7 - 16 keV & 29.43 \\
        \end{tabular}
        \caption{\label{tb:nicer_conversions} Continued.}
    \end{table}

    \begin{table}
    \ContinuedFloat
    \centering
        \setlength{\tabcolsep}{10pt}
        \begin{tabular}{c c c c}
        ObsID & \textit{NICER} & \textit{RXTE} & Conversion \\
        \vspace*{+0.1cm}
        & Energy Band & Energy Band & Factor \\
        \hline
        \vspace*{+0.5cm}
        & 0.5 - 10 keV & 2 - 16 keV & 1.353 \\    
        
        & 2 - 3.5 keV & 2 - 3.5 keV & 0.244 \\
        & 3.5 - 6 keV & 3.5 - 6 keV & 1.448 \\
        7050030113 & 6 - 9.7 keV & 6 - 9.7 keV  & 6.153 \\
        & 9.7 - 12 keV & 9.7 - 16 keV & 29.66 \\
        \vspace*{+0.5cm}
        & 0.5 - 10 keV & 2 - 16 keV & 1.424 \\

        & 2 - 3.5 keV & 2 - 3.5 keV & 0.385 \\
        & 3.5 - 6 keV & 3.5 - 6 keV & 1.490 \\
        7050030114 & 6 - 9.7 keV & 6 - 9.7 keV  & 5.499 \\
        & 9.7 - 12 keV & 9.7 - 16 keV & 21.32\\
        \vspace*{+0.5cm}
        & 0.5 - 10 keV & 2 - 16 keV & 1.991 \\

        & 2 - 3.5 keV & 2 - 3.5 keV & 0.291 \\
        & 3.5 - 6 keV & 3.5 - 6 keV & 1.505 \\
        7050030115 & 6 - 9.7 keV & 6 - 9.7 keV  & 6.560 \\
        & 9.7 - 12 keV & 9.7 - 16 keV & 36.80 \\
        \vspace*{+0.5cm}
        & 0.5 - 10 keV & 2 - 16 keV & 2.666 \\

        & 2 - 3.5 keV & 2 - 3.5 keV & 0.248 \\
        & 3.5 - 6 keV & 3.5 - 6 keV & 1.456 \\
        7050030121 & 6 - 9.7 keV & 6 - 9.7 keV  & 6.231 \\
        & 9.7 - 12 keV & 9.7 - 16 keV & 30.90 \\
        \vspace*{+0.5cm}
        & 0.5 - 10 keV & 2 - 16 keV & 1.571 \\

        & 2 - 3.5 keV & 2 - 3.5 keV & 0.274 \\
        & 3.5 - 6 keV & 3.5 - 6 keV & 1.508 \\
        7050030122 & 6 - 9.7 keV & 6 - 9.7 keV  & 6.772 \\
        & 9.7 - 12 keV & 9.7 - 16 keV & 41.11 \\
        \vspace*{+0.5cm}
        & 0.5 - 10 keV & 2 - 16 keV & 2.823 \\

        & 2 - 3.5 keV & 2 - 3.5 keV & 0.236 \\
        & 3.5 - 6 keV & 3.5 - 6 keV & 1.432 \\
        7050030123 & 6 - 9.7 keV & 6 - 9.7 keV  & 6.010 \\
        & 9.7 - 12 keV & 9.7 - 16 keV & 17.49 \\
        \vspace*{+0.5cm}
        & 0.5 - 10 keV & 2 - 16 keV & 1.160 \\

        & 2 - 3.5 keV & 2 - 3.5 keV & 0.312 \\
        & 3.5 - 6 keV & 3.5 - 6 keV & 1.506 \\
        7050030124 & 6 - 9.7 keV & 6 - 9.7 keV  & 6.379 \\
        & 9.7 - 12 keV & 9.7 - 16 keV & 33.54 \\
        \vspace*{+0.5cm}
        & 0.5 - 10 keV & 2 - 16 keV & 2.587 \\

        & 2 - 3.5 keV & 2 - 3.5 keV & 0.536 \\
        & 3.5 - 6 keV & 3.5 - 6 keV & 1.809 \\
        7050030125 & 6 - 9.7 keV & 6 - 9.7 keV  & 6.543 \\
        & 9.7 - 12 keV & 9.7 - 16 keV & 37.85 \\
        \vspace*{+0.5cm}
        & 0.5 - 10 keV & 2 - 16 keV & 5.635 \\

        & 2 - 3.5 keV & 2 - 3.5 keV & 0.258 \\
        & 3.5 - 6 keV & 3.5 - 6 keV & 1.487 \\
        7050030126 & 6 - 9.7 keV & 6 - 9.7 keV  & 6.590 \\
        & 9.7 - 12 keV & 9.7 - 16 keV & 37.32 \\
        \vspace*{+0.5cm}
        & 0.5 - 10 keV & 2 - 16 keV & 2.199 \\

        & 2 - 3.5 keV & 2 - 3.5 keV & 0.349 \\
        & 3.5 - 6 keV & 3.5 - 6 keV & 1.405 \\
        7050030128 & 6 - 9.7 keV & 6 - 9.7 keV  & 5.017 \\
        
        \end{tabular}
        \caption{\label{tb:nicer_conversions} Continued.}
    \end{table}

    \begin{table}
    \ContinuedFloat
    \centering
        \setlength{\tabcolsep}{10pt}
        \begin{tabular}{c c c c}
        ObsID & \textit{NICER} & \textit{RXTE} & Conversion \\
        \vspace*{+0.1cm}
        & Energy Band & Energy Band & Factor \\
        \hline
        & 9.7 - 12 keV & 9.7 - 16 keV & 16.11 \\
        \vspace*{+0.5cm}
        & 0.5 - 10 keV & 2 - 16 keV & 1.387 \\

        & 2 - 3.5 keV & 2 - 3.5 keV & 0.323 \\
        & 3.5 - 6 keV & 3.5 - 6 keV & 1.471 \\
        7050030139 & 6 - 9.7 keV & 6 - 9.7 keV  & 5.838 \\
        & 9.7 - 12 keV & 9.7 - 16 keV & 25.32 \\
        \vspace*{+0.5cm}
        & 0.5 - 10 keV & 2 - 16 keV & 1.938 \\

        & 2 - 3.5 keV & 2 - 3.5 keV & 0.239 \\
        & 3.5 - 6 keV & 3.5 - 6 keV & 1.422 \\
        7050030141 & 6 - 9.7 keV & 6 - 9.7 keV  & 5.867 \\
        & 9.7 - 12 keV & 9.7 - 16 keV & 25.49 \\
        \vspace*{+0.5cm}
        & 0.5 - 10 keV & 2 - 16 keV & 1.136 \\

        & 2 - 3.5 keV & 2 - 3.5 keV & 0.569 \\
        & 3.5 - 6 keV & 3.5 - 6 keV & 1.602 \\
        7050030143 & 6 - 9.7 keV & 6 - 9.7 keV  & 3.931 \\
        & 9.7 - 12 keV & 9.7 - 16 keV & 6.982 \\
        \vspace*{+0.5cm}
        & 0.5 - 10 keV & 2 - 16 keV & 1.187 \\

        & 2 - 3.5 keV & 2 - 3.5 keV & 0.250 \\
        & 3.5 - 6 keV & 3.5 - 6 keV & 1.376 \\
        7050030144 & 6 - 9.7 keV & 6 - 9.7 keV  & 5.316 \\
        & 9.7 - 12 keV & 9.7 - 16 keV & 15.10 \\
        \vspace*{+0.5cm}
        & 0.5 - 10 keV & 2 - 16 keV & 0.966 \\

        & 2 - 3.5 keV & 2 - 3.5 keV & 0.288 \\
        & 3.5 - 6 keV & 3.5 - 6 keV & 1.390 \\
        7050030145 & 6 - 9.7 keV & 6 - 9.7 keV  & 5.234 \\
        & 9.7 - 12 keV & 9.7 - 16 keV & 18.14 \\
        \vspace*{+0.5cm}
        & 0.5 - 10 keV & 2 - 16 keV & 1.197 \\

        & 2 - 3.5 keV & 2 - 3.5 keV & 0.558 \\
        & 3.5 - 6 keV & 3.5 - 6 keV & 1.912 \\
        7050030146 & 6 - 9.7 keV & 6 - 9.7 keV  & 7.469 \\
        & 9.7 - 12 keV & 9.7 - 16 keV & 59.46 \\
        \vspace*{+0.5cm}
        & 0.5 - 10 keV & 2 - 16 keV & 11.02 \\

        & 2 - 3.5 keV & 2 - 3.5 keV & 0.363 \\
        & 3.5 - 6 keV & 3.5 - 6 keV & 1.277 \\
        7050030147 & 6 - 9.7 keV & 6 - 9.7 keV  & 4.179 \\
        & 9.7 - 12 keV & 9.7 - 16 keV & 8.699 \\
        \vspace*{+0.5cm}
        & 0.5 - 10 keV & 2 - 16 keV & 0.970 \\

        & 2 - 3.5 keV & 2 - 3.5 keV & 0.545 \\
        & 3.5 - 6 keV & 3.5 - 6 keV & 1.695 \\
        7050030148 & 6 - 9.7 keV & 6 - 9.7 keV  & 4.815 \\
        & 9.7 - 12 keV & 9.7 - 16 keV & 15.30 \\
        \vspace*{+0.5cm}
        & 0.5 - 10 keV & 2 - 16 keV & 1.229 \\

        & 2 - 3.5 keV & 2 - 3.5 keV & 0.240 \\
        & 3.5 - 6 keV & 3.5 - 6 keV & 1.336 \\
        7050030149 & 6 - 9.7 keV & 6 - 9.7 keV  & 5.014 \\
        & 9.7 - 12 keV & 9.7 - 16 keV & 15.83 \\
        \vspace*{+0.5cm}
        & 0.5 - 10 keV & 2 - 16 keV & 0.760 \\

        & 2 - 3.5 keV & 2 - 3.5 keV & 0.296 \\
        & 3.5 - 6 keV & 3.5 - 6 keV & 1.418 \\
        
        \end{tabular}
        \caption{\label{tb:nicer_conversions} Continued.}
    \end{table}

\begin{table}
\ContinuedFloat
\centering
    \setlength{\tabcolsep}{10pt}
    \begin{tabular}{c c c c}
    ObsID & \textit{NICER} & \textit{RXTE} & Conversion \\
    \vspace*{+0.1cm}
    & Energy Band & Energy Band & Factor \\
    \hline
        7050030150 & 6 - 9.7 keV & 6 - 9.7 keV  & 5.453 \\
        & 9.7 - 12 keV & 9.7 - 16 keV & 20.53 \\
        \vspace*{+0.5cm}
        & 0.5 - 10 keV & 2 - 16 keV & 1.398 \\

        & 2 - 3.5 keV & 2 - 3.5 keV & 0.247 \\
        & 3.5 - 6 keV & 3.5 - 6 keV & 1.369 \\
        7050030151 & 6 - 9.7 keV & 6 - 9.7 keV  & 5.264 \\
        & 9.7 - 12 keV & 9.7 - 16 keV & 18.35 \\
        \vspace*{+0.5cm}
        & 0.5 - 10 keV & 2 - 16 keV & 0.911 \\

        & 2 - 3.5 keV & 2 - 3.5 keV & 0.212 \\
        & 3.5 - 6 keV & 3.5 - 6 keV & 1.188 \\
        7050030152 & 6 - 9.7 keV & 6 - 9.7 keV  & 4.172 \\
        & 9.7 - 12 keV & 9.7 - 16 keV & 8.451 \\
        \vspace*{+0.5cm}
        & 0.5 - 10 keV & 2 - 16 keV & 0.397 \\

        & 2 - 3.5 keV & 2 - 3.5 keV & 0.208 \\
        & 3.5 - 6 keV & 3.5 - 6 keV & 1.207 \\
        7050030153 & 6 - 9.7 keV & 6 - 9.7 keV  & 4.280 \\
        & 9.7 - 12 keV & 9.7 - 16 keV & 9.330 \\
        \vspace*{+0.5cm}
        & 0.5 - 10 keV & 2 - 16 keV & 0.391 \\

        & 2 - 3.5 keV & 2 - 3.5 keV & 0.266 \\
        & 3.5 - 6 keV & 3.5 - 6 keV & 1.456 \\
        8050030101 & 6 - 9.7 keV & 6 - 9.7 keV  & 6.082 \\
        & 9.7 - 12 keV & 9.7 - 16 keV & 28.62 \\
        \vspace*{+0.5cm}
        & 0.5 - 10 keV & 2 - 16 keV & 1.688 \\

        & 2 - 3.5 keV & 2 - 3.5 keV & 0.129 \\
        & 3.5 - 6 keV & 3.5 - 6 keV & 1.074 \\
        8050030102 & 6 - 9.7 keV & 6 - 9.7 keV  & 3.837 \\
        & 9.7 - 12 keV & 9.7 - 16 keV & 5.597 \\
        \vspace*{+0.5cm}
        & 0.5 - 10 keV & 2 - 16 keV & 0.046 \\

        & 2 - 3.5 keV & 2 - 3.5 keV & 0.194 \\
        & 3.5 - 6 keV & 3.5 - 6 keV & 1.279 \\
        8050030107 & 6 - 9.7 keV & 6 - 9.7 keV  & 4.769 \\
        & 9.7 - 12 keV & 9.7 - 16 keV & 13.49 \\
        \vspace*{+0.5cm}
        & 0.5 - 10 keV & 2 - 16 keV & 0.289 \\

        & 2 - 3.5 keV & 2 - 3.5 keV & 0.238 \\
        & 3.5 - 6 keV & 3.5 - 6 keV & 1.465 \\
        8050030108 & 6 - 9.7 keV & 6 - 9.7 keV  & 6.428 \\
        & 9.7 - 12 keV & 9.7 - 16 keV & 34.23 \\
        \vspace*{+0.5cm}
        & 0.5 - 10 keV & 2 - 16 keV & 1.496 \\

        & 2 - 3.5 keV & 2 - 3.5 keV & 0.128 \\
        & 3.5 - 6 keV & 3.5 - 6 keV & 0.963 \\
        8050030109 & 6 - 9.7 keV & 6 - 9.7 keV  & 3.474 \\
        & 9.7 - 12 keV & 9.7 - 16 keV & 2.478 \\
        \vspace*{+0.5cm}
        & 0.5 - 10 keV & 2 - 16 keV & 0.107 \\

        & 2 - 3.5 keV & 2 - 3.5 keV & 0.169 \\
        & 3.5 - 6 keV & 3.5 - 6 keV & 1.348 \\
        8050030110 & 6 - 9.7 keV & 6 - 9.7 keV  & 6.230 \\
        & 9.7 - 12 keV & 9.7 - 16 keV & 42.28 \\
        \vspace*{+0.5cm}
        & 0.5 - 10 keV & 2 - 16 keV & 0.199 \\

        & 2 - 3.5 keV & 2 - 3.5 keV & 0.218 \\
        \end{tabular}
        \caption{\label{tb:nicer_conversions} Continued.}
    \end{table}

\begin{table}
\ContinuedFloat
\centering
    \setlength{\tabcolsep}{10pt}
    \begin{tabular}{c c c c}
    ObsID & \textit{NICER} & \textit{RXTE} & Conversion \\
    \vspace*{+0.1cm}
    & Energy Band & Energy Band & Factor \\
    \hline
    & 3.5 - 6 keV & 3.5 - 6 keV & 1.282 \\
    8050030111 & 6 - 9.7 keV & 6 - 9.7 keV  & 4.703 \\
    & 9.7 - 12 keV & 9.7 - 16 keV & 12.94 \\
    \vspace*{+0.5cm}
    & 0.5 - 10 keV & 2 - 16 keV & 0.514 \\

    & 2 - 3.5 keV & 2 - 3.5 keV & 0.201 \\
    & 3.5 - 6 keV & 3.5 - 6 keV & 1.424 \\
    8050030112 & 6 - 9.7 keV & 6 - 9.7 keV  & 6.784 \\
    & 9.7 - 12 keV & 9.7 - 16 keV & 52.23 \\
    \vspace*{+0.5cm}
    & 0.5 - 10 keV & 2 - 16 keV & 0.877 \\

    & 2 - 3.5 keV & 2 - 3.5 keV & 0.371 \\
    & 3.5 - 6 keV & 3.5 - 6 keV & 1.288 \\
    8050030113 & 6 - 9.7 keV & 6 - 9.7 keV  & 4.203\\
    & 9.7 - 12 keV & 9.7 - 16 keV & 8.918 \\
    \vspace*{+0.5cm}
    & 0.5 - 10 keV & 2 - 16 keV & 1.010 \\

    & 2 - 3.5 keV & 2 - 3.5 keV & 0.252 \\
    & 3.5 - 6 keV & 3.5 - 6 keV & 1.418 \\
    8050030114 & 6 - 9.7 keV & 6 - 9.7 keV  & 5731 \\
    & 9.7 - 12 keV & 9.7 - 16 keV & 23.76 \\
    \vspace*{+0.5cm}
    & 0.5 - 10 keV & 2 - 16 keV & 1.225 \\

    & 2 - 3.5 keV & 2 - 3.5 keV & 0.206 \\
    & 3.5 - 6 keV & 3.5 - 6 keV & 1.326 \\
    8050030115 & 6 - 9.7 keV & 6 - 9.7 keV  & 5.086 \\
    & 9.7 - 12 keV & 9.7 - 16 keV & 16.48 \\
    \vspace*{+0.5cm}
    & 0.5 - 10 keV & 2 - 16 keV & 0.418 \\

    & 2 - 3.5 keV & 2 - 3.5 keV & 0.232 \\
    & 3.5 - 6 keV & 3.5 - 6 keV & 1.436 \\
    8050030116 & 6 - 9.7 keV & 6 - 9.7 keV  & 6.082 \\
    & 9.7 - 12 keV & 9.7 - 16 keV & 28.55 \\
    \vspace*{+0.5cm}
    & 0.5 - 10 keV & 2 - 16 keV & 1.093 \\

    & 2 - 3.5 keV & 2 - 3.5 keV & 0.242 \\
    & 3.5 - 6 keV & 3.5 - 6 keV & 1.480 \\
    8050030117 & 6 - 9.7 keV & 6 - 9.7 keV  & 6.618 \\
    & 9.7 - 12 keV & 9.7 - 16 keV & 37.85 \\
    \vspace*{+0.5cm}
    & 0.5 - 10 keV & 2 - 16 keV & 1.784 \\

    & 2 - 3.5 keV & 2 - 3.5 keV & 0.202 \\
    & 3.5 - 6 keV & 3.5 - 6 keV & 1.309 \\
    8050030118 & 6 - 9.7 keV & 6 - 9.7 keV  & 4.962 \\
    & 9.7 - 12 keV & 9.7 - 16 keV & 15.27 \\
    \vspace*{+0.5cm}
    & 0.5 - 10 keV & 2 - 16 keV & 0.364 \\

    & 2 - 3.5 keV & 2 - 3.5 keV & 0.240 \\
    & 3.5 - 6 keV & 3.5 - 6 keV & 1.470 \\
    8050030119 & 6 - 9.7 keV & 6 - 9.7 keV  & 6.498 \\
    & 9.7 - 12 keV & 9.7 - 16 keV & 35.53 \\
    \vspace*{+0.5cm}
    & 0.5 - 10 keV & 2 - 16 keV & 1.596 \\
    
    \end{tabular}
    \caption{\label{tb:nicer_conversions} Continued.}
\end{table}
        
\bsp	
\label{lastpage}
\end{document}